\documentclass[12pt]{article}
\pdfoutput=1
\usepackage{latexsym, graphicx} 
\usepackage{amsmath}
\usepackage{amssymb}
\usepackage[utf8]{inputenc}
\numberwithin{equation}{section}

\renewcommand{\title}[1]{\vbox{\center\bf{\Large{#1}}}\vspace{5mm}}
  \renewcommand{\author}[1]{\vbox{\center#1}\vspace{5mm}}
  \newcommand{\address}[1]{\vbox{\center\em#1}}
  \newcommand{\email}[1]{\vbox{\center\tt#1}\vspace{5mm}}

\newcommand{\cO}{\mathcal{O}}

\begin{document}
\begin{titlepage}
\begin{center}
\hfill \\
\hfill \\
\vskip 1cm
\title{Lyapunov growth in quantum spin chains}
\author {Ben~Craps,$^{1}$ Marine~De~Clerck,$^{1}$ Djunes~Janssens,$^{1}$ Vincent Luyten,$^{1}$
Charles Rabideau$^{1,2}$}
\address{\vspace{2mm}
$^{1}$Theoretische Natuurkunde, Vrije Universiteit Brussel (VUB) and\\
The International Solvay Institutes, Pleinlaan 2, B-1050 Brussels, Belgium \vspace{1mm}\\
$^{2}$David Rittenhouse Laboratory, University of Pennsylvania, \\ Philadelphia, PA 19104, USA \\}
\end{center}

\begin{abstract}
The Ising spin chain with longitudinal and transverse magnetic fields is often used in studies of quantum chaos, displaying both chaotic and integrable regions in its parameter space. However, even at a strongly chaotic point this model does not exhibit Lyapunov growth of the commutator squared of spin operators, as this observable saturates before exponential growth can manifest itself (even in situations where a spatial suppression factor makes the initial commutator small). We extend this model from the spin 1/2 Ising model to higher spins, demonstrate numerically that a window of exponential growth opens up for sufficiently large spin, and extract a quantity which corresponds to a notion of a Lyapunov exponent. In the classical infinite-spin limit, we identify and compute the appropriate classical analogue of the commutator squared, and show that the corresponding exponent agrees with the infinite-spin limit extracted from the quantum spin chain. 
\end{abstract}

\vfill
\email{Ben.Craps, Marine.Alexandra.De.Clerck, Djunes.Janssens, Vincent.Luyten, Charles.Rabideau@vub.be}
\end{titlepage}

\tableofcontents

\section{Introduction}

In classical physics, chaos is often characterised by an exponential sensitivity to initial conditions during an early time window, 
\begin{align}
\frac{\partial X_i(t)}{\partial X_j(0)} \sim e^{\lambda_L t}\,,  
\label{eqn:exponential_dependence}
\end{align}
where $X_i, X_j$ are generalized coordinates of the system.
In contrast, in quantum mechanics chaos is often characterised by the system obeying random matrix-like behaviour, which is related to late-time dynamics. 
This can manifest itself in the statistical properties of the spectrum of the Hamiltonian. 
For instance, level spacings of chaotic systems are expected to obey a Wigner-Dyson distribution whereas a generic integrable system will obey Poissonian statistics \cite{BGS1,BGS2,BT}. For a recent review, see \cite{1509.06411}.

Recently, interest in another probe of quantum chaos, the commutator squared \cite{Larkin69}
\begin{align}
C(t) = \langle [ W(t), V(0) ]^\dagger [ W(t), V(0) ] \rangle_\beta \equiv \langle |[ W(t), V(0) ]|^2 \rangle_\beta,
\end{align}
 has been revived \cite{1304.6483,1306.0622, 1312.3296,kitaev:lectures,BoundOnChaos}.\footnote{The commutator squared is closely related to another probe of chaos known as the Loschmidt echo, see \cite{review_Loschmidt} for a review. Loschmidt echoes in higher-spin generalisations of spin chains were studied in \cite{1409.4763}, were an exponential sensitivity compatible with a classical Lyapunov exponent was identified. This is closely related to and consistent with the results presented in this work.}
By writing \eqref{eqn:exponential_dependence} in terms of a Poisson bracket
\begin{align}
\frac{\partial X_i(t)}{\partial X_j(0)} = \{ X_i(t) , P_j(0) \}
\end{align}
we expect that for a chaotic system, at least in the semi-classical regime,
\begin{align}
\langle |[ X_i(t), P_j(0) ]|^2 \rangle_\beta \sim 
e^{2\lambda_L t} 
\end{align}
in an appropriate time window.

In \cite{1306.0622,1007.3957,1409.8180}, the commutator squared was studied in an Ising spin chain with longitudinal ($h_z$) and transverse ($h_x$) magnetic fields,
\begin{align}
H= \sum_n \left[ - S_z^{(n)} S_z^{(n+1)}  - h_x S_x^{(n)} - h_z S_z^{(n)} \right]\, ,
\label{Hamiltonian}
\end{align}
where $n$ labels the sites and e.g.\ $S_x^{(n)}$ denotes the first Pauli matrix at site $n$. This is known as the mixed field Ising model. This system is known to be integrable if either $h_x$ or $h_z$ vanishes and is known to display chaotic spectral statistics at $(h_x^*, h_z^*) \equiv (-1.05, 0.5)$ \cite{1007.3957}, which we will refer to as the ``strongly chaotic point''.
The commutator squared
\begin{align}
C(x,t) =  \left\langle \, \left| \left[S_{z}^{(1)}(t),S_{z}^{(1+x)}(0) \right] \right|^2 \right\rangle_{\beta = 0}
\end{align}
vanishes initially, but increases as the operator $S_z^{(1)}(t)$ ``grows'' (i.e., acts nontrivially on a growing number of sites). 
In order to display exponential growth in the commutator squared, we need a sufficiently long time window between its onset (after an initial dissipation time) and saturation (which must occur since the commutator squared is bounded), which in turn requires that the commutator squared includes a small prefactor. 
While the Hamiltonian does not contain a small parameter, the Lieb-Robinson bound ensures that the commutator squared is initially exponentially suppressed in the spatial distance between sites 1 and $n$, so one might have hoped to be able to display exponential growth by starting with sufficiently separated sites. 
However, this hope is not realized and no exponential growth is observed, even for well-separated operators \cite{1701.09147,Khemani:2018sdn,1805.05376}.

In order to find exponential growth, we increase the dimension of the local Hilbert space by associating to each site a spin $j$ representation of $SU(2)$ (where the original model would correspond to $j=1/2$). In this paper we study this system as we increase $j$ towards the large $j$ classical limit.

We start by studying this model at fixed $j$. We adopt a working definition of chaos based on the spectral statistics of the model and map out the ``phase diagram'' of integrability vs chaos as a function of the magnetic fields. 

Having established where our model is chaotic according to this working definition, we turn to the commutator squared.
At small $j$ we find that the commutator squared follows the first term of the Baker-Campbell-Hausdorff (BCH) expansion until it reaches the near saturation regime where it is well described by a ``diffusion'' type approach to saturation \cite{1805.05376}. As $j$ is increased, an exponential regime appears at intermediate times and persists for an increasing amount of time as $j$ is increased. An exponent that can be understood as a Lyapunov exponent is extracted and extrapolated to the infinite $j$ limit.

We then study the classical large $j$ limit of this model. We construct the quantity which corresponds to the classical limit of $C(x,t)$ \cite{1609.01707} and discuss how the growth of this quantity can be interpreted in terms of the divergence of classical trajectories so that it represents a notion of a classical Lyapunov growth. We finally compute the classical Lyapunov exponent defined in this way and find that it matches the infinite-spin quantum extrapolation within a range of magnetic fields where our model is chaotic according to our spectral statistics based definition of chaos.

The plan of our paper is as follows. In section \ref{section: review spin 1/2}, we describe the behaviour of the spectral statistics and the commutator squared of the mixed field Ising model at spin 1/2. In section \ref{sec:higher_spin}, we introduce the higher spin generalisation of this model and define the classical limit that we consider. We also study the classical limit of the commutator squared. In section \ref{sec:quantum}, we present the results of our numerical study of the mixed field Ising model at higher spin, including an analysis of the spectral statistics and the commutator squared. We describe the appearance of a regime of exponential growth and how we extracted an associated Lyapunov exponent. Finally, in section \ref{sec:classical}, we present the results of a numerical study of the classical limit of the model, describe how to extract the classical analogue of the Lyapunov exponent computed in the quantum model and compare the results of the classical analysis to the limit of the quantum analysis.

\section{Spin $1/2$ mixed field Ising model}
\label{section: review spin 1/2}
In this section, we will give a short review of the study of chaos in the Ising spin chain model with external longitudinal and transverse magnetic fields, where the Hamiltonian is given by \eqref{Hamiltonian}. We will focus on open boundary conditions unless otherwise specified.\footnote{Open boundary conditions correspond to the Hamiltonian \begin{align}
H= - \sum_{n=1}^{L-1} S_z^{(n)} S_z^{(n+1)}  - \sum_{n=1}^L \left[ h_x S_x^{(n)} + h_z S_z^{(n)} \right]\, .
\end{align}
}

\subsection{Spectral statistics} 
In this section we will review the map of where to find chaos in the parameter space $(h_x,h_z)$ of the mixed field Ising model. In the limits of $h_x=0$, $|h_x| \rightarrow \infty$ or $|h_z| \rightarrow \infty$ the model becomes trivially integrable as the Hamiltonian is diagonal in the individual spin basis. It is also known to be integrable for $h_z=0$. This can be understood via a Jordan-Wigner transform \cite{10.1017/CBO9780511973765} and is known to manifest itself in the spectral statistics for finite chains with both open and periodic boundary conditions \cite{1007.3957}. 

Away from these integrable lines, a fiducial parameter point exhibiting random matrix-like spectral statistics has been identified at $ h_x=-1.05$ and $h_z=0.5$ \cite{1007.3957}. We will refer to this parameter point as the ``strongly chaotic'' point. As we move towards the limits of large or small magnetic fields where the model is integrable the spectral statistics interpolates between strongly chaotic and integrable behaviour, so that the region where this model is strongly chaotic corresponds to intermediate values of the magnetic field.

As has been already alluded to, chaotic and integrable systems are differentiated by the distinct spectral statistics that they display. One diagnostic of this is the statistics of the spacing between consecutive energy eigenvalues. However there are two subtleties which must be taken into account. First, the influence of the model-dependent density of states must be removed by normalising the differences by the local density of states. Only those normalised fluctuations are conjectured to present universal features. This procedure is known as unfolding the spectrum. Second, if the Hamiltonian under consideration has symmetries, it first needs to be block-diagonalised according to its conserved charges. The unfolding has to be performed separately for the different blocks, because the eigenvalues in different blocks are uncorrelated \cite{1509.06411}. 

Once these have been dealt with we arrive at the definition of
quantum chaos that we will use in this work: that the differences between consecutive energy levels of the unfolded spectrum in each block obeys the applicable Wigner-Dyson distribution chosen depending on whether or not the Hamiltonian has a time reversal symmetry \cite{1509.06411}. Our model exhibits a time-reversal symmetry\footnote{This is most easily seen by noticing that in the usual representations of $SU(2)$ our Hamiltonian is real. This means that complex conjugation provides a time-reversal operation with $T^2=1$.} so that strongly chaotic points are those where the unfolded energy level spacings $\omega$ are well described by the Wigner surmise \cite{1509.06411}
\begin{align} 
P(\omega) = \frac{\pi \omega}{2}\, e^{-\pi \omega^2/4}.
\label{eqn:Wigner}
\end{align}

On the other hand, non-trivial integrable models are those where the unfolded energy level spacing follows Poisson statistics, 
\begin{align}
P(\omega) = e^{-\omega}.
\label{eqn:Poisson}
\end{align}

Some integrable models are trivial in the sense that they have regularly spaced energy levels with a high level of degeneracy. In such models, the unfolding procedure fails. The mixed field Ising model is trivial in this sense when $h_x=0$, $|h_x| \rightarrow \infty$ or $|h_z| \rightarrow \infty$ so that the Hamiltonian is diagonal in the spin basis. We will refer to both non-trivially integrable models with Poisson statistics and these trivially integrable models as integrable.

As a first step, one should therefore identify the symmetries of the Hamiltonian. For open boundary conditions, the Hamiltonian \eqref{Hamiltonian} has a reflection symmetry about the center of the chain for any value of the magnetic fields. For the special case of $h_z = 0$, we find an additional symmetry generator that flips every spin, $ \Pi_{n} S_{x}^{(n)}$.

Once the eigenvalues have been classified according to the conserved charges of the system, we unfold the different spectra using the method detailed in \cite{1808.09173}.
The idea is that the average spacing between eigenvalues is controlled by the local mean density of states: if there are $D(E) \delta E$ eigenvalues within a region $\delta E$ of the spectrum, then the average spacing will have to be ${1}/{D(E)}$. We are instead interested in the fluctuations around this average. If we rescale the difference between consecutive eigenvalues by the local mean density of states, the average difference will be one and we can study the shape of the distribution. In order to determine the local mean density of states we can compute the difference in energies between two states some distance apart in the spectrum. We should take this distance to scale like the total number of states to some power between 0 and 1, for example 1/2, so that in the large-system limit the window is at an intermediate scale and we obtain a well-defined notion of coarse-grained density of states. An explicit implementation of these ideas is described in \cite{1808.09173}. In the end, what we plot is the distribution of normalised spacings $\omega_i$, defined by
\begin{align}
\omega_i^{(raw)}& =  \frac{E_{i+1} - E_{i}}{ E_{i+\Delta} - E_{i-\Delta} }\,, \\ 
\bar \omega &= (N-10\Delta)^{-1} \sum_{i=5\Delta+1}^{N-5\Delta} \omega_i^{(raw)}   \,, \\
\omega_i &= \frac{\omega^{(raw)}_i}{\bar \omega} \,,
\end{align} 
for each $i= 5\Delta+1, \ldots, N- 5\Delta$, where a cut has been introduced at the edges of the spectrum to remove edge effects.\footnote{The unfolding is independent of the exact number multiplying $\Delta$, as long as it removes a non-negligible part of the spectrum at the edges.} $\Delta$ is the largest integer smaller than $\sqrt{N}$ and $N$ is the total number of states in a given block. 

\begin{figure}[t]
	\centering
	\begin{minipage}[b]{0.32\textwidth}
		\centering
		\includegraphics[width=.95\textwidth]{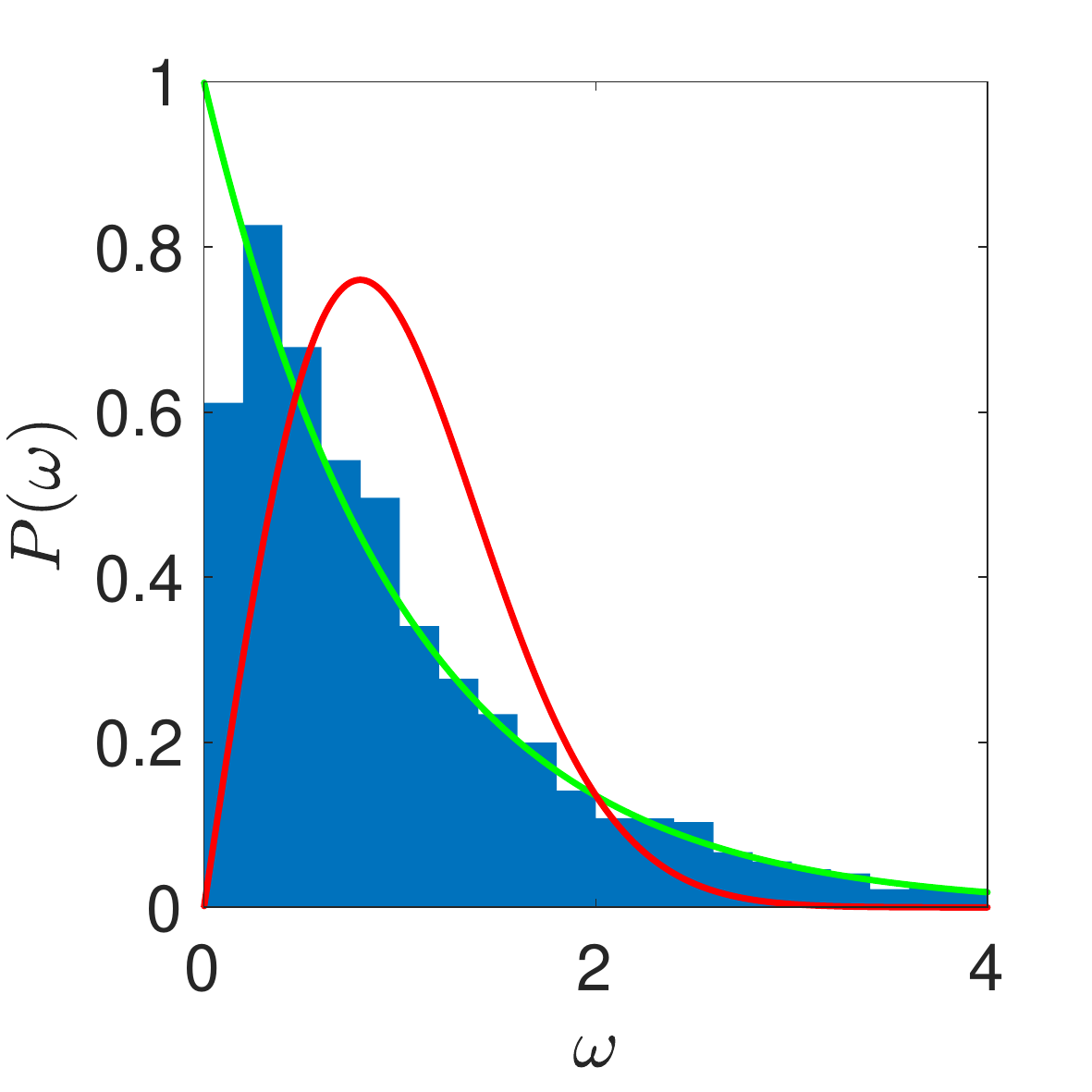}
		$(h_x,h_z)=(-1,0)$		
	\end{minipage}
	\begin{minipage}[b]{0.32\textwidth}
		\centering
		\includegraphics[width=.95\textwidth]{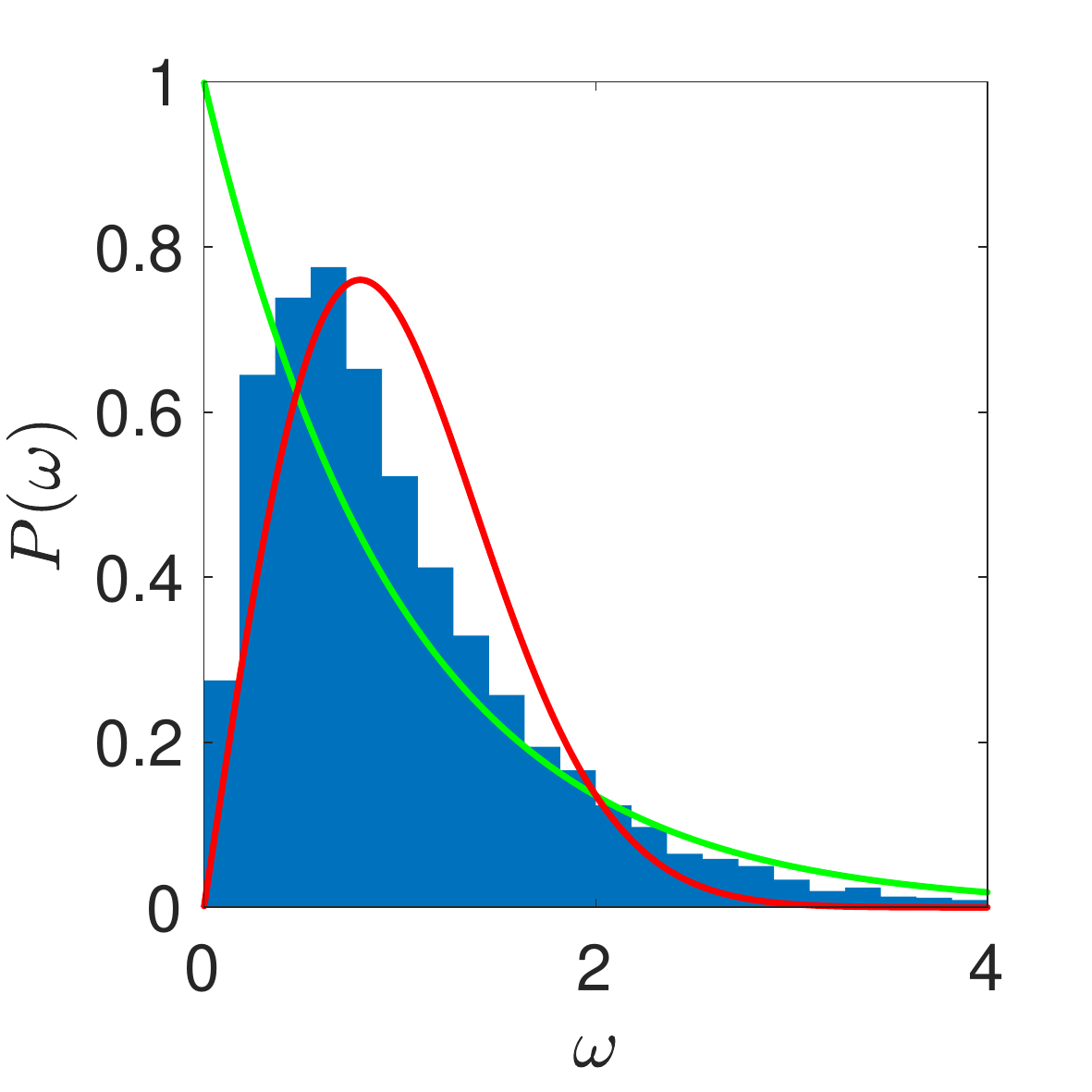}
		$(h_x,h_z)=(-1.002,0.02)$
	\end{minipage}
	\begin{minipage}[b]{0.32\textwidth}
		\centering
		\includegraphics[width=.95\textwidth]{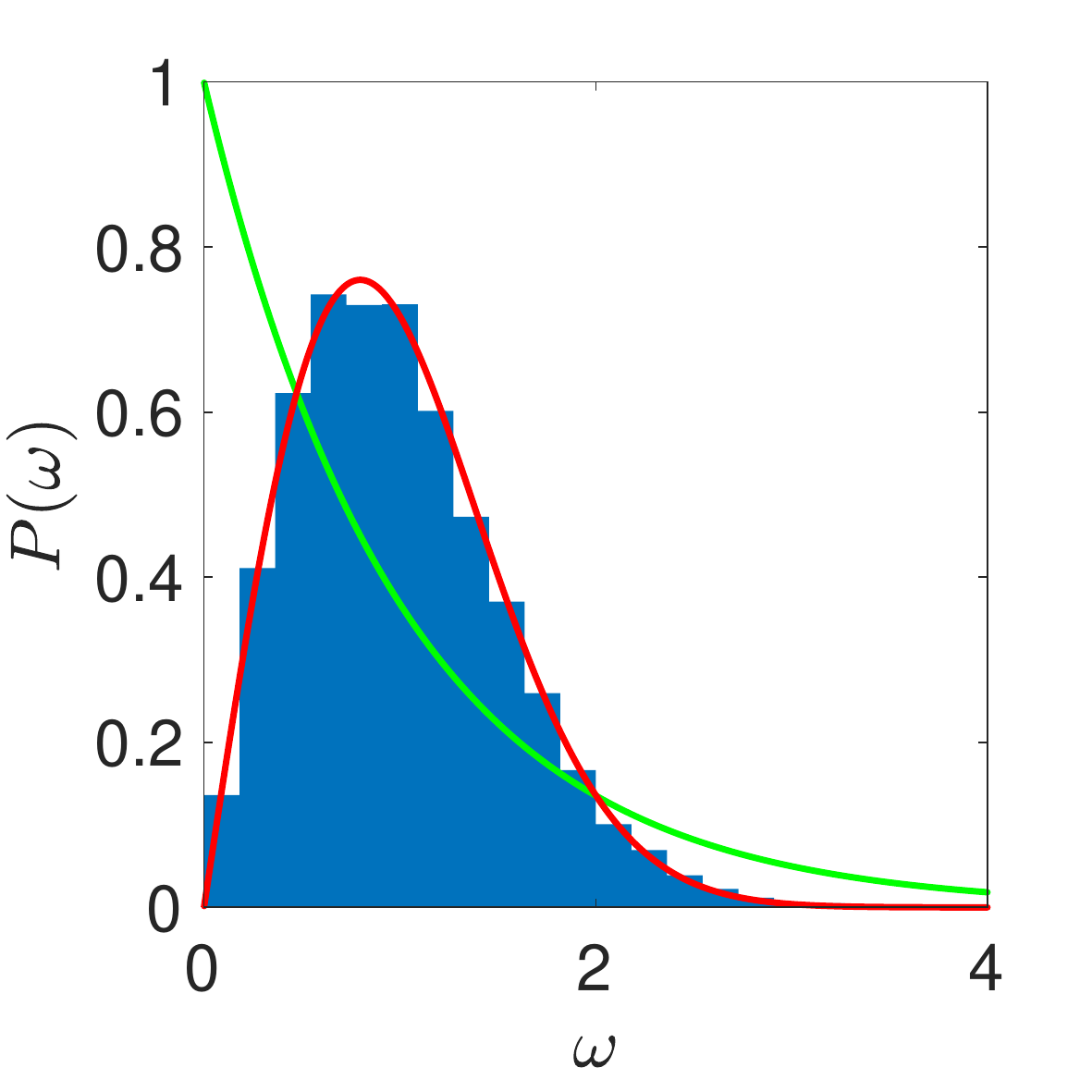}
		$(h_x,h_z)=(-1.05,0.5)$
	\end{minipage}
	\caption{The level spacing distribution for the Ising spin chain ($j = 1/2$) for an $L = 14$ chain at integrable (leftmost) and chaotic (rightmost) points in parameter space follow the Poisson distribution \eqref{eqn:Poisson} and the Wigner surmise \eqref{eqn:Wigner}, respectively.  In between these two parameter points, there is a continuous transition from Poisson to Wigner-Dyson statistics. The middle figure, with magnetic fields very near the integrable line, displays this cross-over.}
    \label{fig: level spacing}
\end{figure}

In figure \ref{fig: level spacing}, we display the cross-over in the level spacing statistics as we move from integrability to the strongly chaotic regime. Similar spectral statistics were found in \cite{1007.3957}. The leftmost figure is a representative of the integrable line $h_z=0$ and can be seen to follow the Poisson distribution except for a deficit in the first bin. A similar deficit at small separations appears in the figures of \cite{1007.3957}. The absence of a peak at the location expected from the Wigner surmise and a close match to the slower decay of the Poisson distribution at large separation allow us to clearly distinguish the level spacing statistics at this integrable point from those expected in the strongly chaotic regime.
The second figure shows the cross-over regime that appears for magnetic fields near the integrable line and the last figure is the strongly chaotic point, studied by \cite{1007.3957}, which follows the Wigner surmise.

\subsection{Commutator squared}
Apart from the spectral statistics, other potential indicators of chaos such as the commutator squared or equivalently the out-of-time-order correlator (OTOC) have been studied in the mixed field Ising model \cite{1409.8180,Hosur:2015ylk}. It was found that the commutator squared displays a sharp growth that spreads to more distant sites. However, the early time behaviour up to saturation was qualitatively similar for both integrable and chaotic values of the magnetic fields. Here we review the study of the time-evolution of the commutator squared in an $L = 8$ spin chain described by the Hamiltonian \eqref{Hamiltonian}, which was identified as chaotic by its spectral statistics, at infinite temperature ($\beta = 0$),
\begin{equation}
C
(x,t) = \left\langle \, \left| \left[S_{z}^{(1)}(t),S_{z}^{(1+x)}(0) \right] \right|^2 \right\rangle_{\beta=0} \,.
\label{eqn:spin12-CS}
\end{equation} 

\begin{figure}[th] 
 \centering
\begin{minipage}[b]{.49\textwidth}
\includegraphics[width=\textwidth]{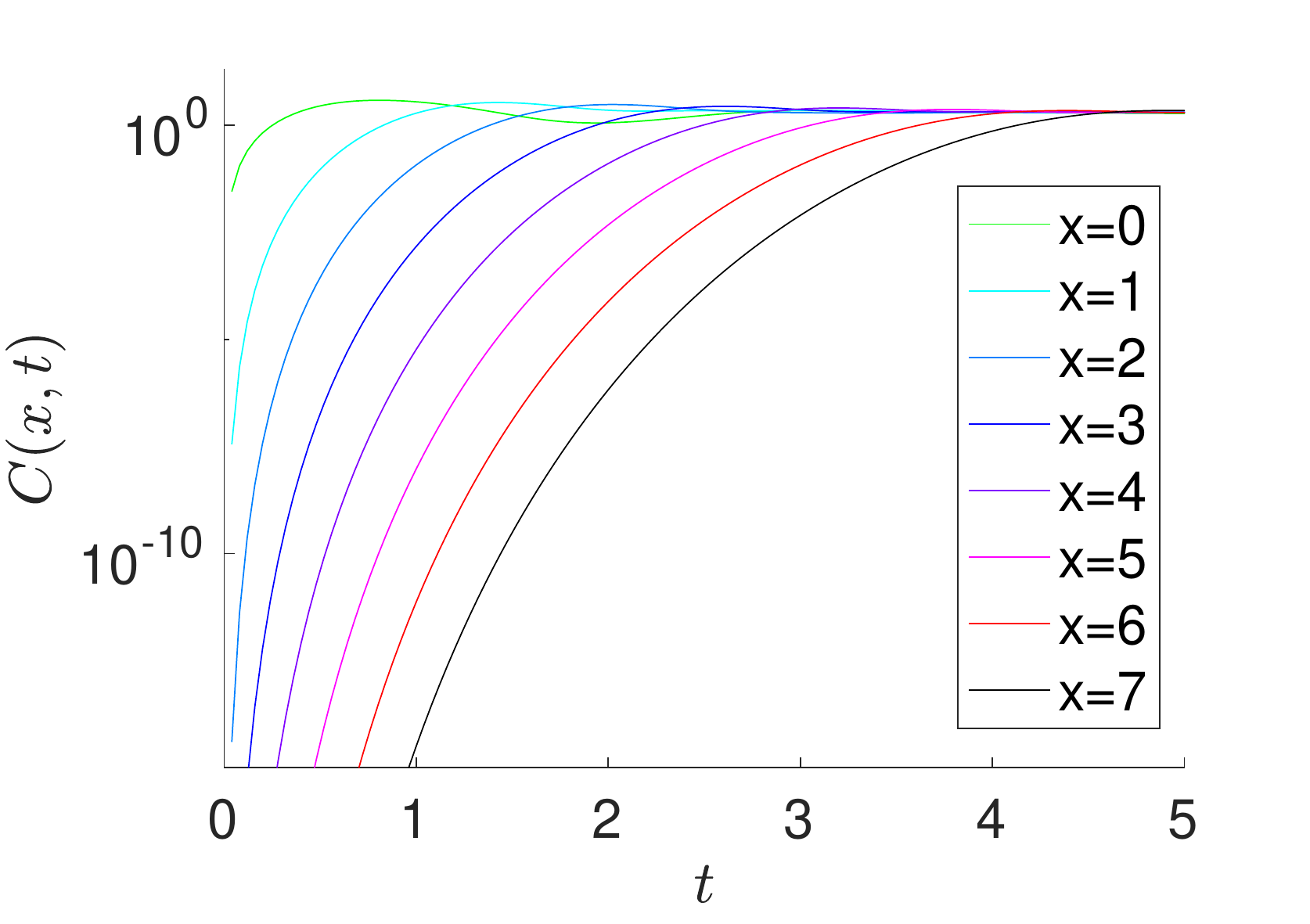}
 \label{fig: spin 1/2 CS}
\end{minipage}
\begin{minipage}[b]{.49\textwidth}
\includegraphics[width=\textwidth]{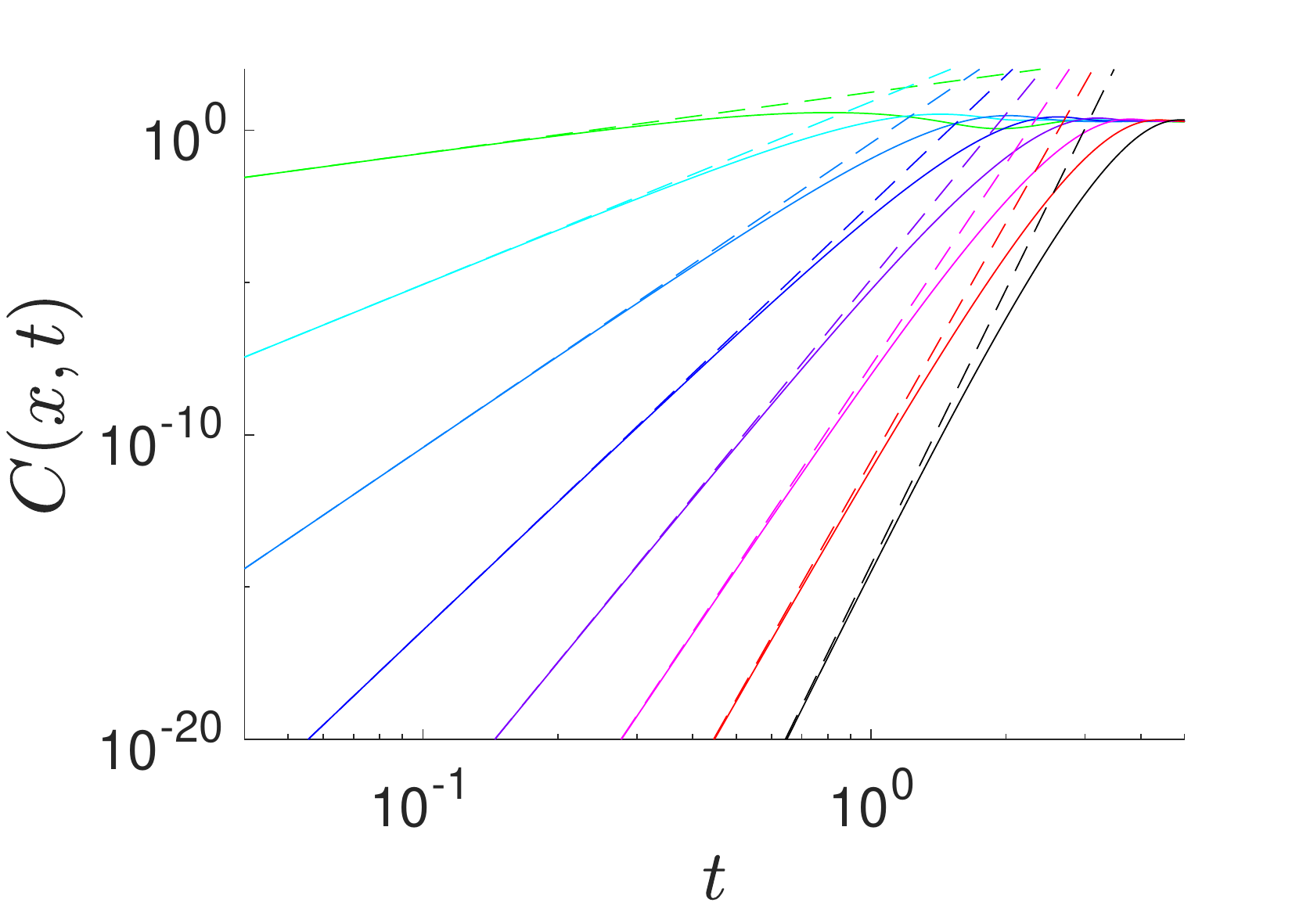}
 \label{fig: BCH}
\end{minipage}
\caption{The commutator squared as a function of time depicted for various positions of the second operator. No exponential regime develops as the distance between the two operators grows. The figure on the left uses a semi-log scale where exponential growth would appear linear, whereas the figure on the right uses a log-log scale to exhibit the early time power law growth. The BCH prediction for this early time power law growth is superimposed (dashed line). }
\end{figure}

\subsubsection{Exponential growth?}
From our classical intuition about this commutator squared, we would expect it to grow exponentially in chaotic systems. More precisely, if a classical limit of the model exists, the corresponding classical system is expected to be chaotic by the BGS conjecture and is therefore expected to exhibit a strong sensitivity to initial conditions  \cite{BGS1,BGS2}. The question is then how robust is this behaviour in the quantum regime? Clearly a region of exponential growth requires that the commutator squared starts out with a large suppression leaving room for the growth. Is this suppression sufficient, such that systems with a small parameter suppressing the commutator squared generically exhibit exponential growth or is the exponential growth solely present in the classical limit?

Large separations between the operators in the commutator squared provide a suppression of the form $\exp{\left(-2\lambda_L \, x/v_B\right)}$, where $v_B$ is known as the butterfly velocity. This was originally conjectured to provide sufficient room for this exponential growth \cite{Hosur:2015ylk}. However, it has been understood that this is not the case and that spin systems with small numbers of degrees of freedom per site do not exhibit exponential growth \cite{1305.2817,1701.09147,Khemani:2018sdn,1805.05376}. 

Figure \ref{fig: spin 1/2 CS} shows the commutator squared for different sites as a function of time. On this semi-log plot, exponential growth of the commutator squared would appear as a linear region. No exponential regime is visible for any site, no matter how far removed from each other. This demonstrates that for this model the large suppression in the commutator squared from the separation between the local operators, which makes the scrambling time parametrically large, is not sufficient to ensure that the commutator squared exhibits a period of exponential growth.

\subsubsection{Early time behaviour}
Given that the commutator squared does not exhibit a period of exponential growth, we would like to gain a better understanding of the behaviour that it does exhibit. We can start by using the BCH expansion to understand the very early time behaviour \cite{1801.01636}. 

For the commutator squared involving two sites separated by $x$ sites, the leading order term in this expansion is
\begin{equation}
C(x,t) \sim \frac{(4 h_x)^{2x+2}}{((2x+1)!)^2}t^{4x+2}.
\end{equation}
This leading order form already goes some way towards explaining the large suppressions we see at large separations.

Figure \ref{fig: BCH} shows the commutator squared on a log-log plot to reveal this early time power law growth with the leading order term overlaid in dashed line. 
For sufficiently early times the commutator squared and the leading order term agree. As the second order term becomes non-negligible, the two curves diverge. However, for all times the commutator squared is below the leading order, evidently power law, value.\footnote{We thank Oleg Evnin for a discussion on this topic.}

\subsubsection{Near saturation behaviour}
It has been argued that the near saturation behaviour should be analysed in terms of the functional form \cite{1802.00801,Khemani:2018sdn}
\begin{align}
 C(x,t) \sim e^{-\lambda \frac{(x-x_0 -v_B t)^{1+p}}{t^p}  } \,,
\label{eqn:SwingleAnsatz}
\end{align}
where $v_B$ is the butterfly velocity chosen so that the saturation of the commutator squared occurs at $t_{sat} = \frac{x-x_0}{v_B} $. Then $x_0$ plays the role of fixing the overall constant in front of the commutator squared and  $p$ is a parameter that describes the shape of the wavefront as the operator spreads. $p=0$ describes a Lyapunov behaviour whereas $p=1$ is known as diffusive behaviour.\footnote{Note that the center of the wavefront is moving ballistically at a constant velocity. The shape of this wavefront is what is spreading out diffusively.}

\begin{figure}[th]
\centering
\includegraphics[height=5.5cm]{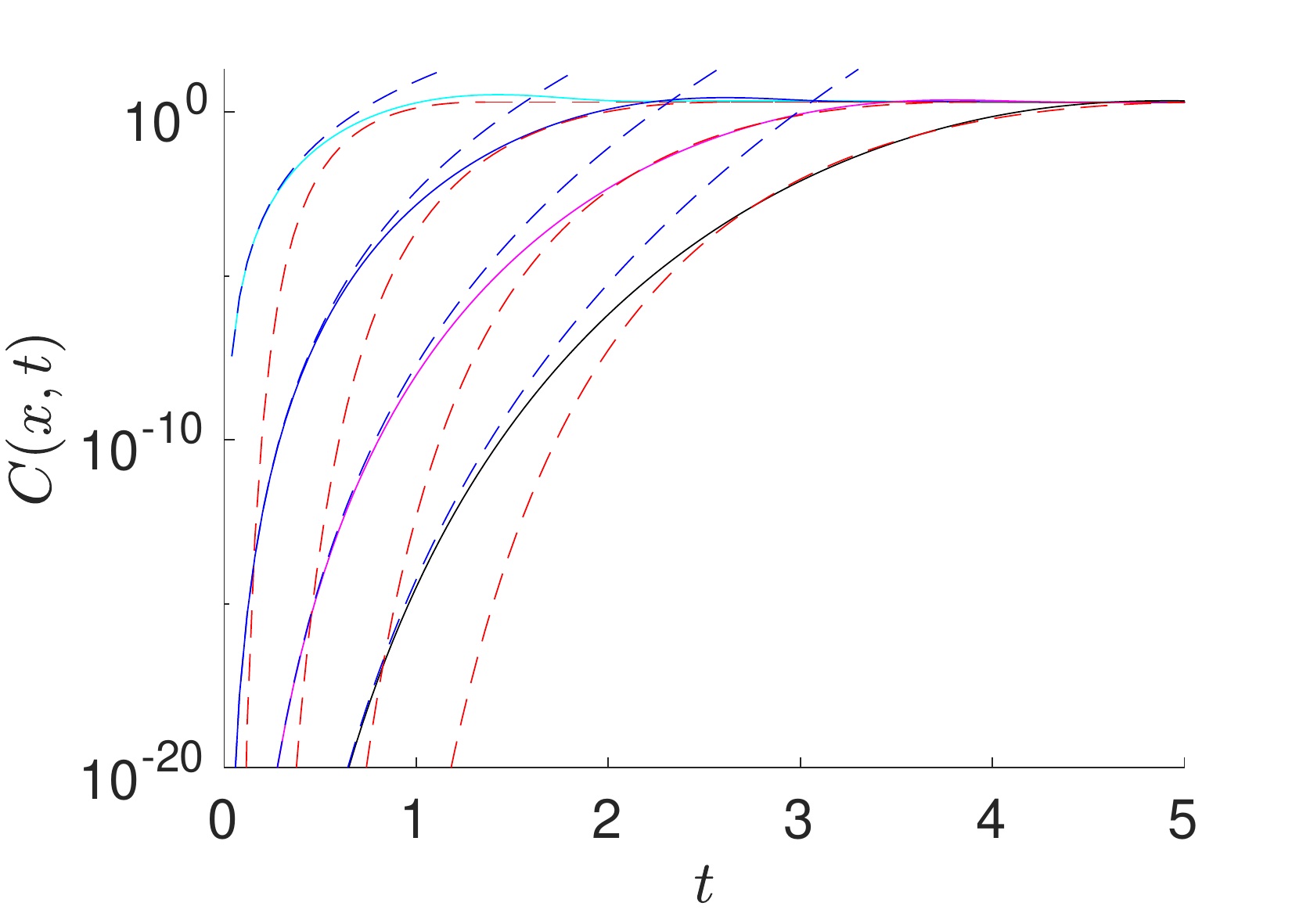}
\caption{The commutator squared for the sites 2, 4, 6 and 8 of an 8-site chain. A fit to the ansatz \eqref{eqn:SwingleAnsatz} with $p=1$ for the near saturation behaviour (red dashed line) and the early time BCH behaviour (blue dashed line) are superimposed.}
\label{fig:saturation_fits}
\end{figure}

 In fact, it was argued in \cite{1805.05376} that in the thermodynamic limit of long chains we should generically expect to see diffusive behaviour with $p=1$. In figure \ref{fig:saturation_fits} we fit the near saturation region for the parameters $v_B, x_0$ and $\lambda$ with $p$ fixed to 1. We see that the early time behaviour is well described by the BCH form, that there is a small cross-over region and then that this diffusive functional form matches on to the near saturation behaviour. As we move to larger separations, the cross-over happens at lower values of the commutator squared.\footnote{The authors of \cite{1805.05376} had to consider very long chains to see the diffusive behaviour with $p=1$. They used a Matrix Product Operator (MPO) approach which works best for small values of the commutator squared, but requires more resources to resolve the near saturation behaviour. The fact that this cross-over into the diffusive near saturation behaviour moves to lower values of the commutator squared as the separation increases explains why such large separations are required to identify this behaviour when focusing only on the regime where the commutator squared is small. With the benefit of hindsight, we can see that this functional form accurately describes the commutator squared even for small chains as long as we look sufficiently close to saturation.}

We conclude that there is no sign of  exponential growth of the commutator squared at the strongly chaotic point. Although the spatial separation of the two operators was expected to play the role of the small parameter needed to open a window for exponential behaviour, the first term in the BCH-expansion fits the early growth very well until the near saturation behaviour described in \cite{1805.05376,1802.00801} takes over, leaving no room for an intermediate exponential regime. In the following sections we will see how, by considering our model with higher representation of the $SU(2)$ living at each site, we will be able to open up a window of exponential growth between the early time BCH and near saturation behaviours.

\section{Higher spin generalisation of the Ising model}
\label{sec:higher_spin}
So far we have worked with the usual mixed field Ising model, which is a Hamiltonian defined for a chain of $L$ spin $1/2$ particles. We will now extend this model in a way such that the model has a classical limit where we can compare classical and quantum indicators of chaos. We will do so by generalising the representation of $SU(2)$ at each site of the chain, replacing the spin $1/2$ particles by spin $j$ particles. A model of this type has been consider in \cite{https://doi.org/10.1007/BF01312650}. The same Hamiltonian still makes sense with the spin matrices replaced by their spin $j$ equivalents. We consider the Hamiltonian
\begin{align}
H= \frac{2}{\sqrt3} \sqrt{j(j+1)} \left[ - \sum_{n=1}^{L-1} S_z^{(n)} S_z^{(n+1)}  - \sum_{n=1}^L \left( h_x S_x^{(n)} + h_z S_z^{(n)} \right) \right]\,,
\label{eqn:higher_spin_H}
\end{align} 
where the prefactor of $\sqrt{j(j+1)}$ is there to ensure that evolution equations have a finite large $j$ limit and the factor of ${2}/{\sqrt3}$ is present to ensure that this Hamiltonian has the conventional normalisation for $j=1/2$ used for example in \cite{1007.3957}. The $S_{a}^{(n)}$ at each site form a spin $j$ representation of $SU(2)$, but in order to fix the relative strengths of the couplings we will now give a prescription for normalising them. 

This model admits a semi-classical limit at large $j$, where the phase space at each site is a fuzzy sphere which goes over to a smooth $S^2$ in the classical limit \cite{10.1088/0264-9381/9/1/008}. The analogy to the sphere can be understood by normalising the spin matrices to have a constant Casimir,\footnote{We use 3 for this constant in order to match onto the conventions of \cite{1007.3957} for spin 1/2.} 
\begin{align}
\sum_{a=x,y,z} S_a^{(n)} S_a^{(n)} = 3\,,
\label{eqn:Casimir}
\end{align} 
which is the equation of a 2-sphere of radius $\sqrt3$.
As we increase $j$, the dimension of the representation increases and this fuzzy quantum sphere goes over to a smooth classical space. With this normalisation the commutation relations are
\begin{align}
[ S_{a}^{(n)},  S_{b}^{(m)} ] = \frac{\sqrt{3}\, i}{\sqrt{j(j+1)}} \delta_{n,m} \epsilon^{abc}  S_{c}^{(n)} \,,
\end{align}
so that the right hand side vanishes in the large $j$ classical limit.

\subsection{Classical limit}
We will now identify the classical system at the endpoint of this limit. 
The usual correspondence principle associates the Poisson bracket to $(-i)$ times the commutator (in units where $\hbar=1$). Equivalent classical dynamics are obtained by simultaneously rescaling the Hamiltonian and the Poisson bracket,
\begin{align}
H_{cl} &\equiv \lim_{j \to \infty} \frac{H}{\sqrt{j(j+1)}} \,, \\
\{\cdot , \cdot \} & \equiv  \lim_{j\to\infty} - i \sqrt{j(j+1)} \; [\cdot,\cdot]\,, \\
\partial_t \cO &= 
 \{\cO , H_{cl} \} \,.
\end{align}

The phase space consists of a 2-sphere at each site on our chain. We can choose more familiar coordinates on the sphere at each site in terms of angles $(\theta^{(n)},\phi^{(n)})$ 
\begin{align}
 S_x^{(n)} &= \sqrt3 \sin\theta^{(n)} \cos\phi^{(n)}\,, \\
 S_y^{(n)} &= \sqrt3 \sin\theta^{(n)} \sin\phi^{(n)}\,, \\
 S_z^{(n)} &= \sqrt3 \cos\theta^{(n)} \,.
 \label{eqn:phase_space_coordinates}
\end{align}
Since the commutators between spins at different sites vanish and the commutators between spins at a given site only involve spins at that site, the Poisson bracket must have the form
\begin{align}
\{ F , G \} =\sum_{n=1}^L  \omega^{-1}(\theta^{(n)},\phi^{(n)})
 \left( \frac{\partial F}{\partial \theta^{(n)}} \frac{\partial G }{\partial \phi^{(n)}} 
 -  \frac{\partial F}{\partial \phi^{(n)}} \frac{\partial G}{\partial \theta^{(n)}} \right) \,.
\end{align}
 It remains to identify the function $\omega^{-1}$. 
By using the correspondence principle rewritten in terms of the angular coordinates
\begin{gather}
-i \sqrt{j(j+1)} [ S_x^{(n)} , S_y^{(n)}] 
=\sqrt{3}S_z^{(n)}
\xrightarrow[j\to \infty]{} 
\{ S_x^{(n)} , S_y^{(n)} \} = \sqrt{3} S_z^{(n)} \,, \\
\qquad \qquad \implies 
\{ \sqrt3 \sin\theta^{(n)} \cos\phi^{(n)} , \sqrt3 \sin\theta^{(n)} \sin\phi^{(n)} \} 
= 3 \cos\theta^{(n)} \,, 
\end{gather}
we can determine that the symplectic form is given by
\begin{gather}
\omega^{-1}(\theta^{(n)},\phi^{(n)})  = \csc \theta^{(n)} \,, \\
\boldsymbol{\omega} = \sum_{n=1}^L \sin\theta^{(n)} \; d\theta^{(n)} \wedge d\phi^{(n)} \,.
\end{gather}
This is the canonical $SU(2)$ invariant symplectic form on the sphere. The final Poisson bracket is
\begin{align}
 \{ F , G \} =\sum_{n=1}^L  \csc\theta^{(n)} 
 \left( \frac{\partial F}{\partial \theta^{(n)}} \frac{\partial G }{\partial \phi^{(n)}} 
 -  \frac{\partial F}{\partial \phi^{(n)}} \frac{\partial G}{\partial \theta^{(n)}} \right) \,. \label{eqn:bracket}
\end{align}

The equations of motion governing this system are
\begin{align}
\dot \theta^{(n)} &= \{  \theta^{(n)} , H \} = 2 h_x\sin\phi^{(n)} \,, \\
\dot \phi^{(n)} &= \{ \phi^{(n)} , H \} = -2 \left( h_z+ \sqrt{3}\cos\theta^{(n-1)}+\sqrt{3}\cos\theta^{(n+1)}-h_x\cot\theta^{(n)}\cos\phi^{(n)} \right)\,.
\end{align}
The inconvenient factors of $\sqrt{3}$ come from the fact that we normalised our Hamiltonian so that it matches with the conventions in the literature for spin 1/2.

It is important to realise that the phase space of the classical model is given by $\bigotimes_{n=1}^L S^2$, so that each site contributes a single conjugate pair to the phase space. This is not a lattice of classical particles moving on a sphere along with their conjugate angular momenta. 

This limit is not quite the usual notion of a classical limit that might be studied to understand the correspondence between the classical and quantum physics of single particle systems, since the number of degrees of freedom at each site is taken to be large and therefore the Hilbert space and not only the Hamiltonian changes as we take $j$ large.\footnote{The feature that the size of the Hilbert space increases is shared by large $N$ limits, but there are a number of differences from gauge theory large $N$ limits: the spins live in a representation of $SU(2)$ but the Hamiltonian is not $SU(2)$ invariant so there is no symmetry to gauge and the dimension of the representation, not the rank of a gauge group, is getting large.}

\subsection{Commutator squared in the classical limit}
\label{sec:classical_commutator_squared}
 In the spin 1/2 case the commutator squared was defined in \eqref{eqn:spin12-CS}. This definition can be straightforwardly extended to the higher spin representations
  \begin{equation}
C^{(j)}
(x,t) \equiv \left\langle \, \left| \left[S_{z}^{(1)}(t),S_{z}^{(1+x)}(0) \right] \right|^2 \right\rangle_\beta \,,
\end{equation} 
where we have added a label $(j)$ indicating which representation we are considering.

The classical analogue of the commutator squared with a finite large spin limit is\footnote{See also \cite{Schuckert:2019oao} for a similar approach to the one described here for studying many-body chaos in a classical lattice model.}
\begin{align}
C^{(cl)}(x,t) &\equiv \lim_{j\to \infty} j(j+1) C^{(j)}(x,t) \\
&= \lim_{j\to \infty}  j(j+1) \left\langle \, \left| \left[S_{z}^{(1)}(t),S_{z}^{(1+x)}(0) \right] \right|^2 \right\rangle_\beta  \\
&=\left\langle \, \left| \left\{ S_{z}^{(1)}(t),S_{z}^{(1+x)}(0) \right\} \right|^2 \right\rangle_\beta 
 \,. \label{eqn:classical_CS}
\end{align}

This Poisson bracket is related to the sensitivity of our system to a perturbation of its initial conditions
\begin{align}
\{ S_z^{(1)} (t) , S_z^{(n)}(0) \}
= 3 \frac{\partial \cos\theta^{(1)}(t)} {\partial \phi^{(n)}(0)}\,. \label{eqn:Poisson_to_derivative}
\end{align}
The quantity in \eqref{eqn:classical_CS} should then be interpreted as an average over phase space, weighted by an appropriate Boltzmann factor. Phase space consists of the different initial conditions for all of the dynamical variables $ \theta^{(n)}(0) , \phi^{(n)}(0)$ and the derivative in \eqref{eqn:Poisson_to_derivative} corresponds to the dependence of a particular phase space coordinate at a later time to a change of initial conditions. 

In practice it can be computed using a Monte Carlo approach. First, randomly choose an initial condition $\theta^{(n)}(0) , \phi^{(n)}(0)$. Then evolve the system for a time $t$ with both this initial condition as well as an initial condition where $\phi^{(n)}(0)$ at the $n^{\textrm{th}}$ site has been deformed to  $\phi^{(n)}(0) + \epsilon$. The derivative can then be written as 
\begin{align}
\frac{\partial \cos\theta^{(1)}(t)} {\partial \phi^{(n)}(0)} = \lim_{\epsilon \to 0} \frac{\cos\theta^{(1)}_{\epsilon}(t) - \cos\theta^{(1)}_{0}(t) }{\epsilon}\,,
\label{eqn:IC_deriv}
\end{align}
where $\cos\theta^{(1)}_{\epsilon}(t)$ denotes the value after time evolution for a time $t$ with perturbed initial conditions.

In practice the numerical accuracy of the calculation puts a lower bound on $\epsilon$. Since $|\cos\theta^{(1)}(t)|\leq 1$, the numerator is bounded by 2. The finite $\epsilon$ approximation to \eqref{eqn:IC_deriv} starts off at zero, since the perturbation in the initial condition of $\phi^{(n)}$ leaves $\theta^{(1)}$ unchanged. It can then grow until it reaches the bound at order $\epsilon^{-1}$ at which point it will generically saturate.

Finally, we should average this numerical estimate for the derivative over different randomly chosen initial conditions. Once we have computed this approximation to $C^{(cl)}$, for chaotic systems we should find a region of exponential growth from $C^{(cl)} \sim O(1)$ to $O(\epsilon^{-2})$. By fitting a line to a semi-log plot of this quantity we can extract the Lyapunov exponent
\begin{align}
\frac{\partial \cos\theta^{(1)}(t)} {\partial \phi^{(n)}(0)} \sim e^{\lambda_L t} \implies C^{(cl)} (x,t) \sim e^{2 \lambda_L t} \,.
\label{eqn:classical_Lyapunov}
\end{align}

Let us now compare this to the quantum story. From the correspondence principle, we expect that in the large $j$ limit, the commutator squared should also grow exponentially
\begin{align}
 j(j+1) C^{(j)} (x,t) \xrightarrow[j\to\infty]{} C^{(cl)}(x,t) \implies C^{(j\gg1)}(x,t) \sim \frac{1}{j(j+1)} e^{2 \lambda_L t} \,.
 \end{align}
Since the commutator squared is bounded\footnote{This can be seen by writing $C^{(j)} (x,t)$ as an inner product
\begin{align}
C^{(j)} (x,t) &= \big( \langle \psi| - \langle \psi'| \big) \big(| \psi\rangle - | \psi' \rangle \big)\,, \\
| \psi\rangle &= S_z^{(1)}(t)S_z^{(1+x)}(0) |\beta \rangle \,, \\
| \psi' \rangle &= S_z^{(1+x)}(0)S_z^{(1)}(t) |\beta \rangle \,,
\end{align}
where we have replaced the thermal trace by an inner product in a unit norm purifying state $|\beta \rangle$. The thermofield double is an example of such a state. By using the Cauchy-Schwarz inequality and the fact that the spin operators normalised as in \eqref{eqn:Casimir} are bounded with maximal eigenvalue $\sqrt{3 j / (j+1)}$, we find that the commutator squared is bounded by 
\begin{align}
 \big| C^{(j)} \big| \leq \frac{36 j^2}{(j+1)^2}\,.
\end{align}
} 
we know that this growth must eventually break down. The suppression by $j(j+1)$ implies that it can persist for a time 
\begin{align}
\lambda_L \Delta t_{exp} \sim \frac12 \log j(j+1) \,.
\end{align}
As $j$ is taken to be large, this regime of exponential growth will persist for longer. In our classical analysis, the use of a finite perturbation $\epsilon$ to estimate the derivative by the initial condition lead to an analogous breakdown of the exponential growth at a scale $\epsilon^{-2}$, where $\epsilon^{-1}$ plays the role of $j$ in this analogy.

In order to study the exponential growth, it will be most convenient to plot $ j(j+1)C^{(j)}(x,t)$ since this is the quantity that goes over to a finite classical curve. This is in contrast to studies of the near saturation behaviour related to the growth of operators, where $C^{(j)}(x,t)$ itself is the quantity of interest.

A quantity analogous to our \eqref{eqn:classical_CS} has been studied in a single particle context \cite{1609.01707}, where its correspondence to the classical limit of the commutator squared computed in the corresponding quantum system was exhibited explicitly. In this work, we aim to exhibit a similar correspondence in a many-body context. Although the formalism we present here applies in the general many-body context, our numerical study will be restricted to very small systems. 

In \cite{1609.01707}, it was emphasised that this quantity does not correspond to the ``standard'' Lyapunov exponent defined by 
\begin{align}
\Bigg\langle \lim_{t\rightarrow\infty}  \frac{1}{t} \log \left|\frac{\partial z(t)}{\partial z(0)} \right| \Bigg\rangle \,,
\end{align}
where $z$ are coordinates on phase space.
None the less, we would argue that $\lambda_L$ in \eqref{eqn:classical_Lyapunov} can be thought of as a Lyapunov exponent of the system. It differs from this standard exponent in two ways. 

First, the standard exponent uses a metric on phase space for defining the distance between two trajectories. Since generic phase spaces do not come equipped with a metric this is a choice that must be made. In our case, there is a natural $SU(2)$ invariant metric on phase space that could be used. The commutator squared consists of a different way to measure the divergence between two trajectories.

The second difference, which was emphasised in \cite{1609.01707}, is that the averaging over phase space is performed differently in the two cases. In the definition of the standard Lyapunov exponent, an exponent is extracted at each point in phase space and then averaged. The growth rate of the commutator squared involves averaging the trajectories and then extracting an exponent from the averaged trajectory. Since these trajectories may be growing exponentially at different rates, the average over phase space will tend to pick out the trajectory that grows at the fastest rate. For intermediate times, when $\lambda_{max} t \gg 1$ but well before saturation of the commutator squared, this average over phase space can be computed in a saddle point approximation
\begin{align}
\left\langle e^{\lambda(\theta^{n}_0,\phi^{n}_0) t} \right\rangle \propto e^{\lambda_{max} t}\,.
\end{align}

While the exponent extracted from the growth rate of the commutator squared is not the standard Lyapunov exponent, the choice of this standard was somewhat arbitrary and not necessarily well defined in the general many body setting. Since the classical limit of the growth rate of the commutator squared is also a characterisation of the divergence of classical trajectories, we will continue to refer to it as a Lyapunov exponent as has become standard terminology in the literature on the commutator squared, e.g.~\cite{BoundOnChaos}.
This exponent is also sometimes known as a generalised Lyapunov exponent and has appeared previously in studies of the phase space dependence of chaos \cite{FUJISAKA83,Benzi85}.

\section{Quantum Analysis}
\label{sec:quantum}
In this section we study the quantum mixed field Ising model at finite $j$. We wish to understand if there is an emergence of early-time exponential growth of the commutator squared, diagnosing a classical notion of chaos, when considering a semi-classical limit of the mixed field Ising model. We will focus on chaotic models, as defined by the late-time diagnostic of the spectral analysis, and therefore first map out where the model is integrable or chaotic. We find that the model is only integrable at the trivial $ h_x = 0$ and $| h_x|\rightarrow \infty$ and $| h_z| \rightarrow \infty$ limits and that the integrable line at transverse fields $h_z=0$ only exists for $j=1/2$. Then we turn to the commutator squared. By increasing $j$ we are able to see the appearance of the Lyapunov regime, although we must restrict to very short spin chains 
in order for the numerics to be tractable. By extrapolating this result to large $j$ we extract the classical limit of the Lyapunov exponent.

\subsection{Spectral statistics}
\label{subsection: quantum spectral statistics}
We use the same approach as in the spin 1/2 case in Section \ref{section: review spin 1/2}. First we must determine the symmetries of the model. The reflection symmetry of the chain generalises to higher spin representations. When $h_z=0$, there is still a symmetry under $S_z \rightarrow -S_z$ while keeping $S_x$ fixed, implemented by the unitary
\begin{align}
U = \bigotimes_{n=1}^L e^{ \sqrt{\frac{j(j+1)}{3}} i \pi S_x^{(n)}   }\,.
\end{align}

We present the spectral analysis for higher spin mixed field Ising chains with $L=2$ and $L=6$. The former case is relevant to determine which points in the parameter space represent chaotic models, for our later investigation of the commutator squared. The latter serves as a representative for slightly longer chains which illustrates the non-integrable structure of the $h_z = 0$ line away from the origin and the integrable character of the $| h_x|\rightarrow \infty$ and $| h_z| \rightarrow \infty$ limits for higher spins. For $L=3$, however, we observed Poission statistics for every value of the magnetic fields, which is probably due to a residual symmetry that we were not immediately able to identify. Since the energy eigenvalues in different blocks are uncorrelated, even if the eigenvalues in each block obey Wigner-Dyson statistics the combined distribution will be Poisson. This interpretation is supported by the fact that the $L=3$ chain exhibits exponential growth of the commutator squared in the semi-classical regime for chaotic parameter points, as seen in figure \ref{fig: CSvsSpin L=3}.

\begin{figure}[t]
	\centering
	\begin{minipage}[b]{0.4\textwidth}
		\centering
	\includegraphics[height=4cm]{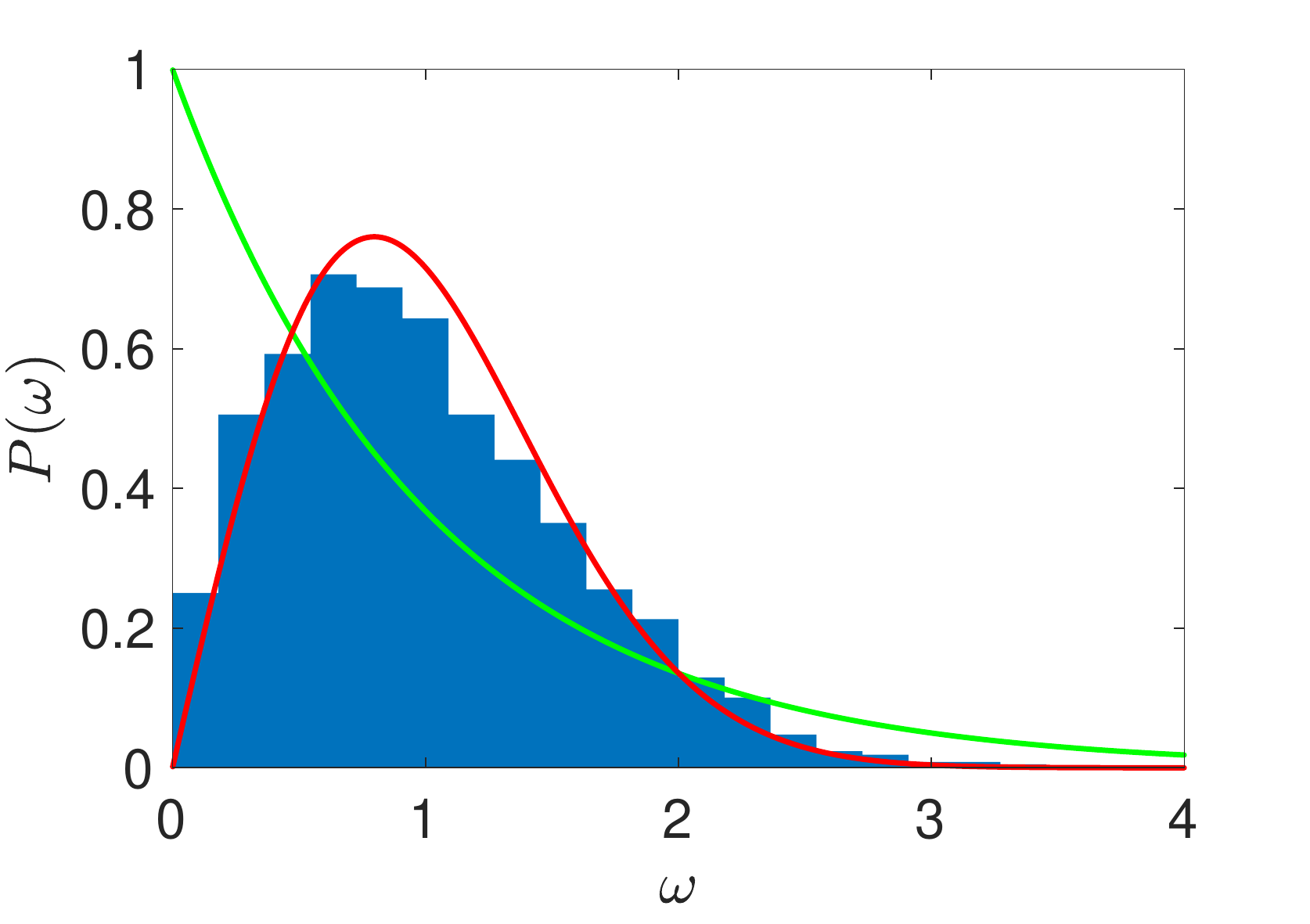} 
	\end{minipage}
	\begin{minipage}[b]{0.4\textwidth}
		\centering
		\includegraphics[height=4cm]{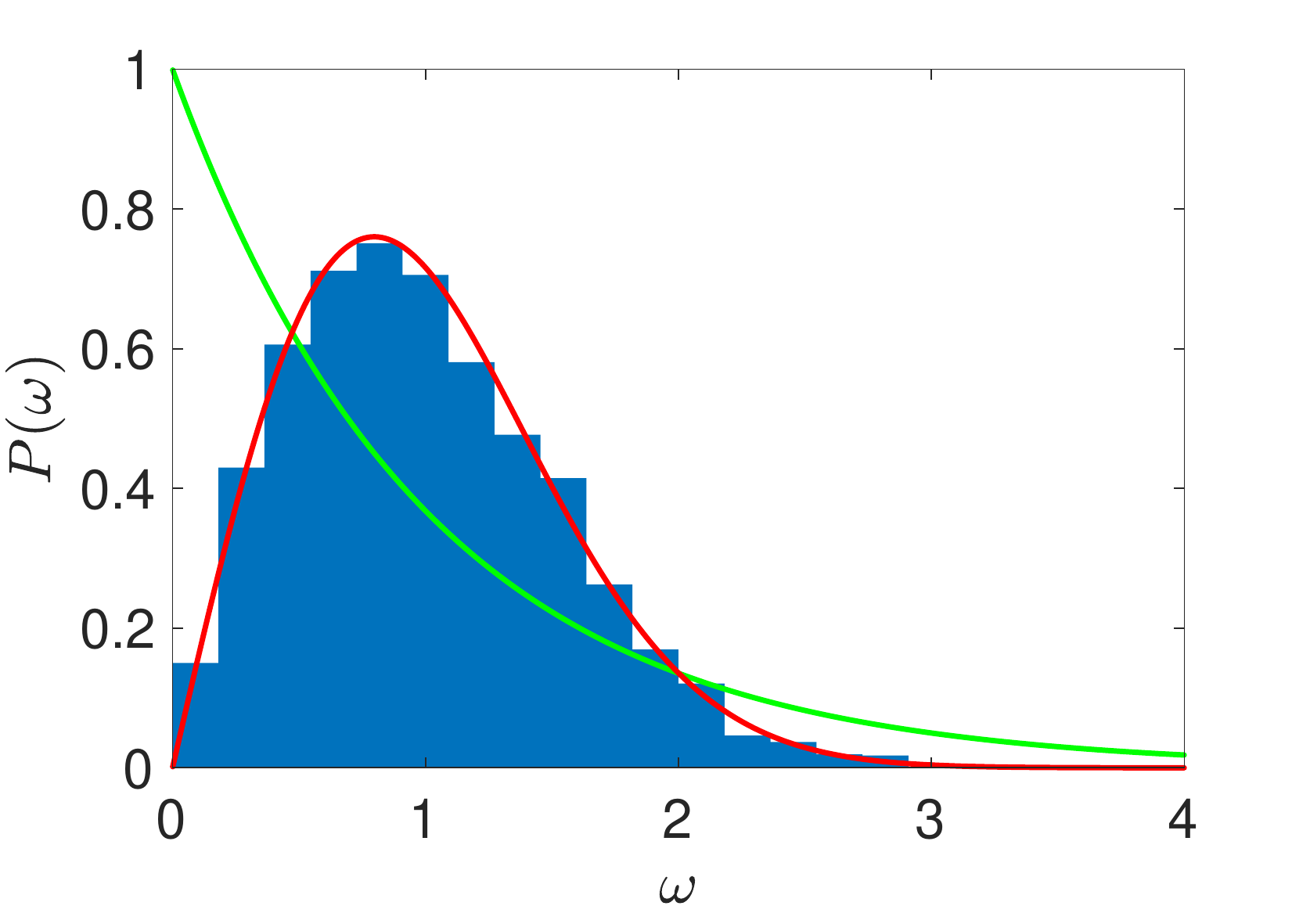} 
		\end{minipage} \\
$(h_x,h_z) = (1,0)$ 
	\caption{The level spacing statistics for a 2-site chain for $j=45$ (left) and a 6-site chain for $j=2$ (right) with a purely transverse magnetic field. For both chain lengths, we see no sign of the integrability at $h_z=0$ present in the $j=1/2$ case.}
	\label{fig: level spacing higher spins a}
\end{figure}

\begin{figure}
\centering
\begin{minipage}[b]{0.3\textwidth}
		\centering
		\includegraphics[width=\textwidth]{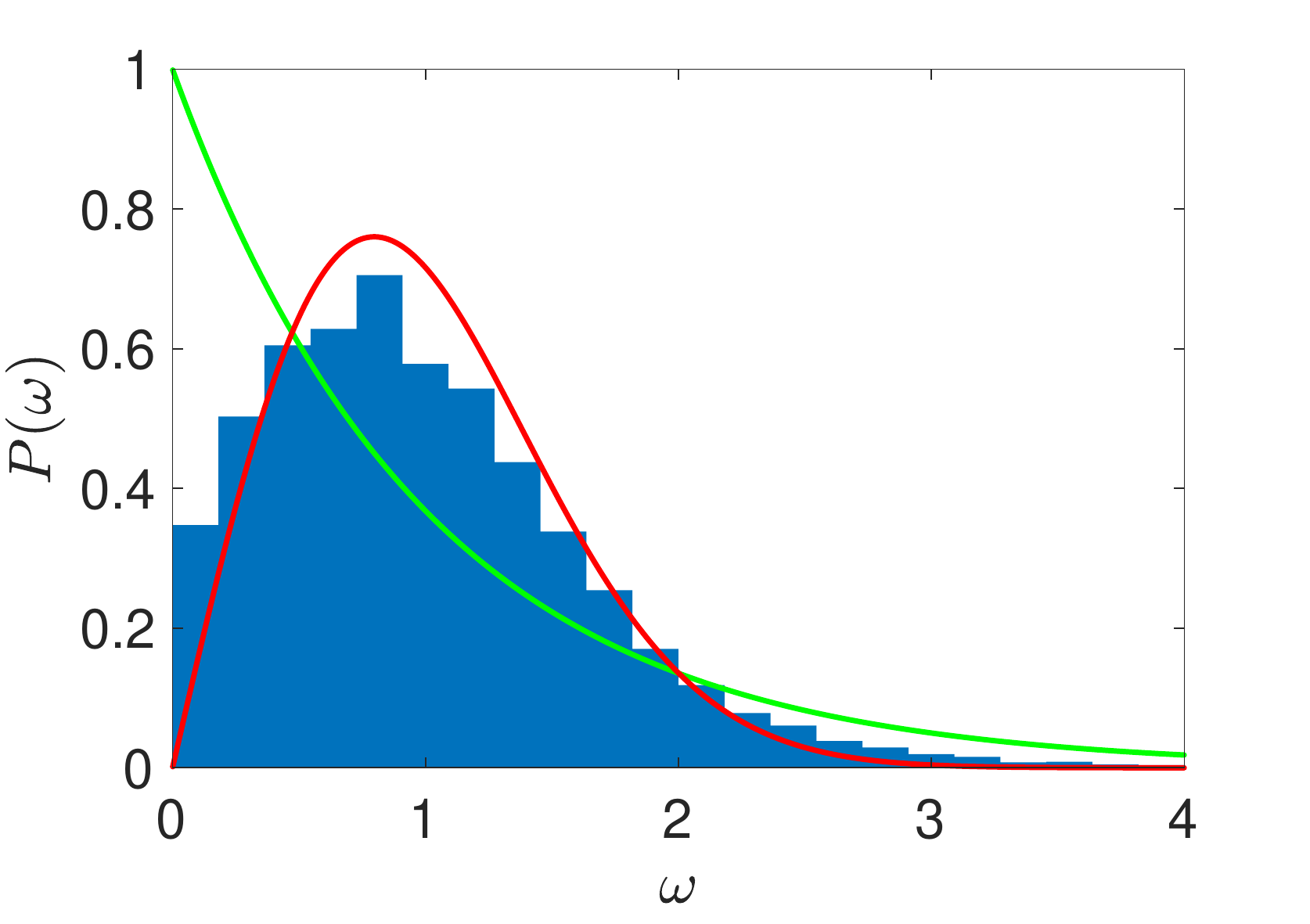}
		$(h_x,h_z) = (0.2h_x^*,h_z^*)$
	\end{minipage}
	\begin{minipage}[b]{0.3\textwidth}
		\centering
		\includegraphics[width=\textwidth]{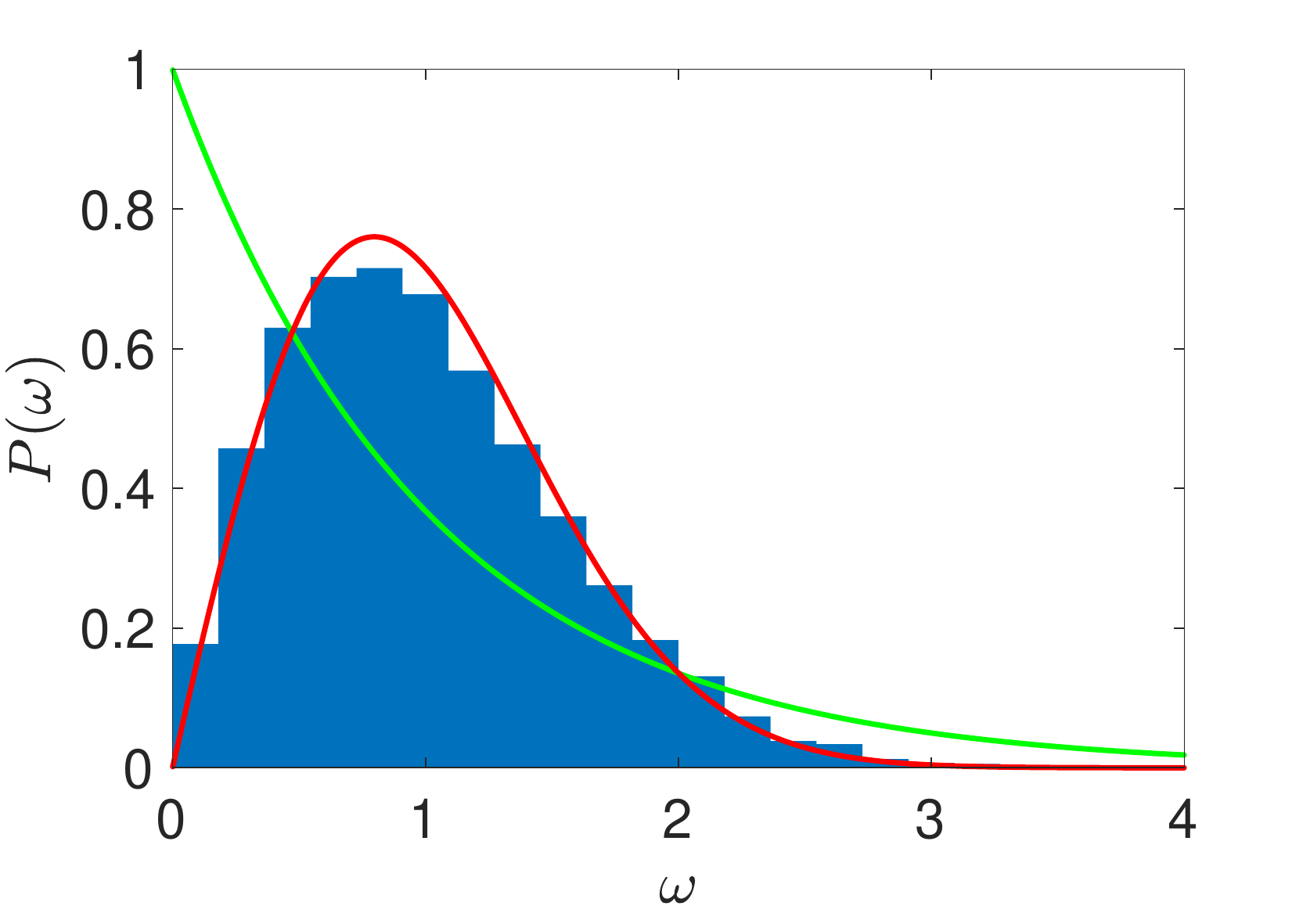}
		$(h_x,h_z) = (h_x^*,h_z^*)$
	\end{minipage}
	\begin{minipage}[b]{0.3\textwidth}
		\centering
		\includegraphics[width=\textwidth]{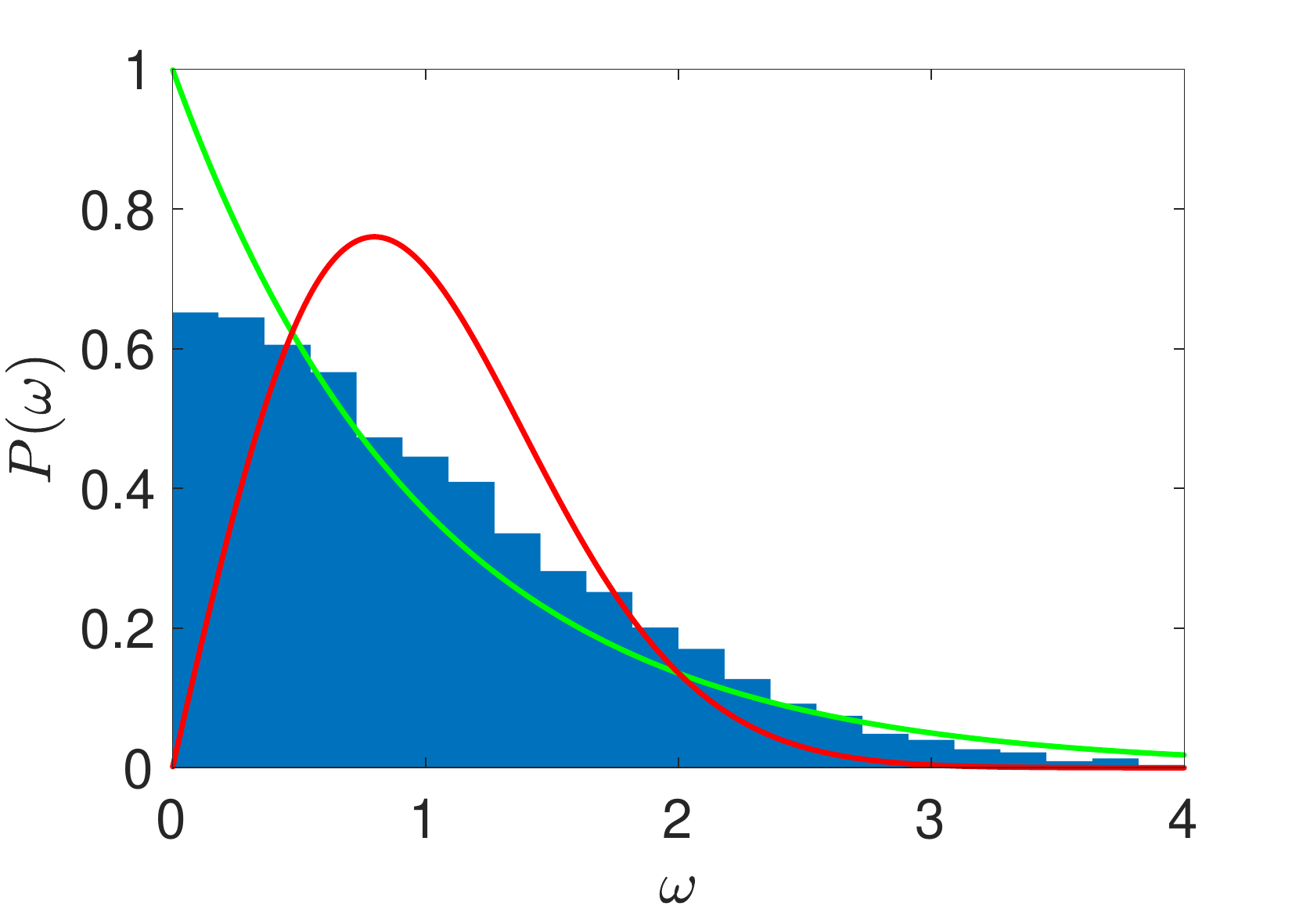}
	$(h_x,h_z) = (2.1h_x^*,h_z^*)$
	\end{minipage}
	\caption{The level spacing statistics for a 2-site chain for $j=45$ as we vary the transverse magnetic field $h_x$. The center figure is at the conventional strongly chaotic point ($h_x^*, h_z^*) = (-1.05 , 0.5)$ where we observe Wigner-Dyson statistics. As we decrease or increase the magnetic field $h_x$ (in the figures to the left and right respectively) we see that the Wigner-Dyson statistics start to break down.}
\label{fig: level spacing higher spins b}
\end{figure}

\begin{figure}
\centering
\begin{minipage}[b]{0.3\textwidth}
		\centering
		\includegraphics[width=\textwidth]{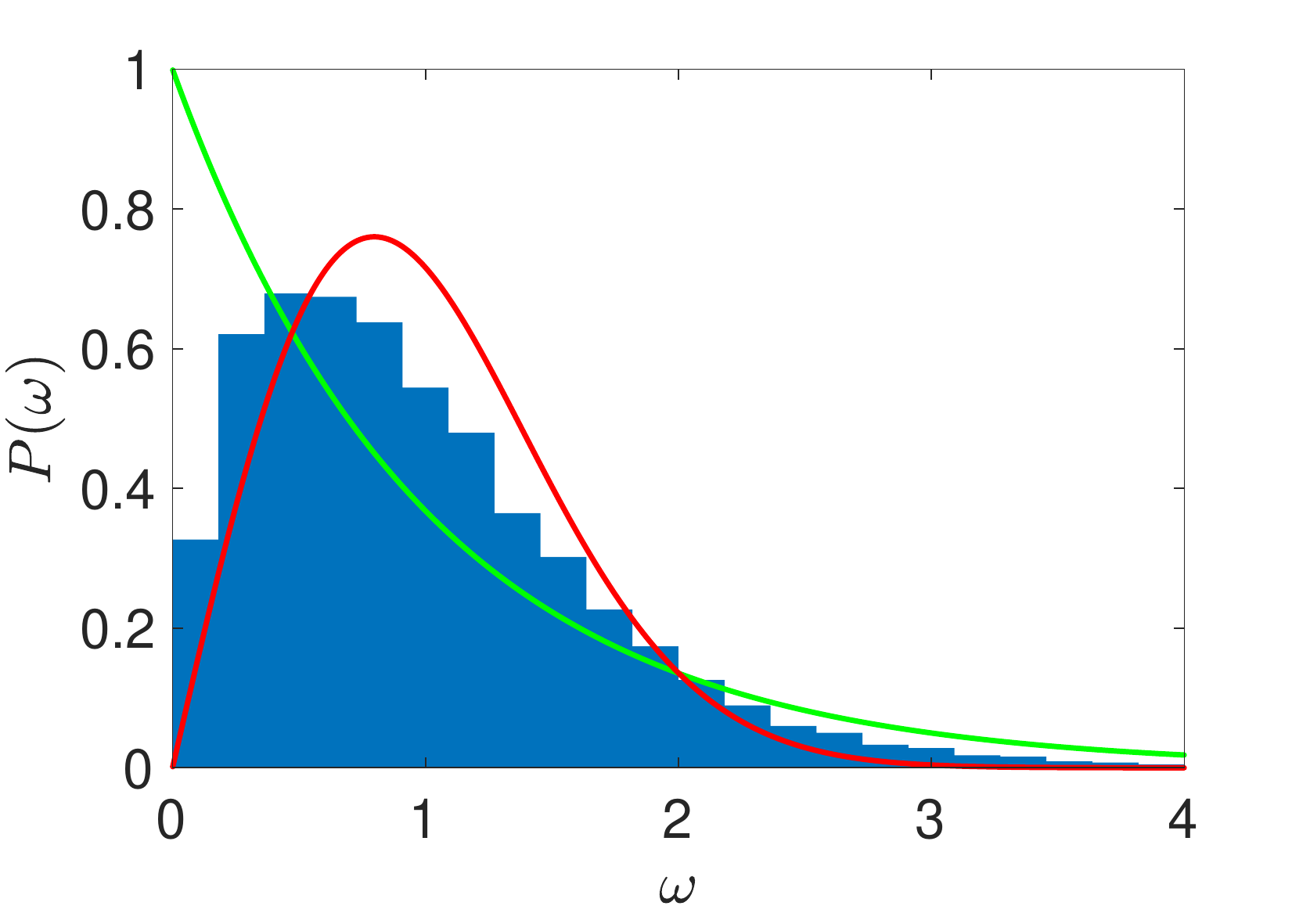}
		$(h_x,h_z) = (0.2h_x^*,h_z^*)$
	\end{minipage}
	\begin{minipage}[b]{0.3\textwidth}
		\centering
		\includegraphics[width=\textwidth]{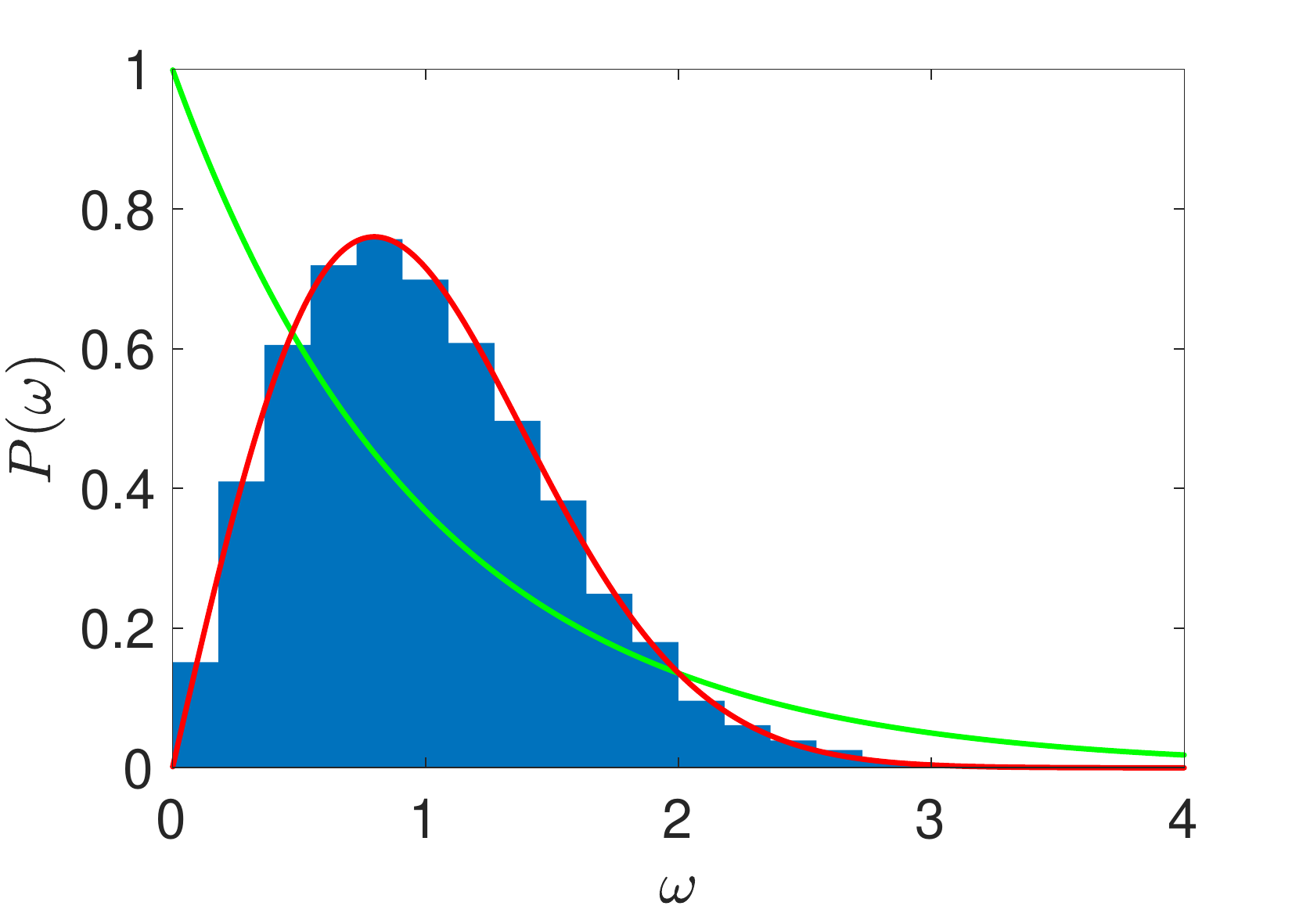}
		$(h_x,h_z) = (h_x^*,h_z^*)$
	\end{minipage}
	\begin{minipage}[b]{0.3\textwidth}
		\centering
		\includegraphics[width=\textwidth]{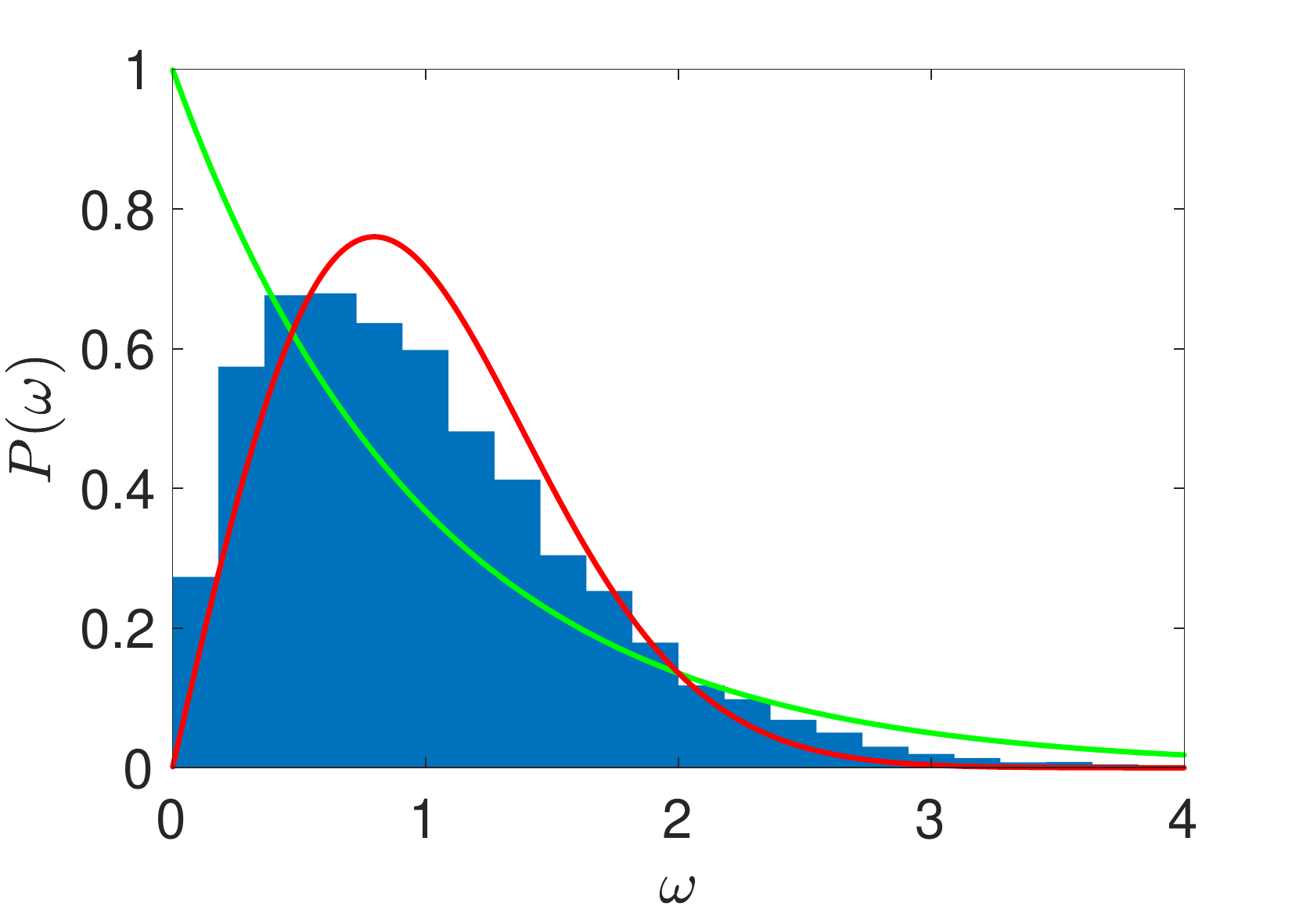}
	$(h_x,h_z) = (5h_x^*,h_z^*)$
	\end{minipage}
	\caption{The level spacing statistics for a 6-site chain for $j=2$ as we vary the transverse magnetic field $h_x$. The center figure is at the conventional strongly chaotic point ($h_x^*, h_z^*) = (-1.05 , 0.5)$ where we observe Wigner-Dyson statistics. As we decrease or increase the magnetic field $h_x$ we see that the Wigner-Dyson statistics start to break down, although the spread in the magnetic fields for which one finds Wigner-Dyson statistics is larger than for the 2-site chain.}
\label{fig: level spacing higher spins c}
\end{figure}

Figure \ref{fig: level spacing higher spins a} demonstrates that the line $h_z = 0$ is no longer integrable.
The strongly chaotic point at 
$(h^{*}_x,h^{*}_z) =(-1,05,0.5)$ 
keeps displaying Wigner-Dyson statistics for higher spins. We observe a breakdown of the Wigner-Dyson statistics as we tune $h_x$ to be small or large while keeping $h_z$ fixed (see figures \ref{fig: level spacing higher spins b} and \ref{fig: level spacing higher spins c}). This cross-over behaviour will be shown to be visible in the study of the commutator squared as well (cf. section \ref{section: Matching classical and quantum Lyapunov exponents}).

\subsection{Commutator Squared}
\label{subsection: quantum_CS}
As reviewed in section \ref{section: review spin 1/2}, the early time behaviour of the commutator squared is not able to distinguish an integrable from a chaotic Hamiltonian in a spin 1/2 chain. In particular, the correlator does not possess the exponential regime expected from the classical picture of chaos. We now show that as the local Hilbert space increases, by taking higher $SU(2)$ representations, an exponential region develops between the early BCH and the near-saturation behaviour. The emergence of an exponential regime is a consequence of the increase of the time to saturation, which grows logarithmically in dimension of the Hilbert space at each site. 
As discussed in the previous section, we expect the exponential growth to be visible for a window of
\begin{align}
\lambda_L \Delta t_{exp} \sim \frac12 \ln j (j+1) \,.
\end{align}
If we want to be able to observe a few e-foldings of exponential growth, we need $j \gtrsim 10$ and are therefore forced to restrict our study to very short chains. In the following, we will detail our investigation for $L=2$ at the strongly chaotic point, while we will briefly discuss the $L=3$ case at the end.

\begin{figure}[th] 
 \centering
\includegraphics[width=.48\textwidth]{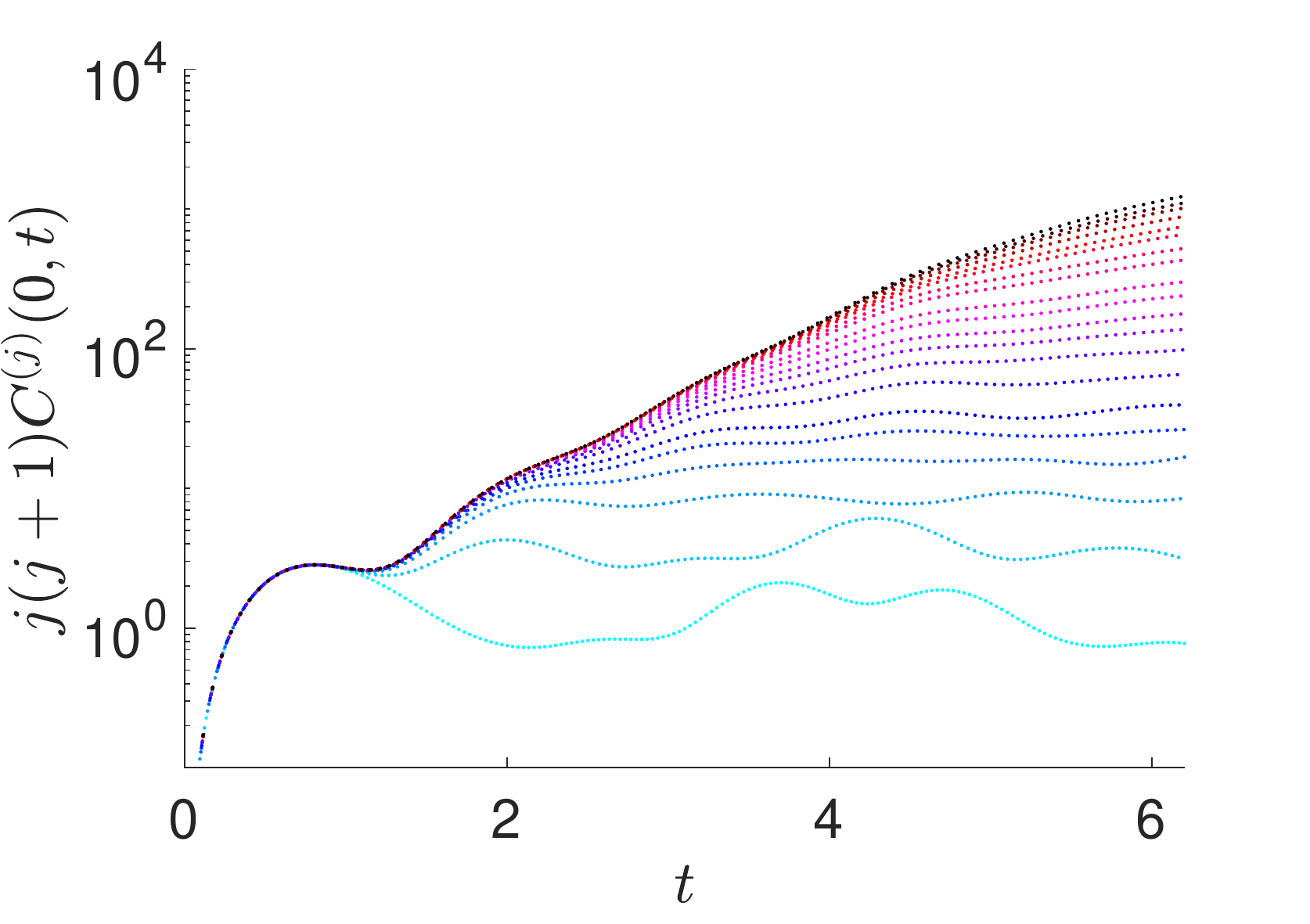}
\includegraphics[width=.48\textwidth]{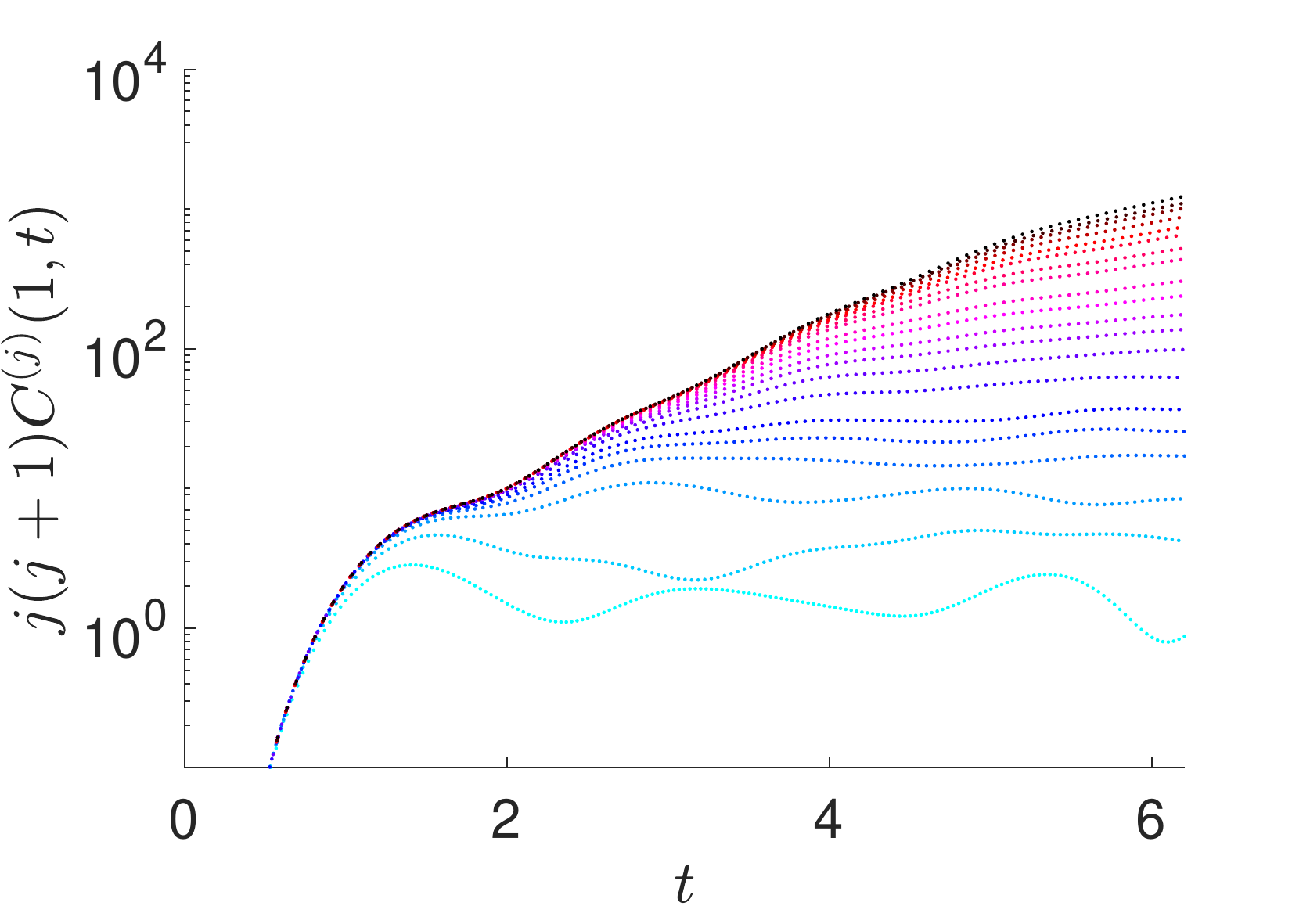}
 \caption{A semi-log plot of the commutator squared is shown as a function of time for various spins, with separation $x=0$ on the left and $x=1$ on the right. The lowermost (light blue) curve corresponds to the spin 1/2 case. The different curves represent increasing spin representations as one moves towards the darkest curve, for which $j=61$. An exponential regime emerges at intermediate values of the commutator squared and grows as we increase the spin.}
 \label{fig: CSvsSpin}
\end{figure}

The smallness of the chain allows
us to use exact diagonalisation of the Hamiltonian $H$ up to $j\sim 61$. In figure \ref{fig: CSvsSpin}, $j(j+1) C^{(j)}(x,t)$
is computed for various spins ranging from $j=1/2$ to $j=61$. 

For the spin 1/2 curve, the early time growth is power law (BCH) as discussed in section \ref{section: review spin 1/2} and immediately followed by the saturation. However, as we increase the local Hilbert space, a region of exponential growth develops before saturation, which is observed to grow with increasing spin.

\subsubsection{Near saturation behaviour}
\label{sec:near_saturation}
We also note the feature that this exponential regime breaks down well below the value to which the commutator squared saturates. 

But first, we need to understand to what value the commutator squared saturates. It can be written in terms of the overlap of the states
\begin{align}
| \psi\rangle &\equiv S_z^{(1)}(t)S_z^{(1+x)}(0) |\beta \rangle \,, \\
| \psi' \rangle &\equiv S_z^{(1+x)}(0)S_z^{(1)}(t) |\beta \rangle \,, \\
C^{(j)} (x,t) &= \big( \langle \psi| - \langle \psi'| \big) \big(| \psi\rangle - | \psi' \rangle \big)\,, \\
&= \langle \psi | \psi \rangle + \langle \psi' | \psi' \rangle - 2 \mathrm{Re} \langle \psi | \psi' \rangle\,,
 \end{align}
where we have replaced the thermal trace by an inner product in a unit norm purifying state $|\beta \rangle$. The thermofield double is an example of such a state.
The first two terms are norms and can be thought of as the expectation value of an operator in a state prepared by inserting an operator at another time
\begin{align}
\langle \psi | \psi \rangle = \left[ \langle \beta | S_z^{(1+x)}(0) \right]  S_z^{(1)}(t) S_z^{(1)}(t) \left[ S_z^{(1+x)}(0) |\beta \rangle \right]\,.
\end{align}
After a dissipation time, we expect the state in brackets to thermalise and be indistinguishable by simple probes from the thermal state. Therefore these terms are expected to decay to the disconnected answer
\begin{align}
\left\langle  S_{z}^{(1+x)}(0) S_{z}^{(1)}(t)  S_{z}^{(1)}(t) S_{z}^{(1+x)}(0) \right\rangle_{\beta = 0} \xrightarrow[t \to \infty]{}
 \left\langle \, (S_{z}^{(1)})^2 \right\rangle_{\beta = 0}
\left\langle \, (S_{z}^{(1+x)})^2 \right\rangle_{\beta = 0} = 1 \,.
\end{align}
On the other hand, the last term involving the overlap between the two states, which is an OTOC, is expected to decay to zero at late times in chaotic systems as a consequence of the butterfly effect \cite{1304.6483,BoundOnChaos}. The state $|\psi \rangle$ can be thought of as being prepared by perturbing the system at time $0$, evolving forward to time $t$ and inserting another perturbation, before evolving the system back to time $0$ in order to compare the state to $|\psi' \rangle$. For chaotic systems and a sufficiently long time $t$, we expect the second perturbation to sufficiently change the trajectory of the evolution such that the first perturbation does not rematerialise when the state is evolved back to time $0$. In other words, the state $|\psi\rangle$ is expected to be indistinguishable from the thermal state for simple operators inserted at time $0$. Since $| \psi'\rangle$ is perturbed at time $0$, the overlap between the two states is expected to be small.

\begin{figure}[th] 
 \centering
\includegraphics[width=.95\textwidth]{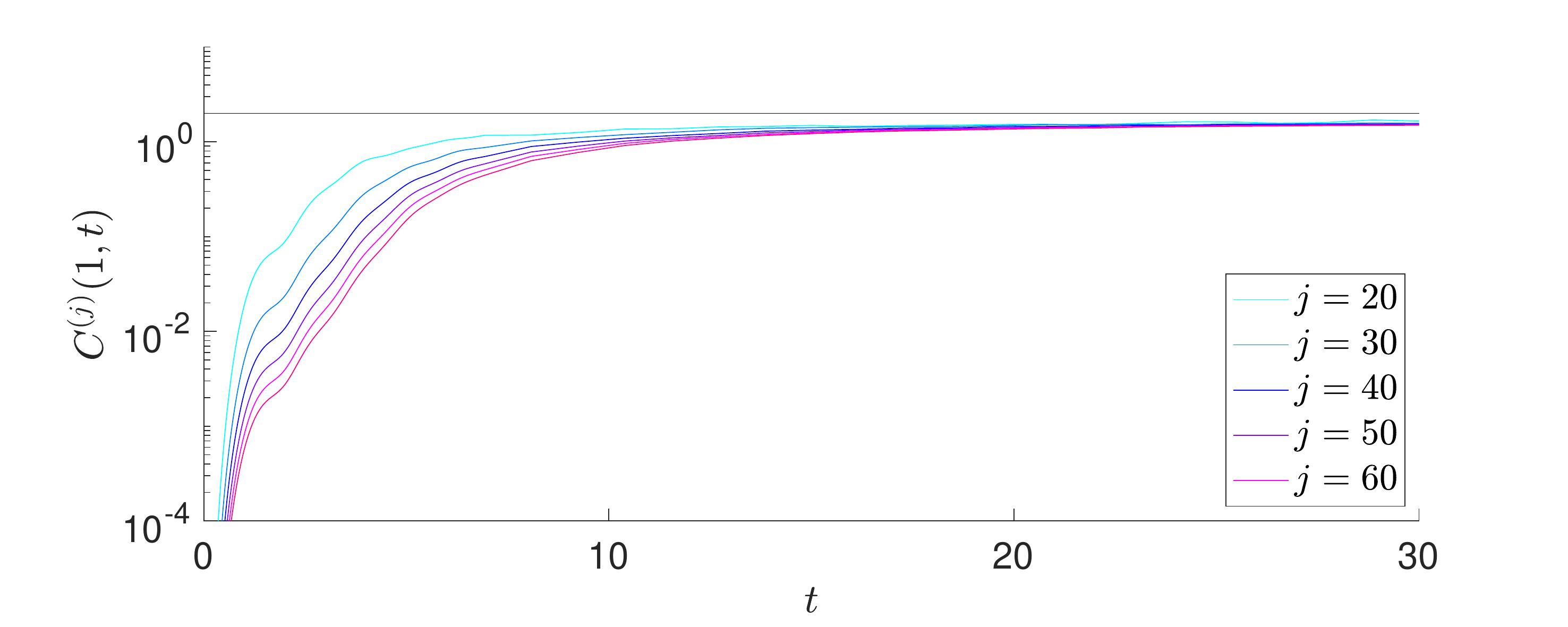}
 \caption{A semi-log plot of the commutator squared displaying its late time saturation for various spins. The horizontal black lines depicts the naive saturation of the commutator squared at the value of 2.}
 \label{fig:late-times}
\end{figure}

These considerations might lead one to expect that
\begin{align}
C^{(j)} (x,t) \xrightarrow[t \to \infty]{}
2 \left\langle \, (S_{z}^{(1)})^2 \right\rangle_{\beta = 0}
\left\langle \, (S_{z}^{(1+x)})^2 \right\rangle_{\beta = 0} = 2 \,.
\end{align} 
However, we find instead that the commutator squared saturates before reaching this value and oscillates around approximately $1.6$ for large spin. This is due to the fact that the OTOC does not strictly go to zero and instead oscillates around a small positive value. Indeed, in \cite{1705.07597} it was understood that for finite size systems with energy conservation, the OTOC saturates to a non-zero value inversely proportional to the system size. 
We will denote the true late time value of the commutator squared as $C_{sat}^{(j)}$.
Figure \ref{fig:late-times} demonstrates the behaviour of the commutator squared beyond the region of exponential growth until  saturation.

In addition to this small deviation in $C_{sat}^{(j)}$, we wish to highlight another feature of the late time behaviour. Denote the approximate value of the commutator squared where the exponential growth breaks down and the near saturation behaviour takes over by $C_{break}^{(j)}$. Then we observe that 
\begin{align}
\Delta_{sat} \equiv |C_{break}^{(j)} - C_{sat}^{(j)}| 
\label{eqn:break}
\end{align}
increases as we increase $j$. This means that the Lyapunov growth breaks down well before saturation and that the separation between these scales is a feature that survives in the classical limit. In \cite{1802.00801}, exponential growth right up to saturation was contrasted with other types of near saturation behaviour. We would like to emphasise that due to the breakdown of the exponential growth observed here, our identification of a region of exponential growth does not have any implications for the near saturation behaviour of the commutator squared and that our results are therefore independent of the analysis presented in that work. We will now focus on the region of exponential growth and leave a more detailed analysis of the near saturation behaviour for future work.

\subsubsection{Extracting a quantum Lyapunov exponent}
\label{sec:quantum_Lyapunov}
\begin{figure}[th] 
 \centering
 \includegraphics[width=.48\textwidth]{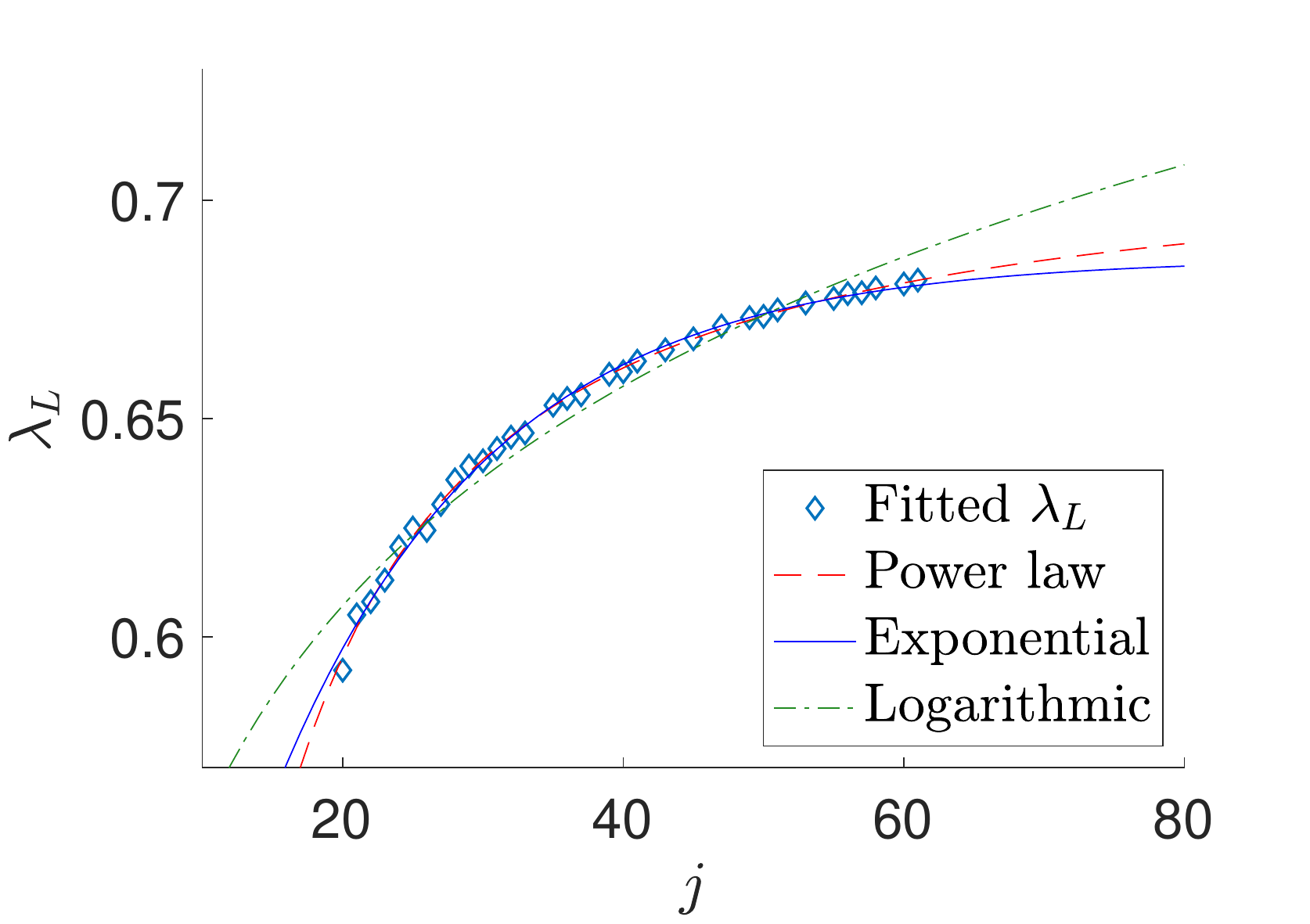}
  \includegraphics[width=.48\textwidth]{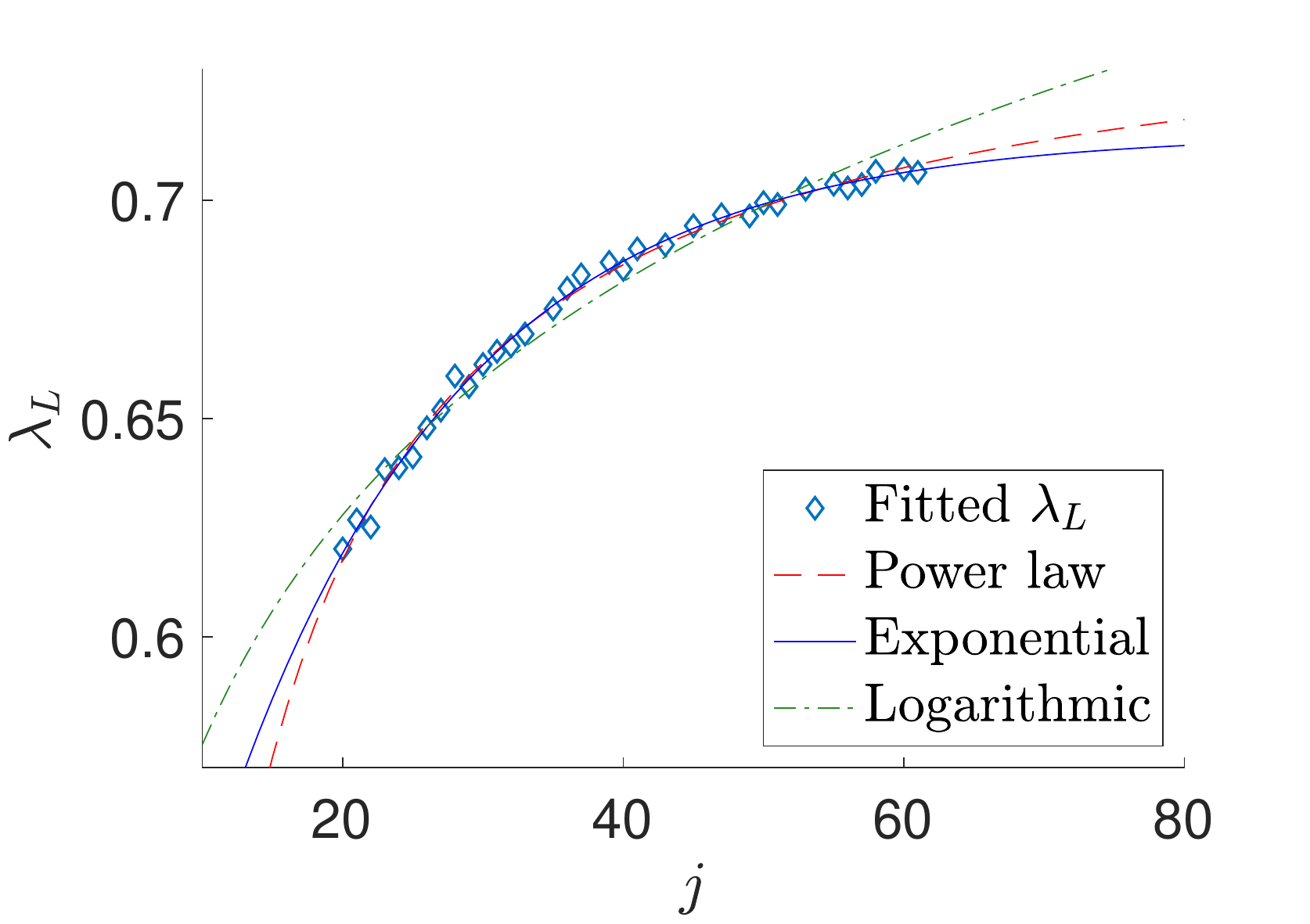}
 \caption{The Lyapunov exponents extracted from the fit to the exponential regime of the commutator squared as a function of the spin $j$ at infinite temperature and at the strongly chaotic point $(h^{*}_x,h^{*}_z)=(-1.05,0.5)$ for the largest time range $[t_i,t_f]$. A power law (red, dashed), an exponential (blue, solid) and a logarithmic (green, dash-dotted) function have been fitted to the data for separations of $x= 0$ (left) and $x=1$ (right). We see that both the power law and exponentials provide a good fit, whereas the logarithm (which grows without bound at large spin) does not. }
 \label{fig: Lyapunov vs Spin} 
\end{figure}

Now that we have observed the qualitative existence of a region of exponential growth in the commutator squared, we would like to give a quantitative estimate of the resulting exponent. 
Although the analysis is performed in the semi-classical regime of the quantum model, we will term this quantity the quantum Lyapunov exponent to distinguish it from the exponent that will be determined in the classical model and to emphasise that it is computed using quantum dynamics. We will fit a line to the Lyapunov region of the semi-log plot of the commutator squared and extract from this the Lyapunov exponent $\lambda_L^{(j)}$ for each choice of the spin $j$, see figure \ref{fig: Lyapunov vs Spin}. As we increase the spin towards the classical limit, we observe that the quantum Lyapunov exponent saturates to a finite value $\lambda_L^\infty$ which we compare to the Lyapunov exponent extracted from a classical analysis in the next section.

However, there are a number of ambiguities that must be fixed in this procedure. In order to quantify the uncertainty introduced by these ambiguities, we will vary the choices we make and produce a distribution of $\lambda_L^\infty$. The variance of this distribution will then give us a rough estimate of the size of these uncertainties.

The first ambiguity comes from the presence of small fluctuations on top of the exponential regime (which appears as a linear regime in the semi-log plot of figure \eqref{fig: CSvsSpin} which make the fit to this exponential regime depend on exactly which time interval $[t_i,t_f]$ we fit to. We will argue at the end of this section that this may be due to edge effects in the 2-sites setting, but in the meantime we would like to estimate the uncertainty in $\lambda_L^\infty$ coming from this ambiguity. To do so, we will choose a variety of different cuts within which to perform the linear fit on the semi-log plot and we will use the variance in $\lambda_L^\infty$ as we vary these cuts as an estimate of the ambiguity introduced from this issue.

\begin{figure}[th]
 \centering
	\begin{minipage}[b]{.9\textwidth}
	      \centering
		\includegraphics[width=0.48\textwidth]{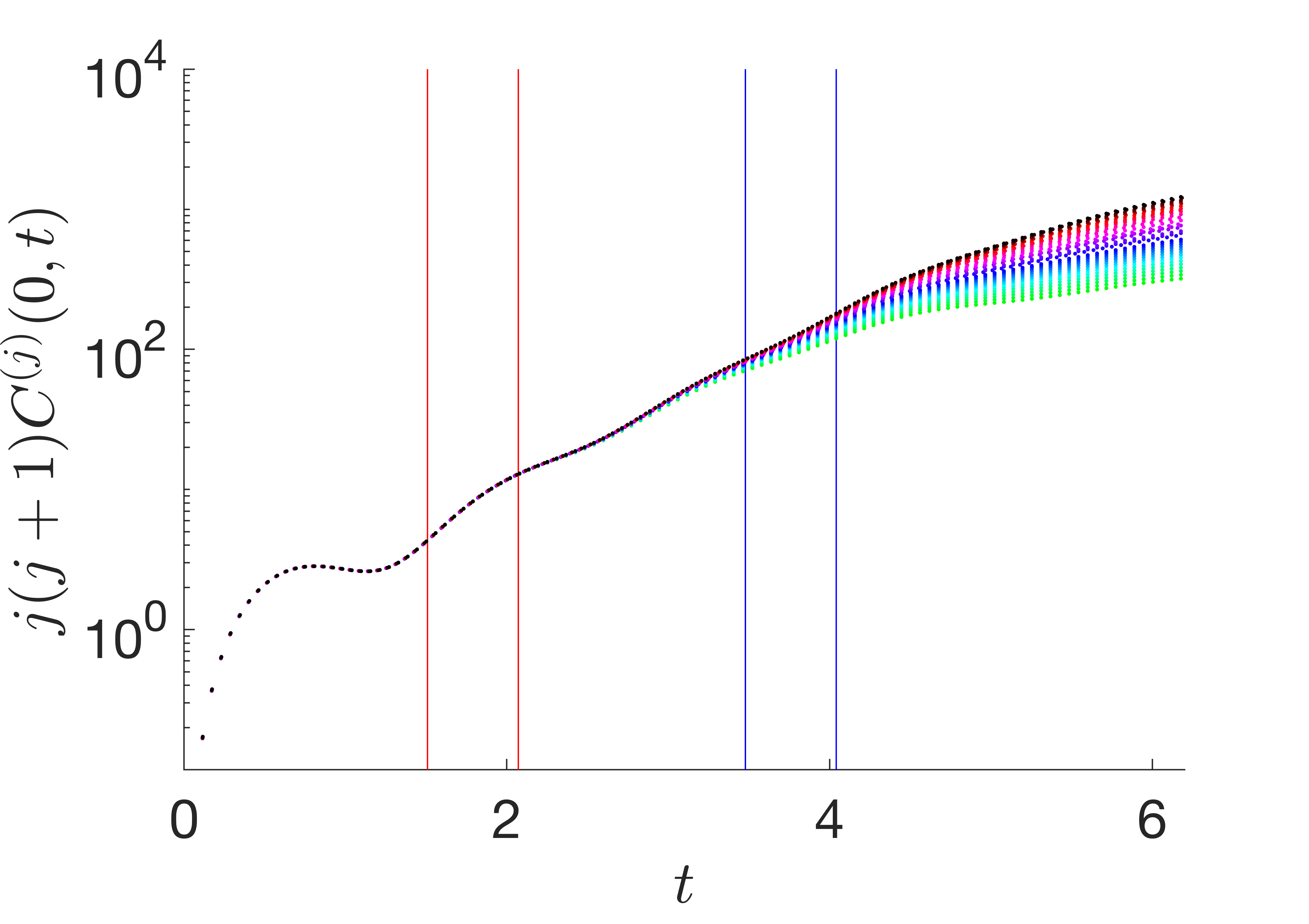}
		\includegraphics[width=0.48\textwidth]{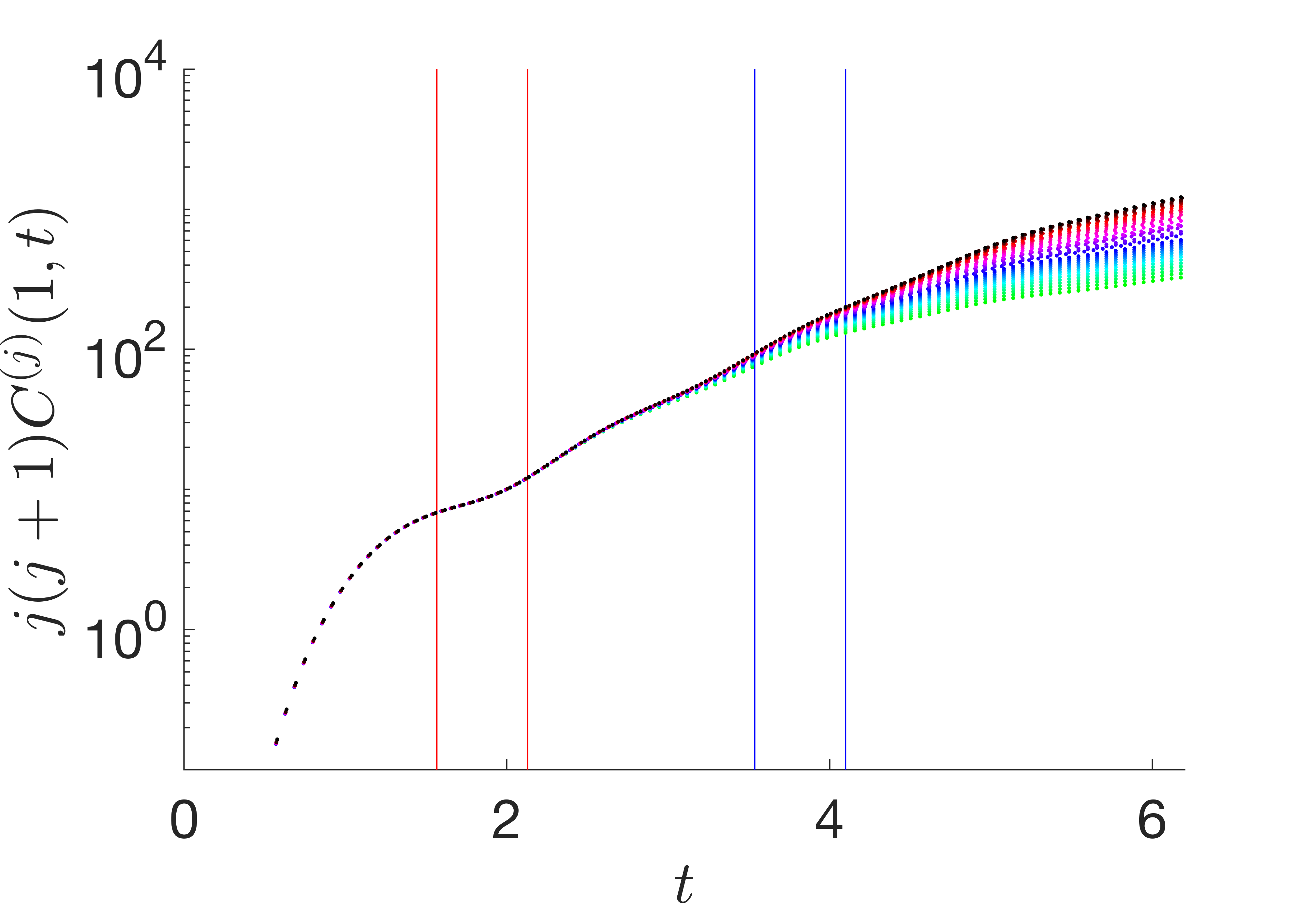}
	\end{minipage}
	\caption{The various time intervals used for the fits of the Lyapunov exponents at the strongly chaotic point. The lower (upper) bounds of the time interval $t_i$ ($t_f$) are taken in between the two vertical red (blue) lines.
	The left figure is for $x=0$ and the right for $x=1$.}
 \label{fig: CSvsSpin fits region}
\end{figure}
	
\begin{figure}[th]	
	\begin{minipage}[b]{.9\textwidth}
		\centering
		\includegraphics[width=0.48\textwidth]{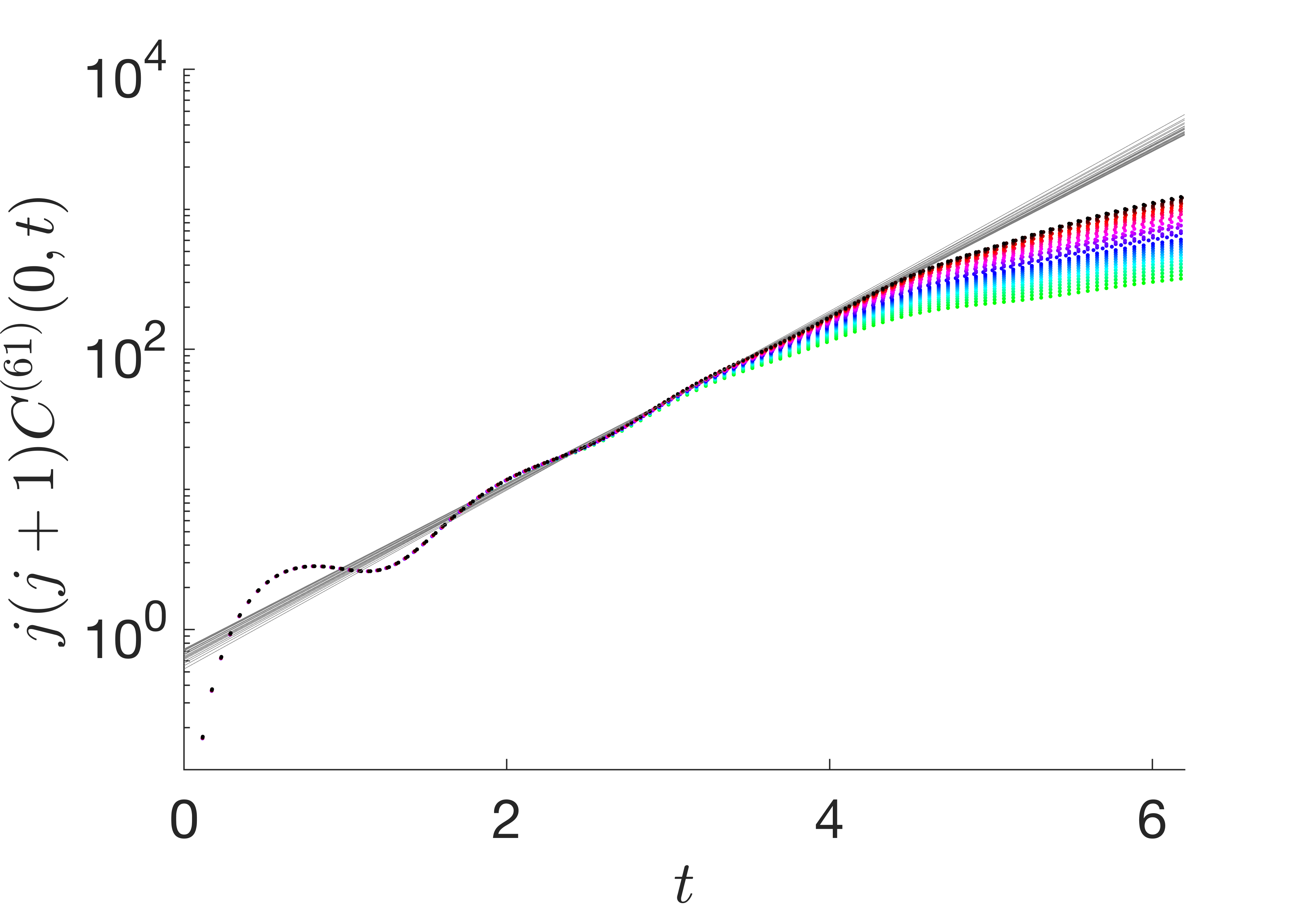}
	    \includegraphics[width=0.48\textwidth]{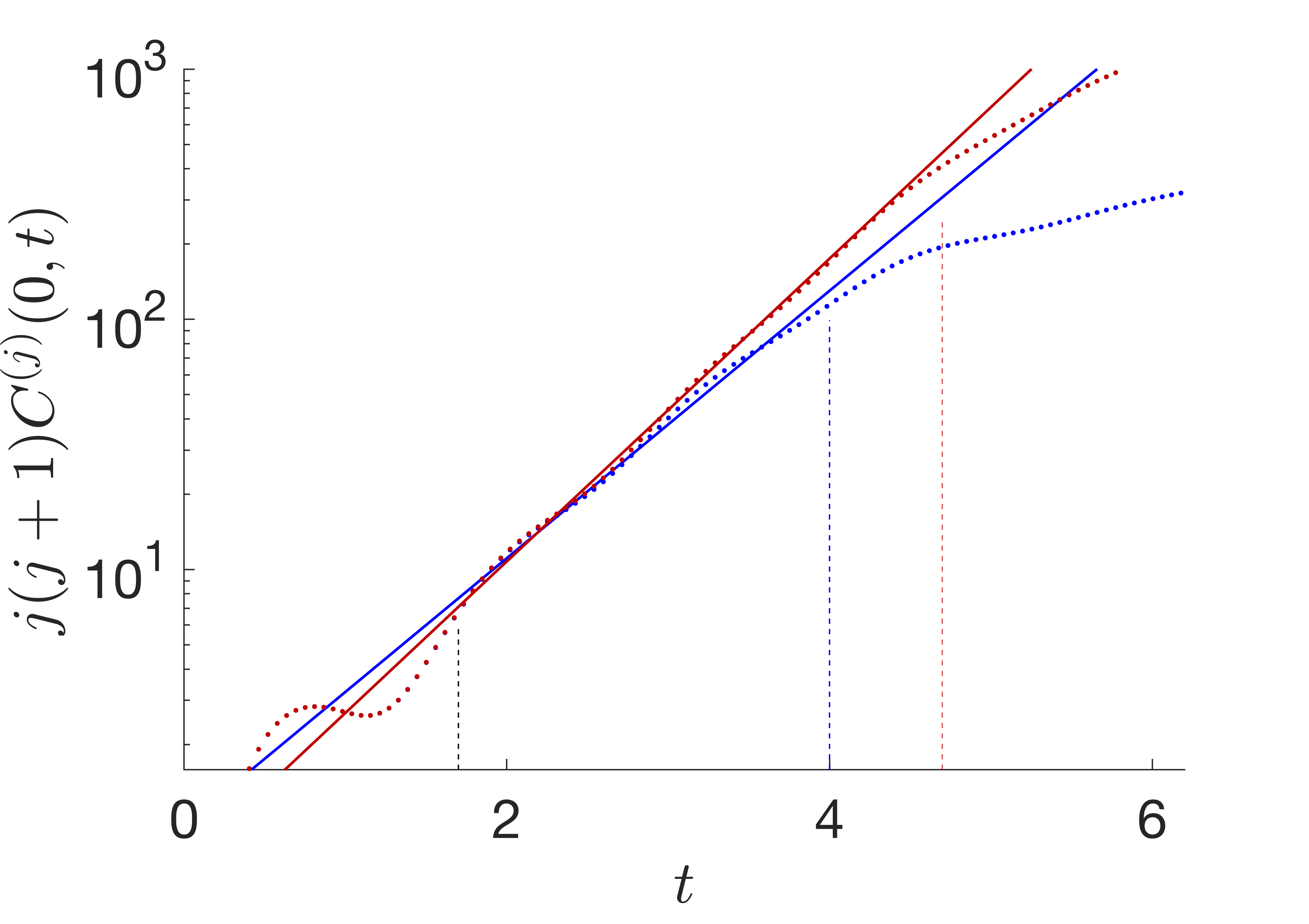}
	\end{minipage}
 \caption{The left figure demonstrates the spread of the linear fits (gray) to the region of exponential growth for the commutator squared at the highest spin ($j=61$) we studied for various choices of the time interval to which we fit. The commutator squared is shown for $j=1, 7, 16, 20, 25, 33, 45$ and $ 61$, to illustrate its approach to these linear fits. The right figure depicts commutator squared and its corresponding fit for $j=20$ in blue and $j=61$ in red. The dashed vertical lines illustrate the time at which the data starts to deviates from the fit. The region where the data fits well to a line corresponds to an exponential growth of roughly 3 and 4 e-foldings for $j=20$ and $61$ respectively.}
  \label{fig: CSvsSpin fits highest}
	\end{figure}

To be specific, we start by determining the time at which the difference between the commutator squared of the highest two spins, normalised by the value for the highest spin, becomes one percent. This time gives a rough estimate as to 
where the exponential growth starts to breakdown
and at the same time seems to be early enough (see figure \ref{fig: CSvsSpin fits region}) to be a valid upper bound for the lower spins as well. This time will therefore be chosen as an upper bound for $t_f$ for all spins. Afterwards, we pick the lower bound $t_i$ at the time where $j(j+1) C^{(j)} (x,t) = 9$, which roughly coincides with the time at which the exponential regime starts. 
We divide this time range into four equal parts, and define within the first and last part five equally spaced initial and final times.
In this way, the commutator squared appears to be in the exponential regime for every one of the 25 constructed time intervals for all the spins we wish to analyse. 

This procedure is intended to account for the small fluctuations, as the range of the initial (final) times $t_i$ ($t_f$) covers roughly one wiggle of the commutator squared. We illustrate this for the strongly chaotic point in figure \ref{fig: CSvsSpin fits region}, together with the corresponding linear fits to the commutator squared restricted to those intervals for the highest spin used in this analysis ($j=61$) in figure \ref{fig: CSvsSpin fits highest} and we comment on the size of the region to which can be fit an exponential growth.

Now that we have defined all these time intervals, we fit a line in each interval to the curve for the commutator squared at each spin. Fixing the time interval, we generate $\lambda_L^{(j)}$ for each spin and extrapolate these to $\lambda_L^\infty$ using a procedure which will be described below. This gives us a distribution of $\lambda_L^\infty$ for all the different choices of the time interval. By taking the mean and standard deviation of this distribution we obtain an estimate of $\lambda_L^\infty$ and the uncertainty on this estimate coming from the ambiguity of the fitting procedure.
 
Next, there is an ambiguity in how to perform the extrapolation of $\lambda_L^{(j)}$ to $\lambda_L^\infty$. The dependence of the Lyapunov exponent on the spin is displayed in figure \ref{fig: Lyapunov vs Spin}. We see that $\lambda_L^{(j)}$ is monotonic in $j$ and that the rate of increase slows as $j \to \infty$, such that the Lyapunov exponent saturates towards a finite value, $\lambda_L^{\infty}$, in the infinite spin classical limit. 
We tried to fit two types of functional forms to $\lambda_L^{(j)}$:  an exponential of the form 
\begin{align}
\lambda_{L}^{(j)} = \lambda^{\infty}_{L,exp} - a_1 \exp(-a_2\sqrt{j(j+1)}) \,,
\end{align}
 and a power law  of the form 
 \begin{align}
 \lambda_{L}^{(j)} = \lambda^{\infty}_{L,pow} - a_1 \left(j(j+1)\right)^{-\frac{a_2}{2}} \,,
 \end{align}
which were both found to provide a good fit. We also tried fitting an unbounded function of the form
\begin{align}
\lambda_L^{(j)} = a_1 + a_2\ln \sqrt{j(j+1)}\,,
\end{align}
but this was not found to provide a good fit, giving further confidence that this quantity saturates in the classical limit.  These three types of fits are compared in figure \ref{fig: Lyapunov vs Spin}. In addition to varying the time interval over which we fit, we will also extrapolate using both of the exponential and the power law functional forms for each time interval when generating our distribution for $\lambda_L^\infty$.

\begin{figure}[th]
\centering
\begin{minipage}{0.45\textwidth}
$x=0$\\
\includegraphics[width=.95\textwidth]{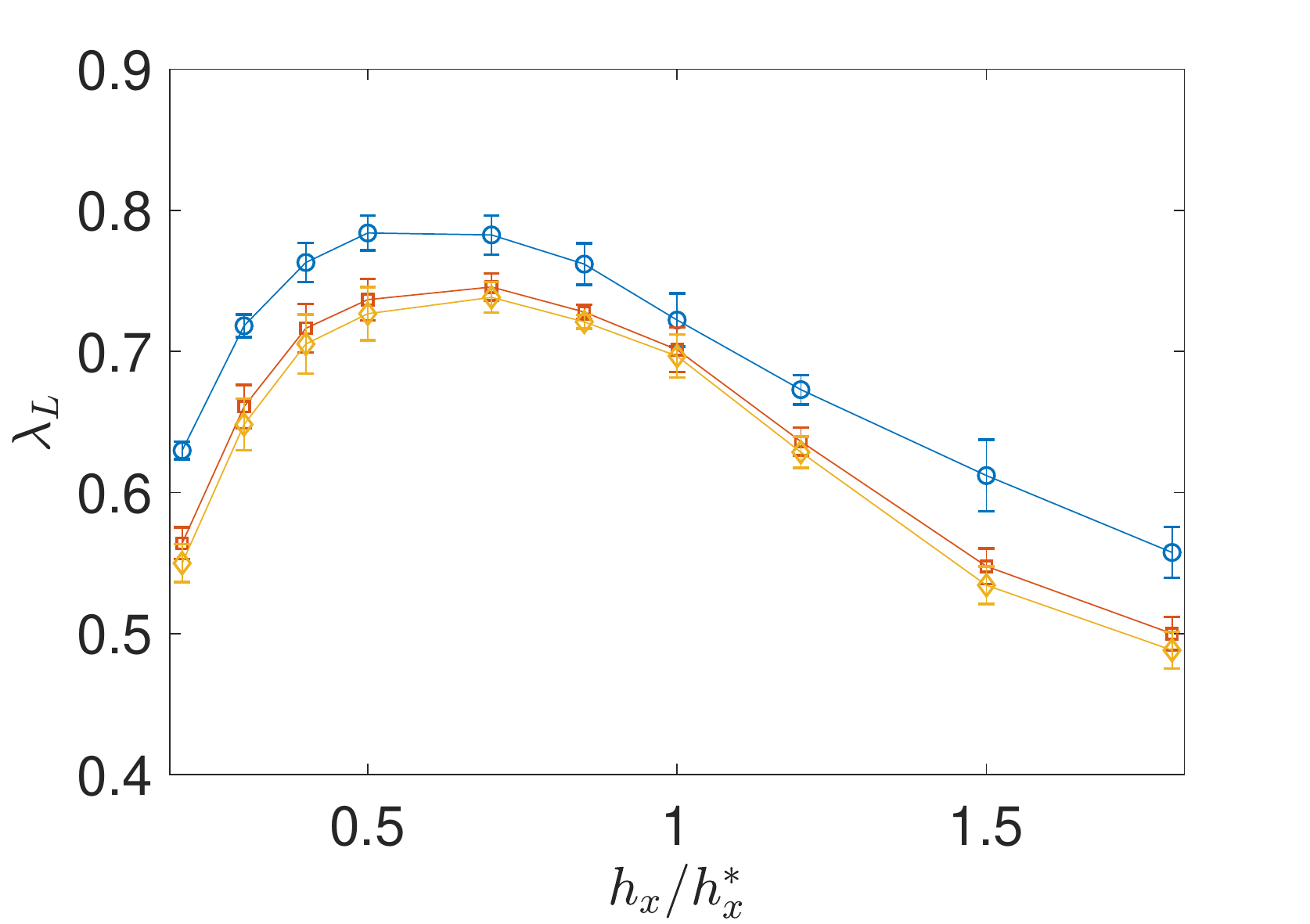}
\end{minipage}
\begin{minipage}{0.45\textwidth}
$x=1$\\
\includegraphics[width=.95\textwidth]{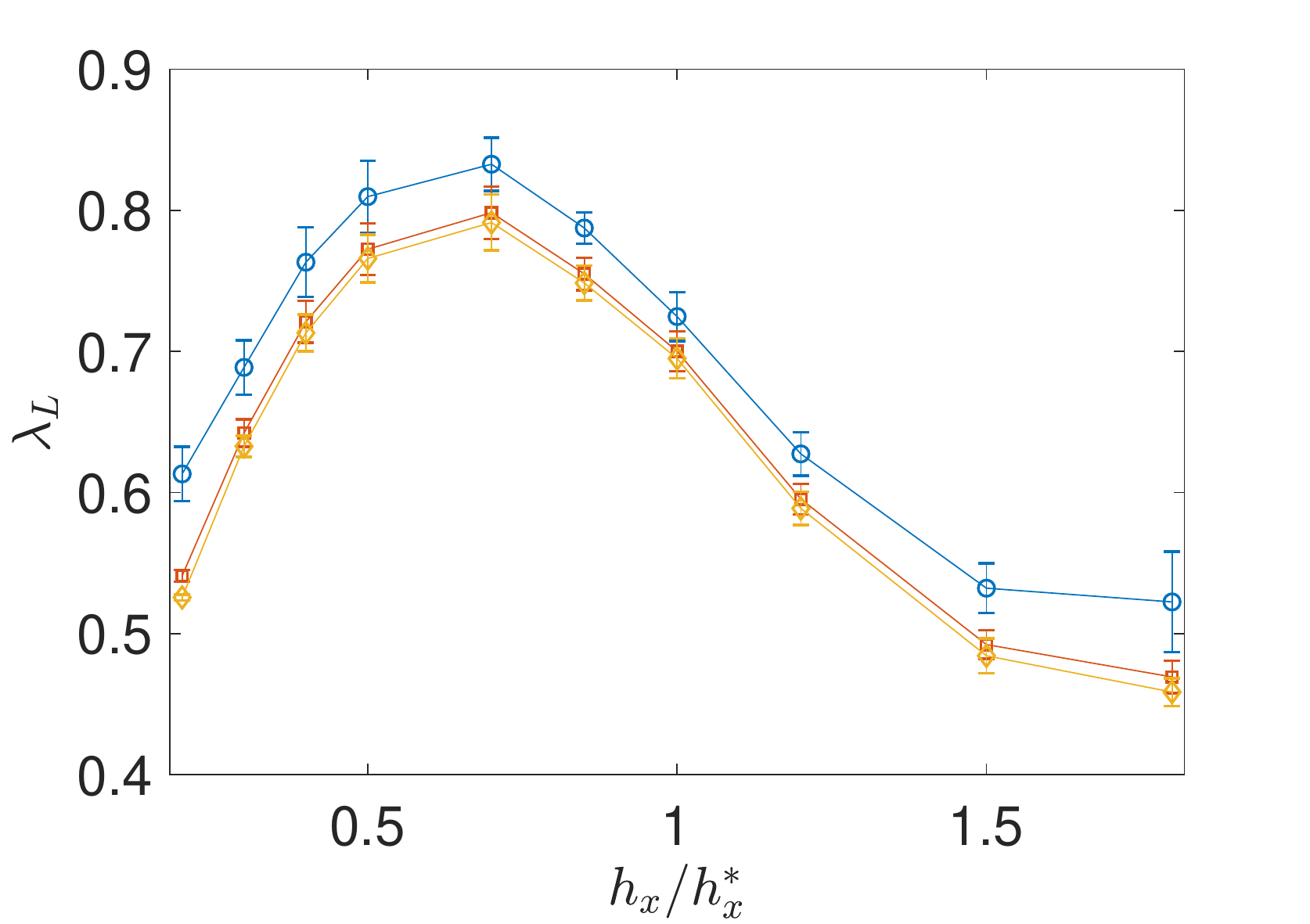}
\end{minipage}
\caption{A comparison between the Lyapunov exponent at the highest computed spin (yellow diamonds) and the ones obtained from a power law (blue circles) or an exponential (red circles) extrapolation to infinite spin. The transverse field $h_x$ is varied around the strongly chaotic value of $h_x^*=-1.05$ and the longitudinal field is fixed at $h_z^*=0.5$. The error bars denote one standard deviation of the distribution of Lyapunov exponents found using the procedure outlined in section \ref{sec:quantum_Lyapunov}.}
\label{fig:lambda_vs_hx_extrapolation}
\end{figure}

We could also simply provide a lower bound on the extrapolated value by looking at the value of $\lambda_L^{(j)}$ for the highest spin we have analysed. In fact, this turns out to be quite close to the result found by the exponential extrapolation, although the power law approach gives a higher value. These three approaches are compared in figure \ref{fig:lambda_vs_hx_extrapolation}. However, we will stick to using the two extrapolation methods described above as we expect $\lambda_L^{(61)}$ to systematically underestimate the infinite spin limit.

\begin{figure}[th] 
 \centering
 \includegraphics[width=0.45\textwidth]{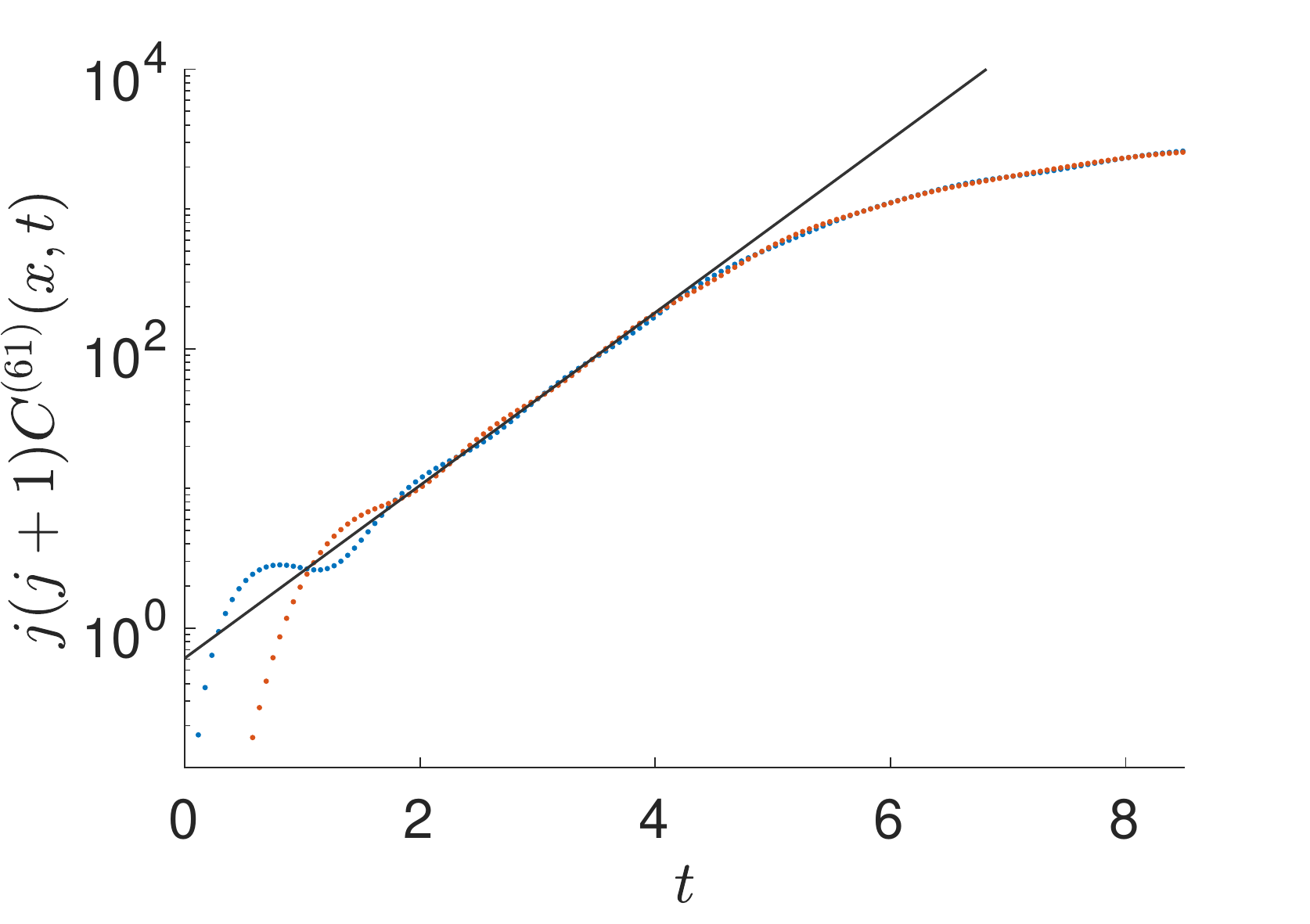}
 \caption{The commutator squared for the highest spin at the strongly chaotic point together with the slope (black) found by averaging the Lyapunov exponents over both extrapolation methods and over the two sites. The blue (orange) curve is for $x=0$ ($x=1$).}
 \label{fig: CSvstfits1}
\end{figure}

We have described two ambiguities in our procedure for extracting $\lambda_L^\infty$: 1) the choice of time interval; and 2) the choice of extrapolating function.  By varying the choice of time interval we produce a distribution for $\lambda_L^\infty$ and compute its mean and standard deviation for each extrapolation separately. In table \ref{tbl:Lyapunov}, we provide the results for the extrapolated Lyapunov exponent using each of the extrapolation methods for both of the sites of our chain at the strongly chaotic point. In figure \ref{fig: CSvstfits1} we show the commutator squared at $j=61$ for the two sites together with the slope set by averaging the Lyapunov exponents over both sites and over the exponential and the power law extrapolation. 
In the next section, we will compare the extrapolated infinite-spin quantum Lyapunov exponents to that obtained from an analysis of the system in the classical limit as we vary the magnetic fields.

\begin{figure}[th] 
 \centering
 \includegraphics[width=0.32\textwidth]{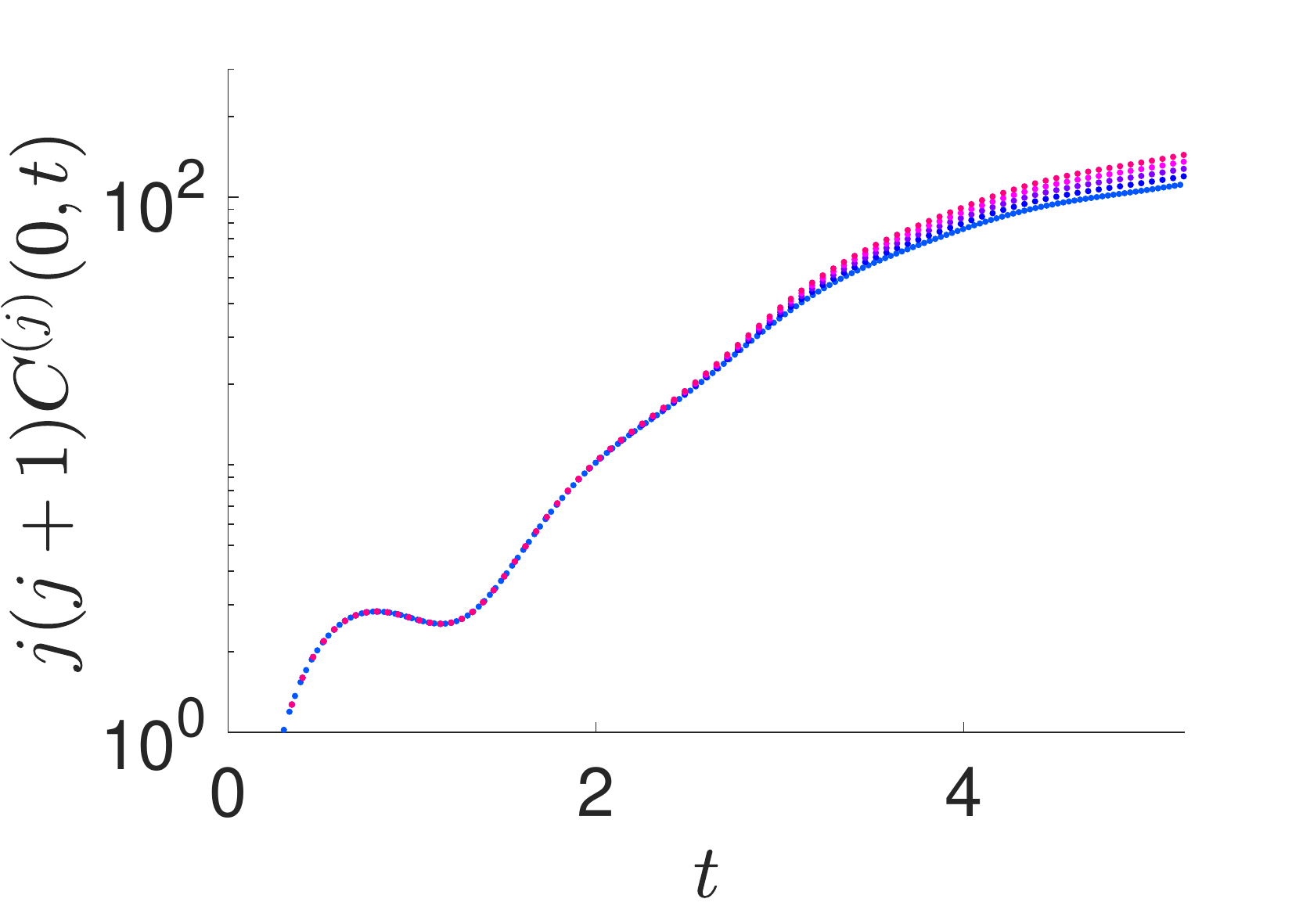}
 \includegraphics[width=0.32\textwidth]{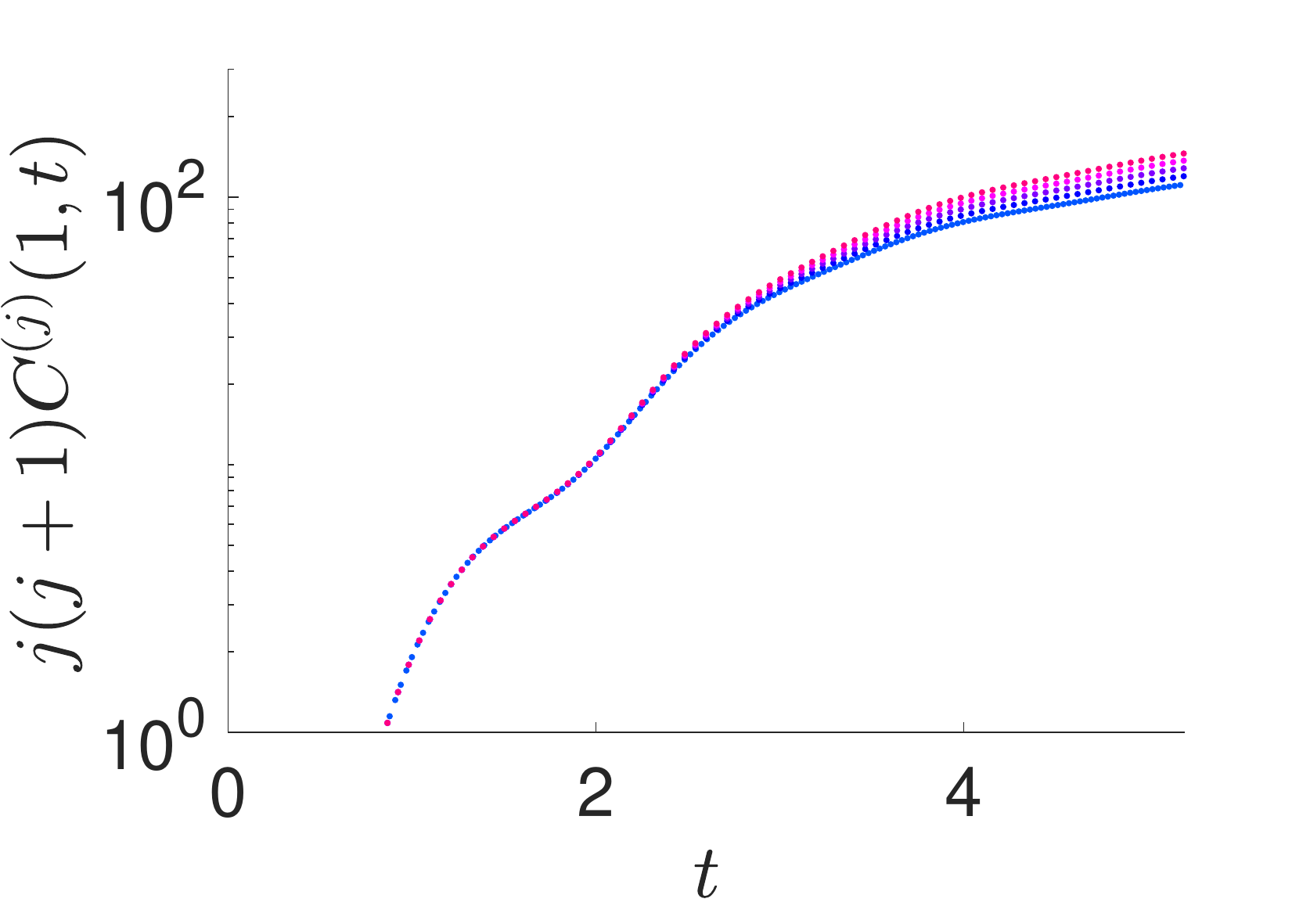}
 \includegraphics[width=0.32\textwidth]{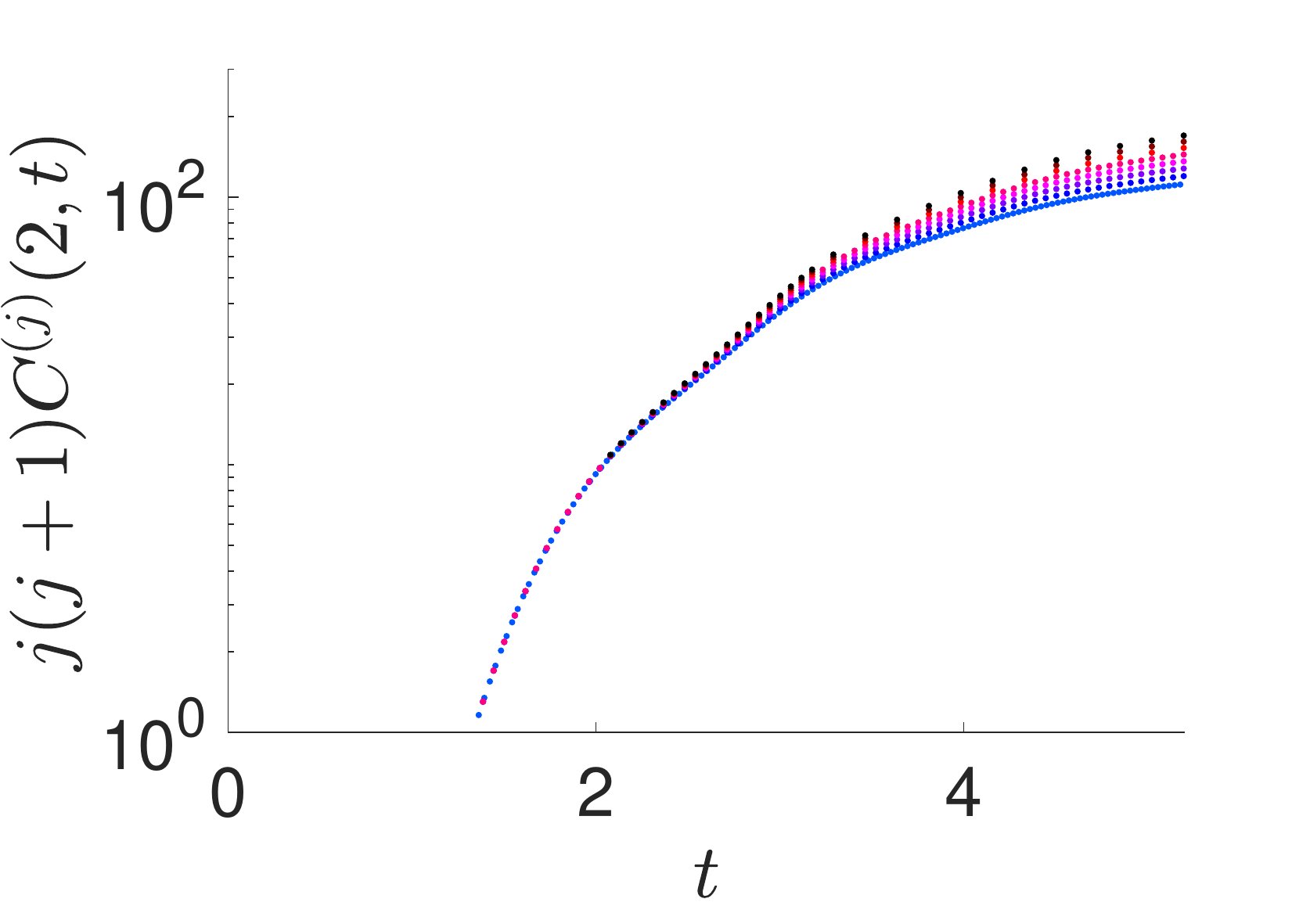}
 \caption{A semi-log plot of the time-dependence of the commutator squared at the strongly chaotic point for a 3-site chain. The small fluctuations on top of the exponential regime that are present when the perturbation is close to the probe site seem to become already almost imperceptible for a distance of 2 between the perturbation and the probe site.}
 \label{fig: CSvsSpin L=3}
\end{figure}

To conclude the semi-classical analysis, we would like to argue that the fluctuations on top of the exponential behaviour are due to edge effects in $L=2$ chains and that we expect them to vanish for larger separation between the two operators in the commutator squared. In particular, by inspecting $L=3$ chains for relatively small $j$ where the exponential regime starts to appear, we notice that for larger spatial separation between the perturbation and probe sites the fluctuations seem to fade away, as seen in figure \ref{fig: CSvsSpin L=3} for $j$ ranging between 10 and 13.5. Although this argument should be taken with care, as we are restricted to very low spin and therefore an exponential regime is barely present, this figure gives some support the statement that the analysis we just performed can be applied to longer chains and that the exponential regime of the commutator squared is not only present in $L=2$ chains.

\section{Classical Analysis}
\label{sec:classical}
In this section, we will implement the procedure described in section \ref{sec:classical_commutator_squared}, to compute the classical analogue of the commutator squared, $C^{(cl)}$, and compare it to the results obtained in the classical limit of our quantum model.

\subsection{Extracting a classical Lyapunov exponent}
\label{sec:classical_lyapunov}
The quantum case includes an average over a thermal ensemble at infinite temperature. The initial conditions in the classical averaging are therefore also selected from a Boltzmann ensemble at infinite temperature, i.e.\ a uniform distribution. Since the phase space is a product a spheres, the appropriate uniform distribution for this space must be used, namely
\begin{align}
P(\theta^{(n)},\phi^{(n)}) = \bigwedge_{n=1}^L \frac{\sin\theta^{(n)}}{4\pi} \; d\theta^{(n)} \wedge d\phi^{(n)}\,.
\end{align}

As discussed in section \ref{sec:classical_commutator_squared}, in practical computations the exponential regime only persists until the saturation of $C^{(cl)}(x,t)$ at $O(\epsilon^{-2})$, where $\epsilon$ is the small perturbation in the initial condition. The exponential regime can be extended by decreasing the size of the initial perturbation. However, in practice the size of the perturbation is limited by the numerical accuracy of the computation and so we will usually use perturbations of order $10^{-6}$.

\begin{figure}[th]
\centering
\begin{minipage}[b]{0.49\textwidth}
		\centering
		\includegraphics[width=0.9\textwidth]{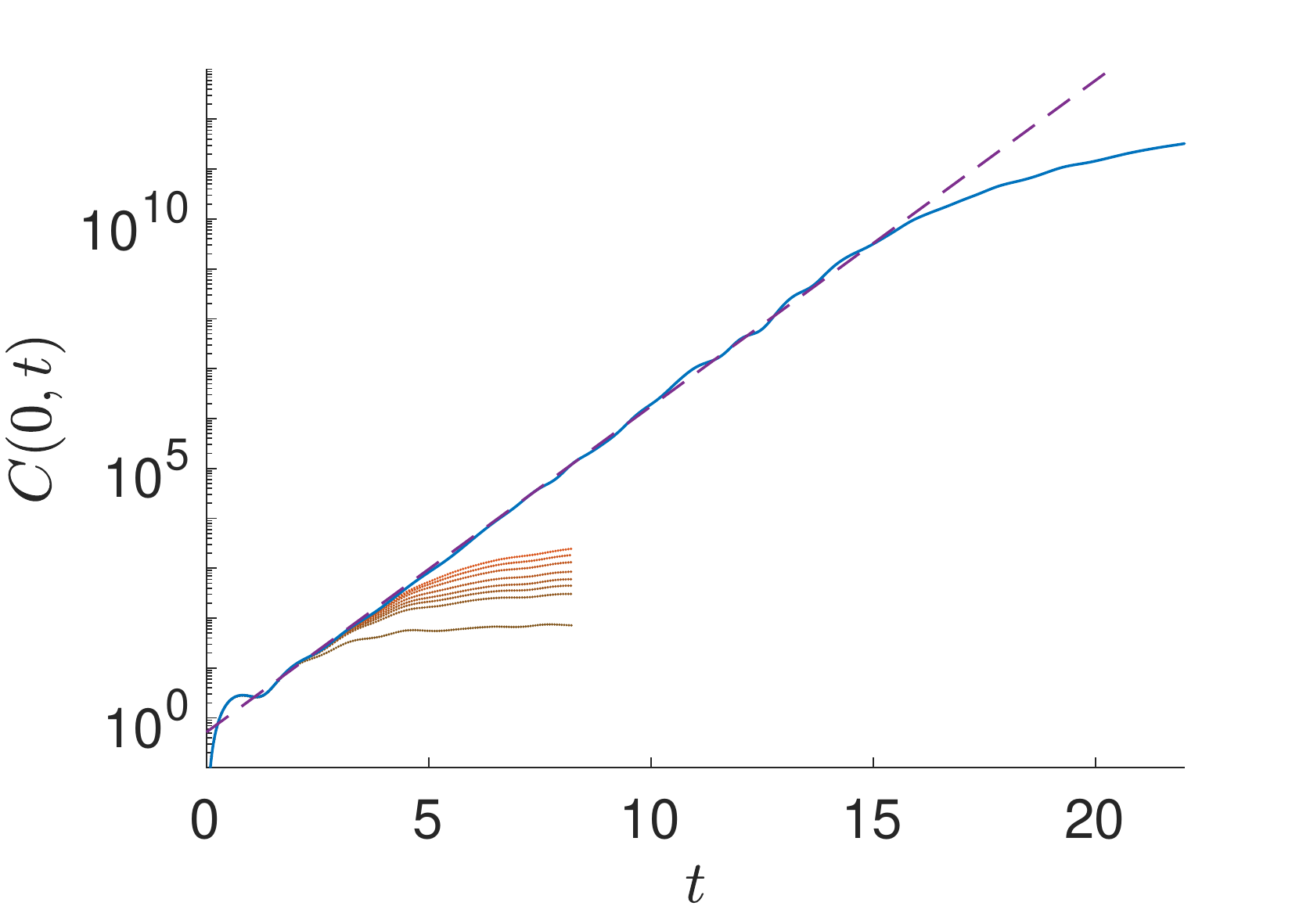}
	\end{minipage}
	\begin{minipage}[b]{0.49\textwidth}
		\centering
		\includegraphics[width=0.9\textwidth]{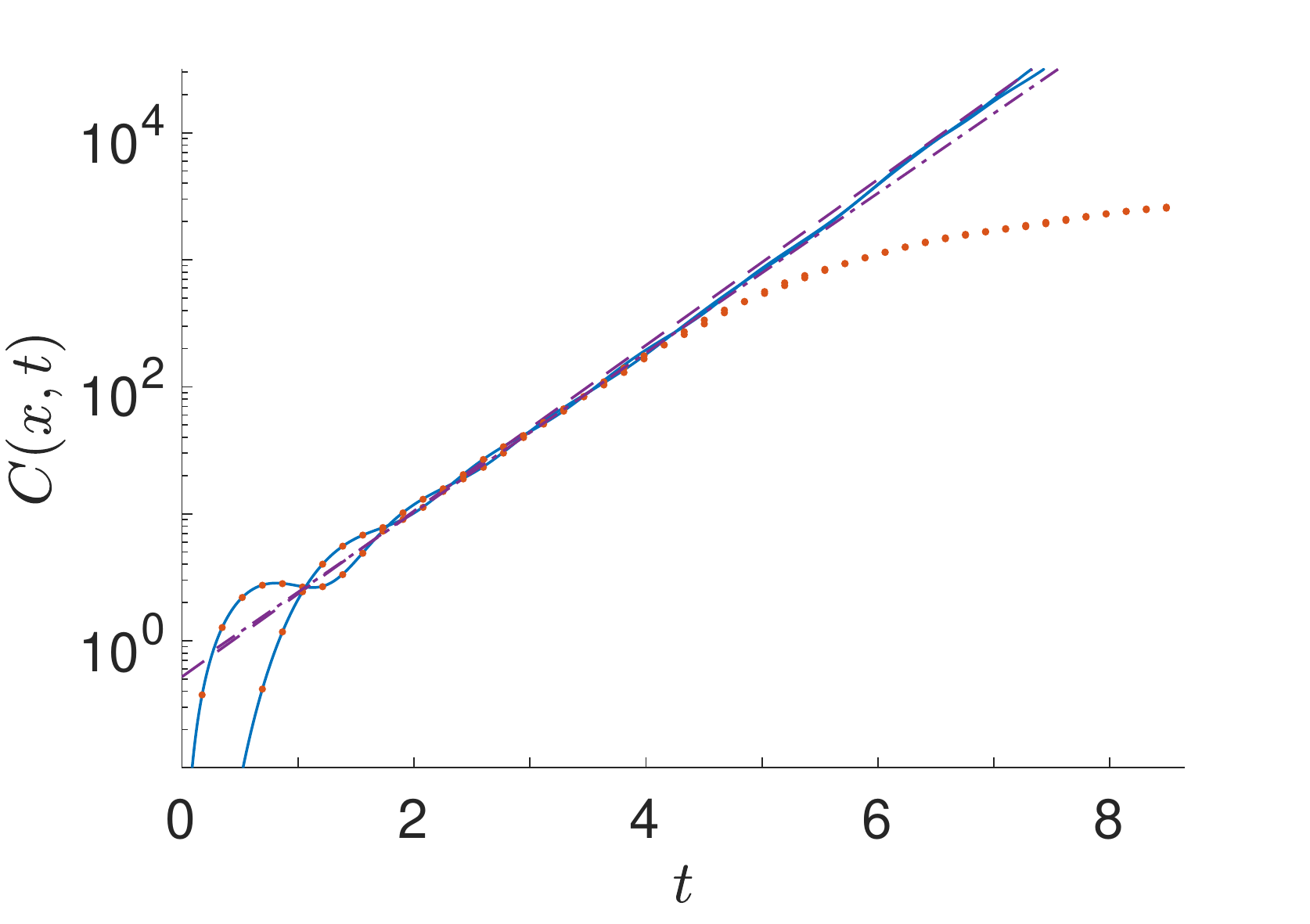}
	\end{minipage}
\caption{The left figure shows the classical version of the commutator squared (blue solid), which is the average over initial conditions of a Poisson bracket squared, between $S_z^{(1)}(t)$ and $S_z^{(1)}(0)$. This is plotted for a 2-site chain at the strongly chaotic point. The classical Lyapunov exponent is extracted from a fit (purple dashed) to the exponential regime. On top of the Poisson bracket, the commutator squared (times the usual factor of $j(j+1)$) is shown for increasing spins in red. On the right figure, the classical (blue solid) and highest spin quantum (red dotted) commutator squared are displayed at early times for both sites, with their respective fits. The slope of the fit to the classical data (purple dashed) is obtained by averaging the Lyapunov exponents at the two sites, while an average of the power law extrapolated Lyapunov exponent at both sites is used as slope to show the fit to the quantum data (purple dashed-dotted).}
\label{Pbvst}
\end{figure}

Figure \ref{Pbvst} shows the result for the average of the Poisson bracket as a function of time in a semi-log plot. The curve has an intermediate regime which consists of small fluctuations superimposed on an overall linear growth. The slope of a linear fit to this regime of the curve is what we will call the classical Lyapunov exponent. At high magnetic fields, some fluctuations persist even when averaging over 
$O(10^5)$ 
initial conditions. This results in the slope being sensitive to the exact endpoint of the time interval on which we fit. Furthermore there is a smooth cross-over between the Lyapunov regime and the subsequent saturation regime, making the exact endpoint of the Lyapunov regime somewhat arbitrary. Therefore we vary the endpoints of the fitting region and compute the mean and standard deviation of the resulting distribution to produce the Lyapunov exponent $\lambda_L$ and an estimate of the uncertainty due to this ambiguity.\footnote{This is another average than the one over initial conditions, the curve in figure \ref{Pbvst} is already the average over curves at many different initial conditions.} The start of the fitting time interval is kept fixed at the time when $C^{(cl)}$ crosses 9, while the end of the fitting time interval is varied.

\begin{table}[t]
\centering
\begin{tabular}{l | cc}
& site 1 & site 2 \\
\hline
$\lambda_{L,pow}^{\infty}$ & $0.722 \pm 0.019$ & $0.725 \pm 0.017$ \\
$\lambda_{L,exp}^{\infty}$  & $0.701 \pm 0.016$ & $0.700 \pm 0.014$ \\
$\lambda_L^{(61)}$              & $0.697\pm 0.015$ &  $0.695 \pm 0.014$ \\
$\lambda_{L}^{classical}$     & $0.752 \pm 0.013$ & $0.750 \pm 0.010$\\
\end{tabular}
\caption{Lyapunov exponents at the strongly chaotic point extrapolated to infinite spin using different methods compared to the result from the classical analysis.}
\label{tbl:Lyapunov}
\end{table}

In table \ref{tbl:Lyapunov}, we summarise our results for the Lyapunov exponent at the strongly chaotic point. We find that the errorbars of the classical exponent and the power law extrapolation overlap, which suggests that the power law approach provides a better estimate than the exponential extrapolation.

On the left of figure \ref{Pbvst}, we present the commutator squared computed at various spins on top of the Poisson bracket. We see that the two match very well at early times until the commutator squared starts to saturate. As we increase $j$, we see that they match for a longer range of time and that the commutator squared is indeed converging to the Poisson bracket. 
On the right of figure \ref{Pbvst}, we present the highest spin commutator squared on top of the Poisson bracket alongside a line with a slope given by the Lyapunov exponents extracted by our procedure outlined so far. We see that at a qualitative level these lines both provide a reasonable fit to the curve. 

The small differences between them are due to the details of how they are extracted. The quantum Lyapunov at fixed spin tends to be lower since the commutator squared tends to deviate from the classical curve towards a lower value as it moves towards saturation. This means that the extracted exponent is dependent on exactly where we cut off the fitting region and that it tends to get biased downwards. The classical Lyapunov is extracted by fitting to the much longer period of exponential growth visible in that case. However, this means it is sensitive to fluctuation that can appear beyond the times reached by the commutator squared. Potential sources of these fluctuations include the residual variance in the Monte Carlo approach to computing the phase space average over trajectories as well as intrinsic fluctuations away from pure exponential growth in the Poisson bracket. These issues speak to the difficulty in extracting a Lyapunov exponent from numerical data and lead to the variances observed in table \ref{tbl:Lyapunov}. None the less, figure \ref{Pbvst} provides convincing evidence that the commutator squared is exhibiting a region of exponential growth that captures the exponential divergence of trajectories in the classical limit of the model.

Figure \ref{Pbvst} also demonstrates that the scale at which the linear growth in the commutator squared breaks down, which we denoted $C_{break}^{(j)}$ in equation \eqref{eqn:break}, should be identified with the scale at which the classical-quantum correspondence breaks down. The time at which this occurs is known as the Ehrenfest time and we could also define a concomitant Ehrenfest scale. The discussion in section \ref{sec:near_saturation} of the near saturation behaviour of the commutator squared can therefore be recast in terms of understanding the behaviour of the system after the Ehrenfest time. This suggests that the reason for the existence of this new scale where the linear growth breaks down may be related to wavepackets spreading as is the case in single-particle models involving coherent states. However, this line of reasoning doesn't lead us immediately to a better understanding in our many-body model, since the correspondence we use is between thermal averages which do not obviously suffer from this issue.

\begin{figure}[th!]
	\centering
	\begin{minipage}[b]{0.49\textwidth}
		\centering
		$x=0$ \\
		\includegraphics[height=5.5cm]{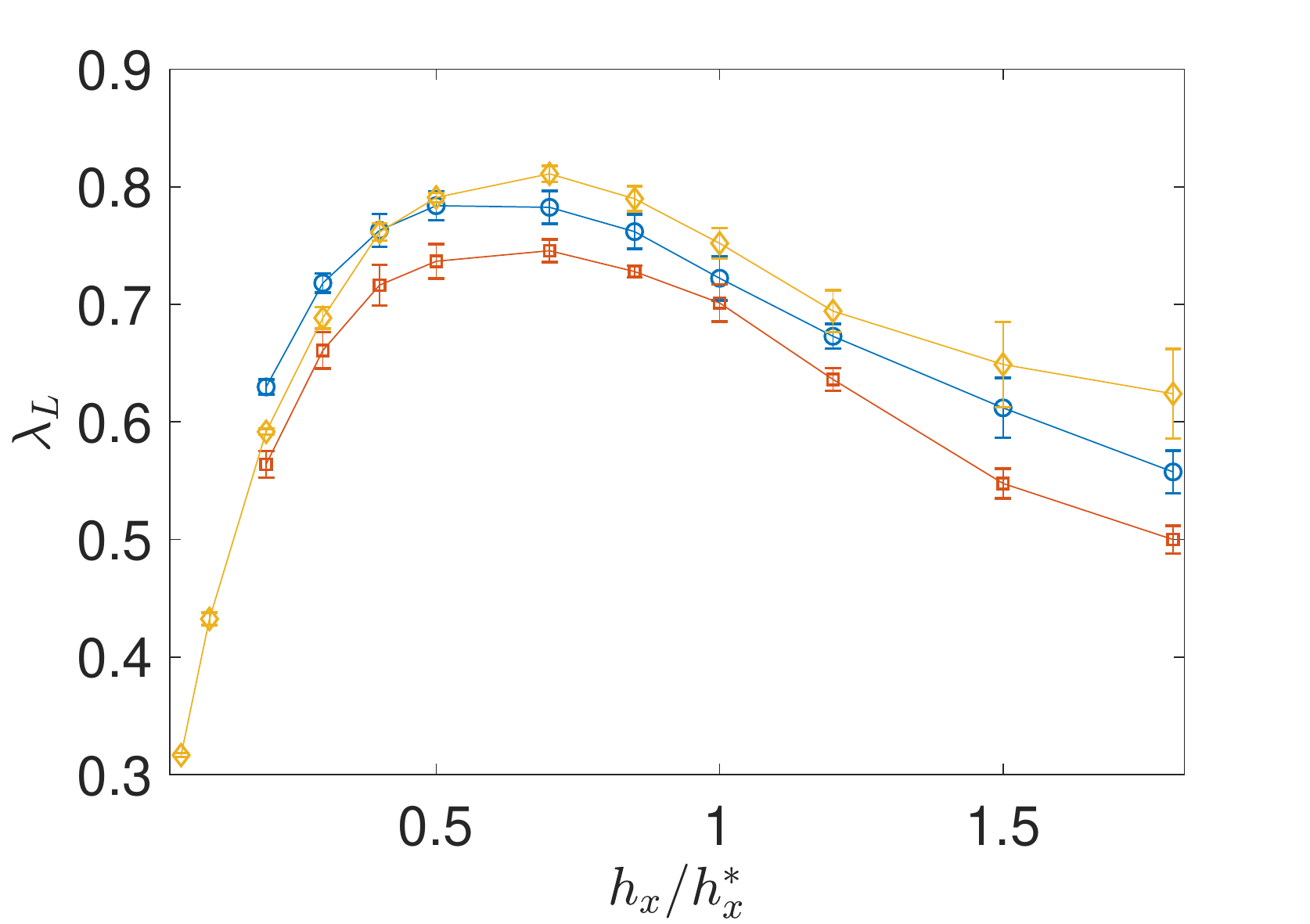}
	\end{minipage}
	\begin{minipage}[b]{0.49\textwidth}
		\centering
		$x=1$\\
		\includegraphics[height=5.5cm]{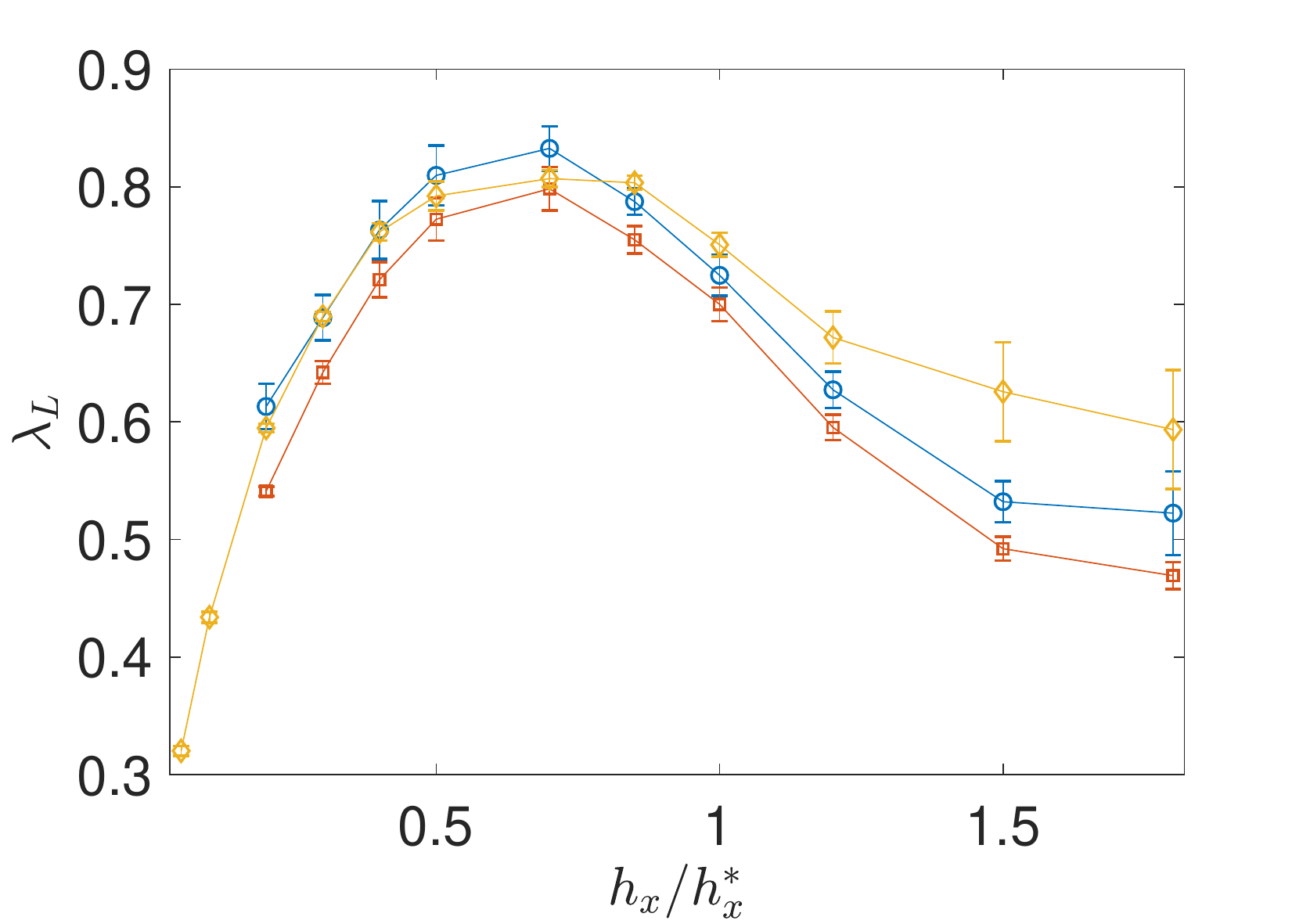}
	\end{minipage} 
	\begin{minipage}[b]{0.65\textwidth}
		\centering
				Average Lyapunov exponents\\
		\includegraphics[height=6cm]{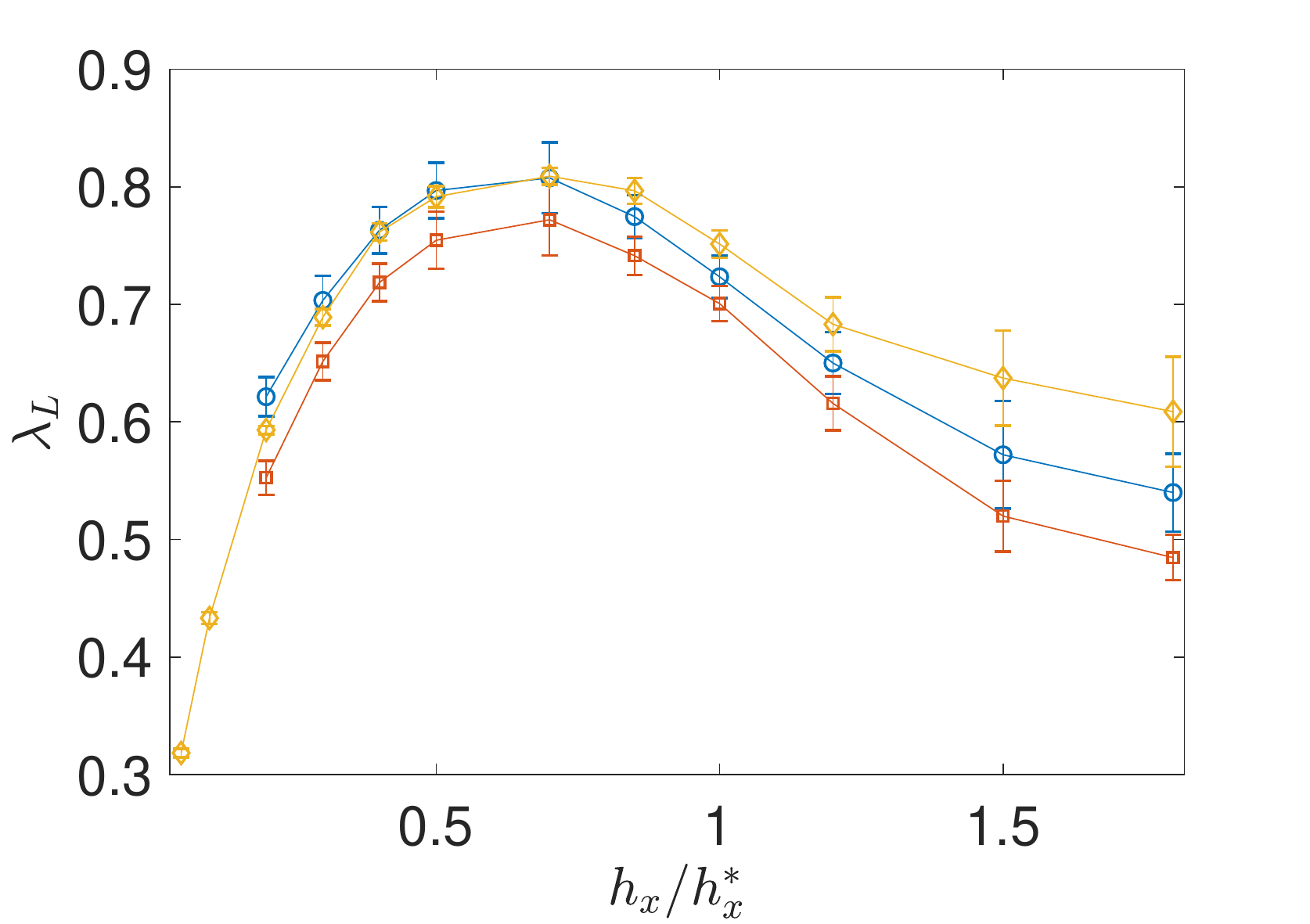}
	\end{minipage}
	\caption{A comparison between the classical Lyapunov exponents (yellow diamonds) and those obtained from the classical limit of the quantum model using the power law extrapolation (blue circles) and the exponential extrapolation (red squares) for a range of transverse field strengths.
The errorbars depict an estimate of the uncertainty involved in extracting these Lyapunov exponents as described in sections \ref{sec:quantum_Lyapunov} and \ref{sec:classical_lyapunov}.  
The longitudinal magnetic field is fixed at $h_z^*=0.5$. We find a good match between the exponents extracted from the power law approach and the classical exponents in the strongly chaotic domain.} 
	\label{fig:lambda_vs_hx_clasvsquant}
\end{figure}

\subsection{Matching classical and quantum Lyapunov exponents}
\label{section: Matching classical and quantum Lyapunov exponents}
We present the results for the Lyapunov exponents, obtained using both a classical and a quantum mechanical approach, over a range of magnetic fields including the previously studied strongly chaotic point in figure \ref{fig:lambda_vs_hx_clasvsquant}. We find that the procedure for extrapolating a classical Lyapunov exponent from the quantum data using a power law extrapolation provides a better estimate than the exponential extrapolation and that the procedure generally works best for parameter points associated to high values of the Lyapunov exponent.

Figure \ref{fig:lambda_vs_hx_clasvsquant} also supports the idea that as we move away from the strongly chaotic region towards the integrable lines (at $h_x=0$ and $h_x \to \infty$) there is a smooth crossover where the Lyapunov exponent tends towards 0 at the integrable lines. The strongest chaos is then the point where the Lyapunov exponent achieves its maximum.\footnote{The point at $(h^{*}_x,h^{*}_z)=(-1.05,0.5)$, which we have referred to as the strongly chaotic point, is not where the Lyapunov exponent achieves its maximum. It is simply a somewhat arbitrarily chosen point in the strongly chaotic regime that is chosen as a convenient point for comparison.} This smooth cross-over between chaos and integrability was observed in the spectral statistics in section \ref{subsection: quantum spectral statistics}.

\begin{figure}[th!]
\centering
\begin{minipage}{0.9\textwidth}
\centering
\includegraphics[width=0.47\textwidth]{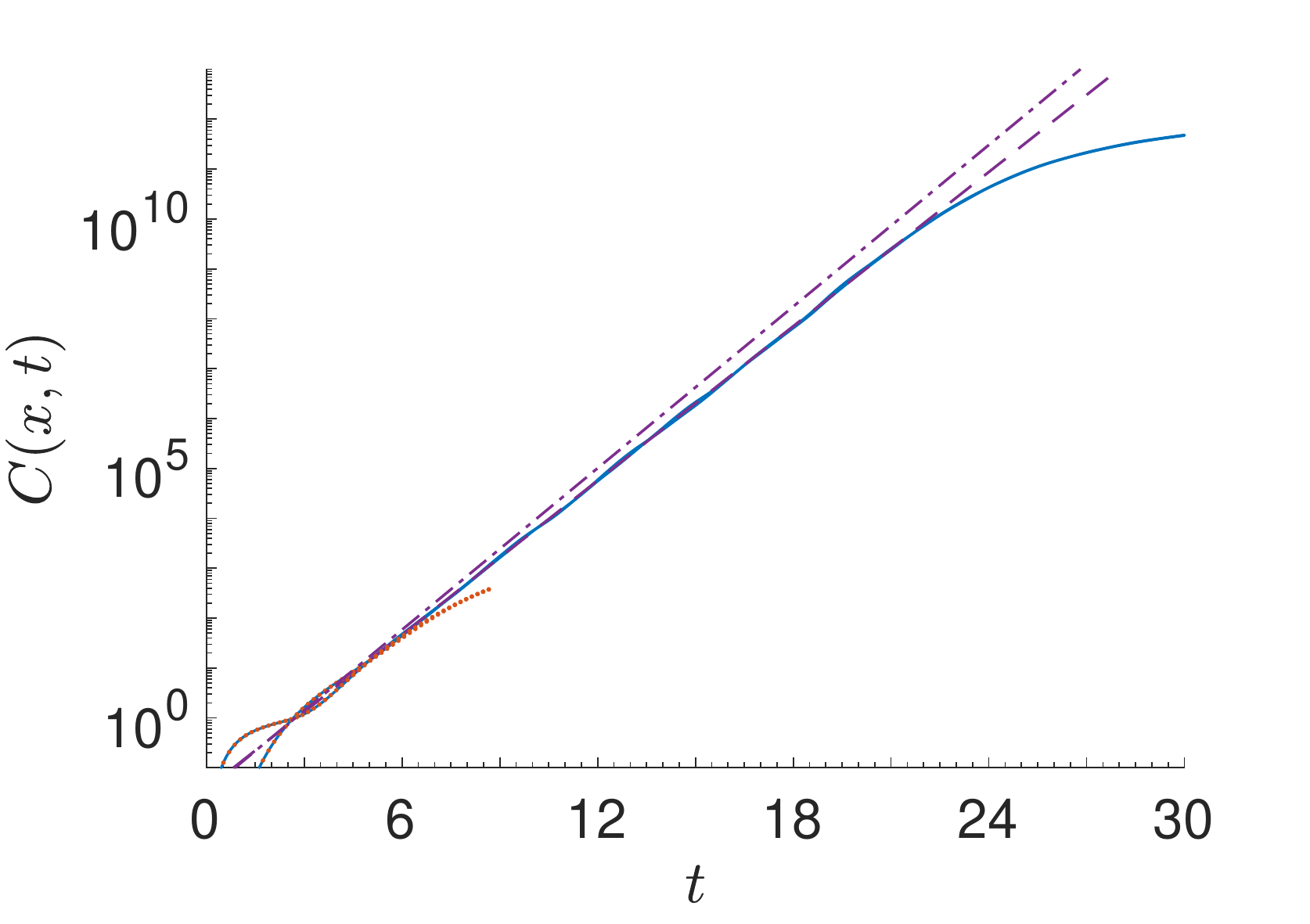}
\includegraphics[width=0.47\textwidth]{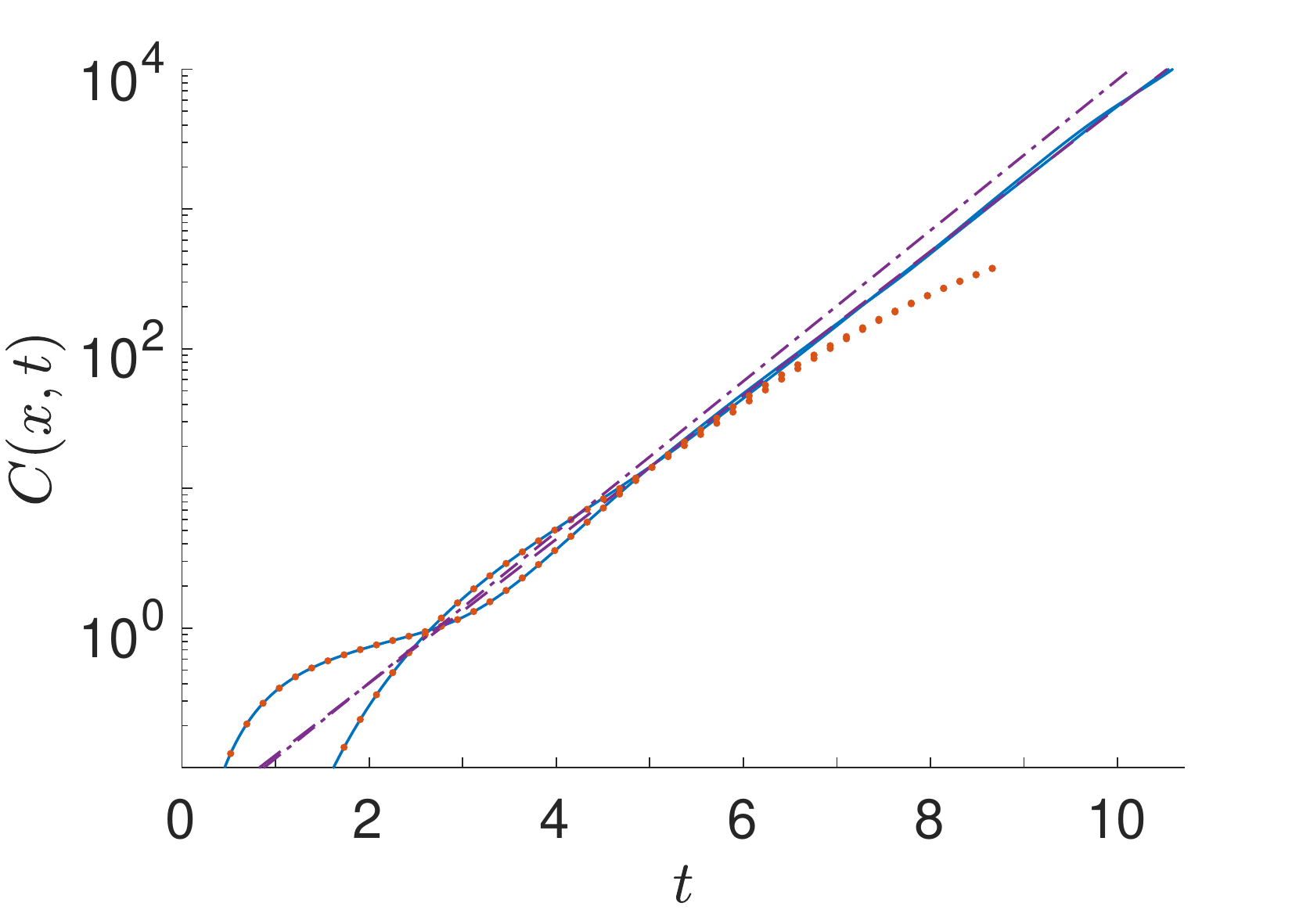}
$(h_x,h_z)=(0.2 h_x^*,h_z^*)$
\end{minipage}
\begin{minipage}{0.9\textwidth}
\centering
\includegraphics[width=0.47\textwidth]{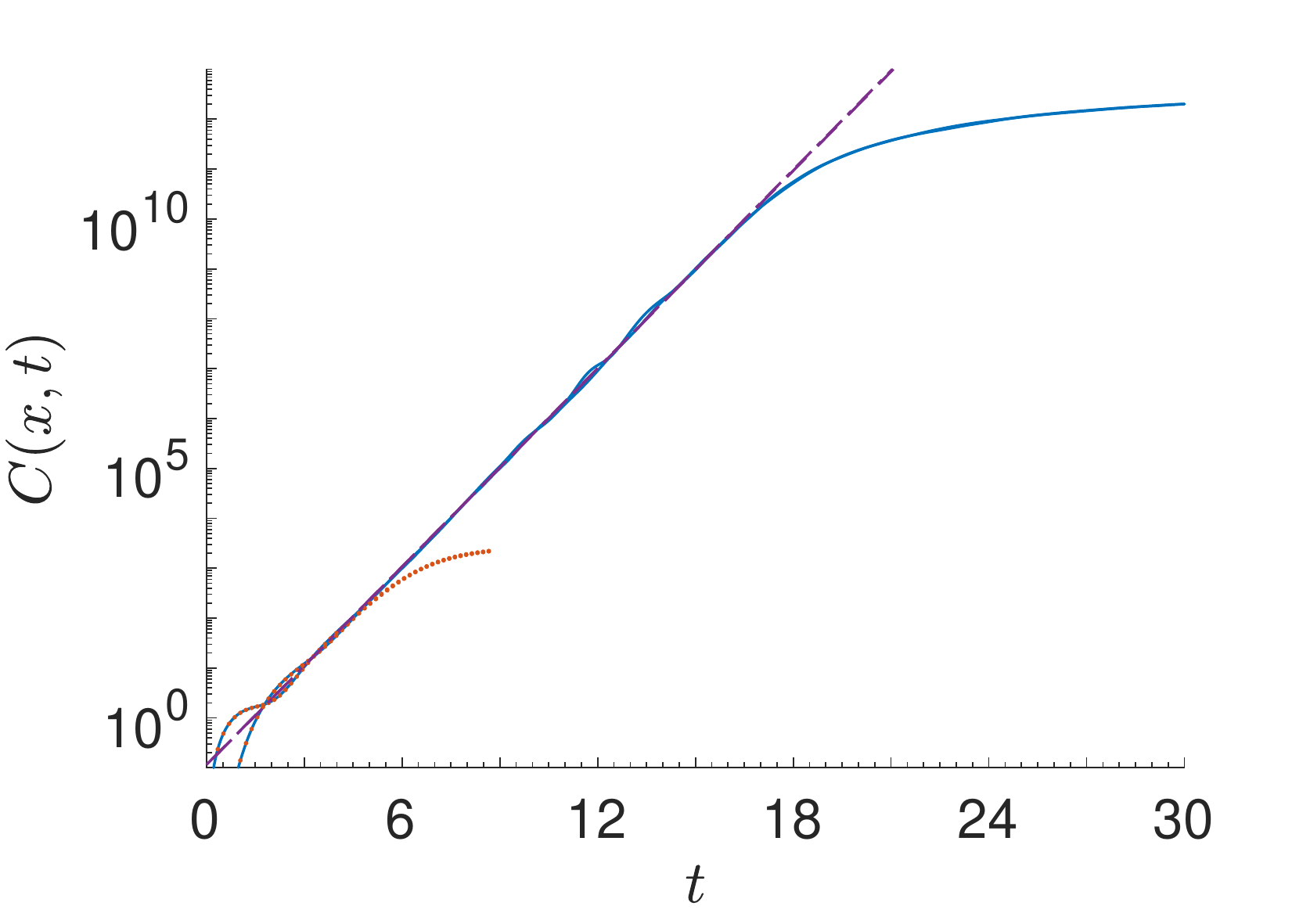}
\includegraphics[width=0.47\textwidth]{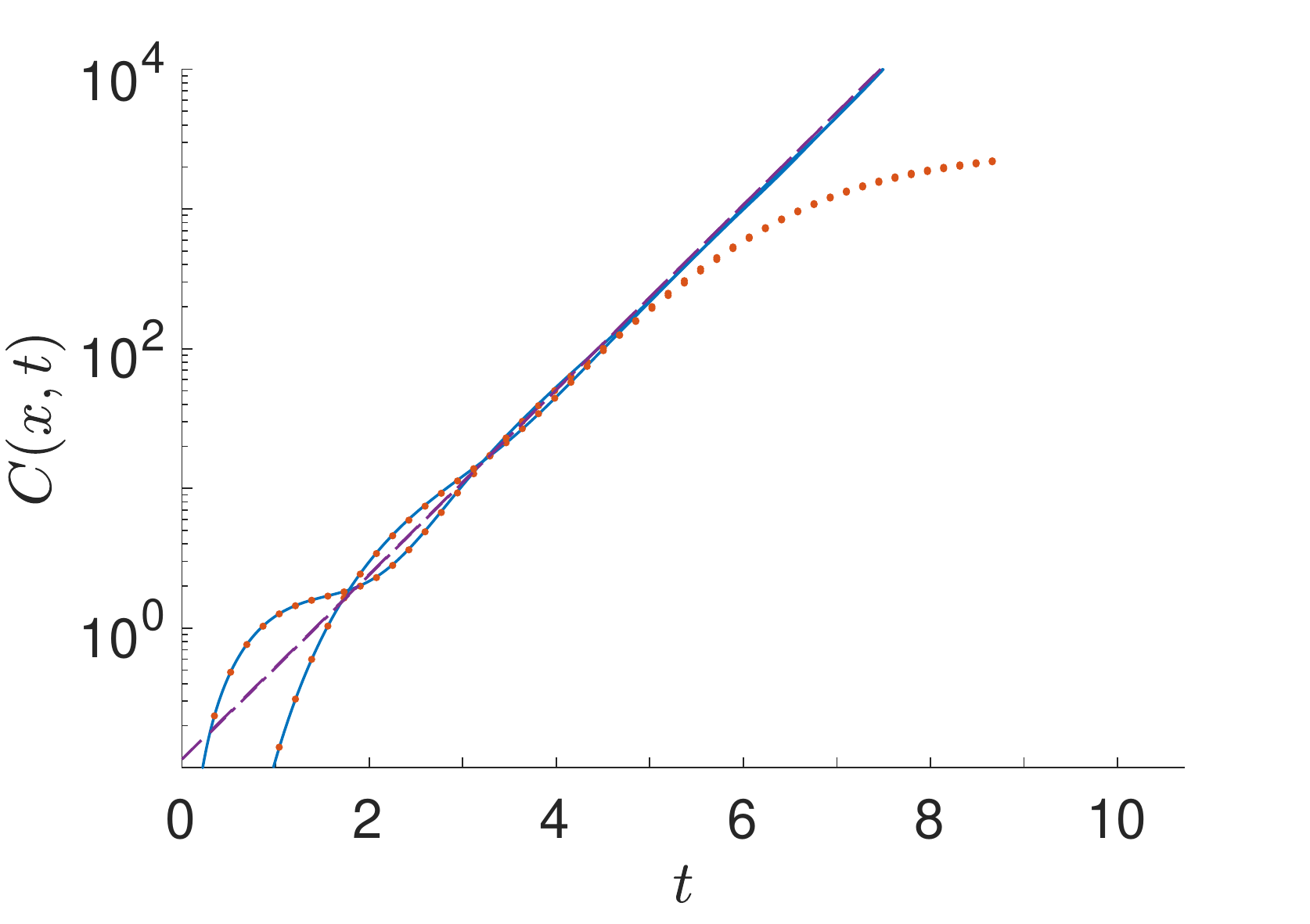}
$(h_x,h_z)=(0.4 h_x^*,h_z^*)$
\end{minipage}
\begin{minipage}{0.9\textwidth}
\centering
\includegraphics[width=0.47\textwidth]{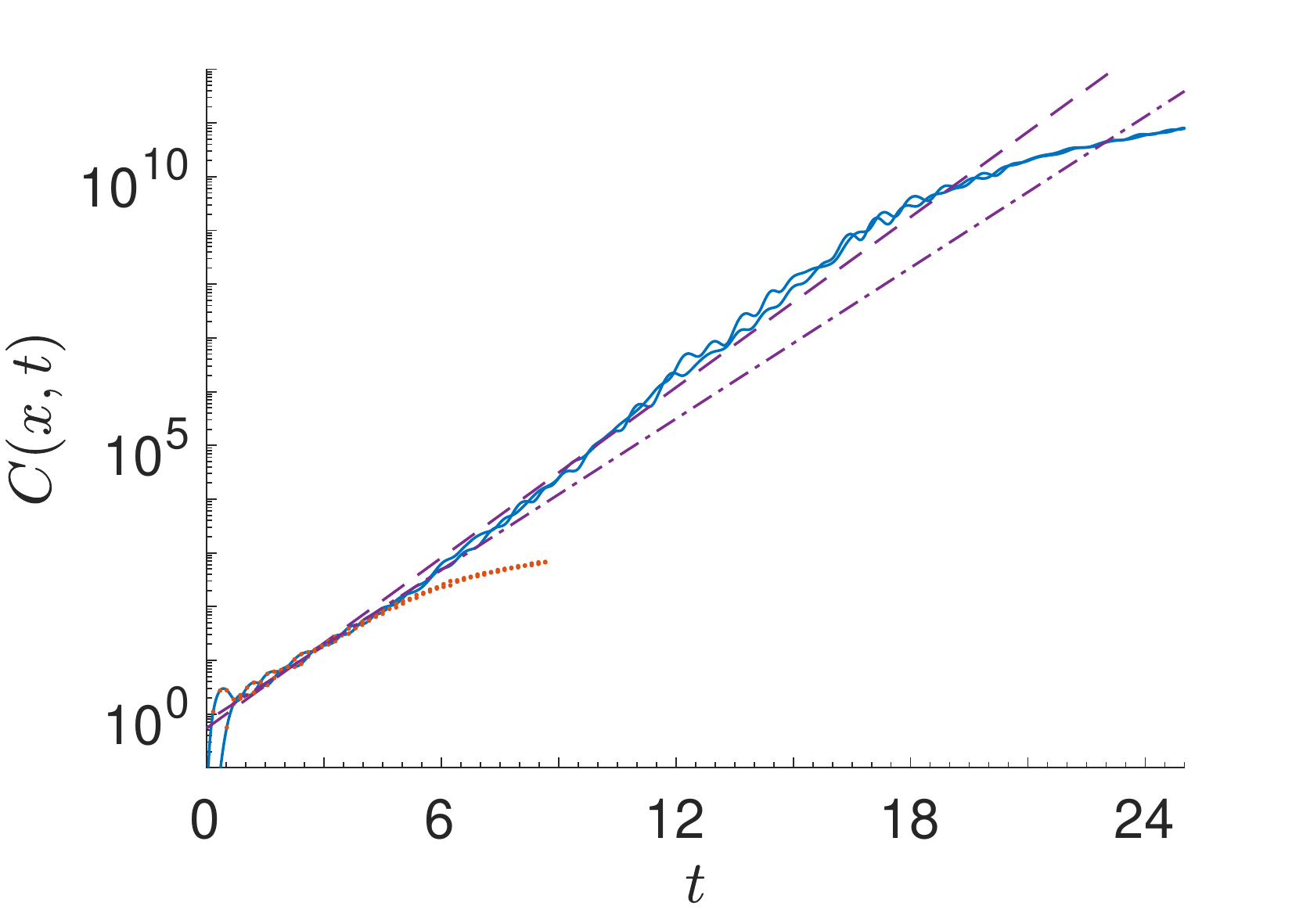}
\includegraphics[width=0.47\textwidth]{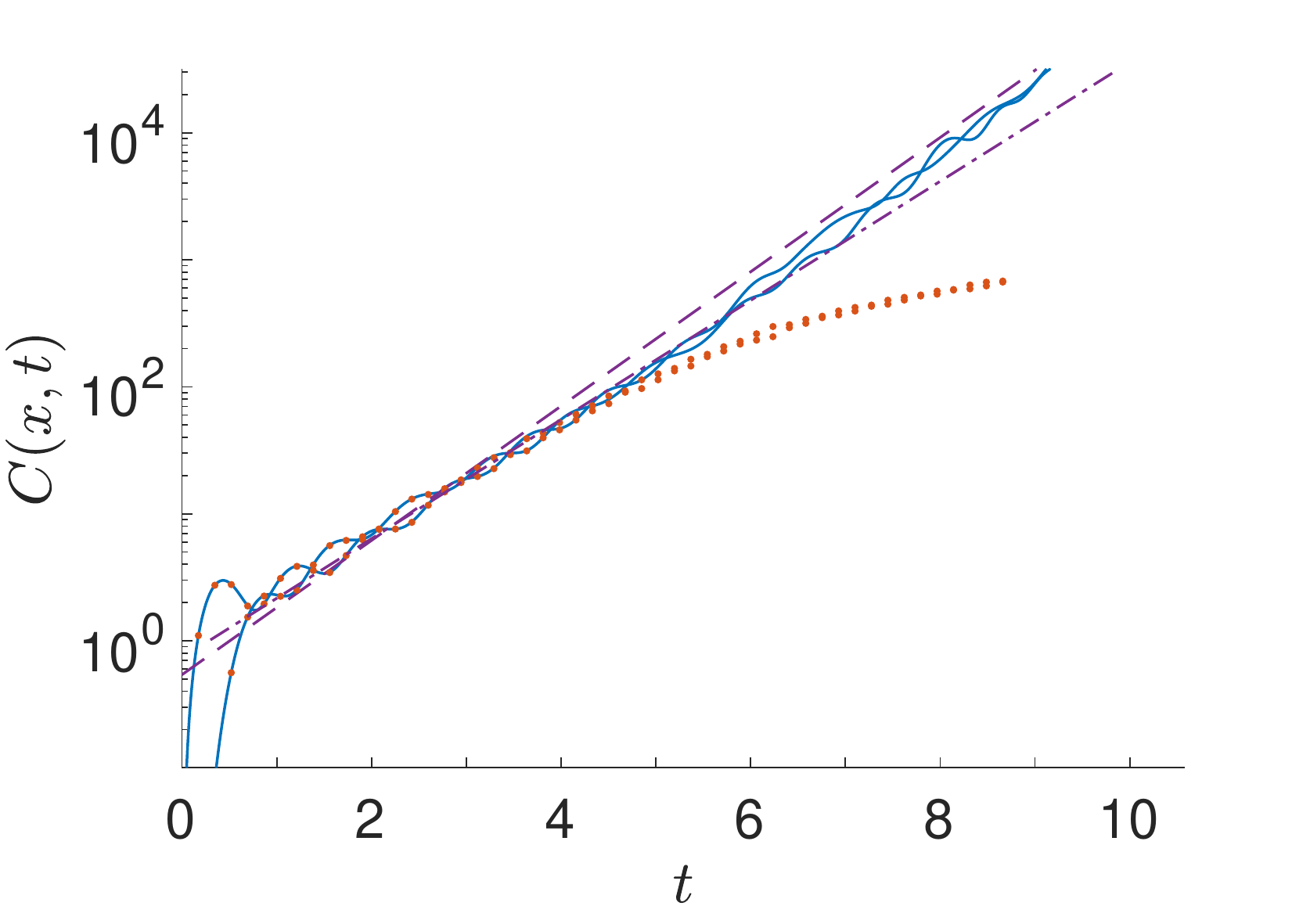}
$(h_x,h_z)=(1.8 h_x^*,h_z^*)$
\end{minipage}
\caption{The classical version of the commutator squared (blue solid) together with the the quantum commutator squared (times the usual factor of $j(j+1)$) at $j=61$ (red dotted), for a 2-site chain with both sites superimposed and for various strengths of the transverse magnetic field in our analysis. The purple dashed and dashed-dotted straight lines represent the fits to the classical and quantum Lyapunov regime, where respectively the average of the classical and the average of the power law extrapolated exponents at the two sites were used as a slope. The left figure shows the time evolution until the saturation of the Poisson bracket while the right figure emphasises the early time part.}
\label{Pbvst2}
\end{figure} 

As we move towards integrability the determination of a Lyapunov exponent is more difficult, as is reflected in the larger error bars, although the reason is different at large and small values of the magnetic field $h_x$.

At large magnetic field, 
the Poisson bracket has progressively larger fluctuations, as can be seen in figure \ref{Pbvst2}. As the deviations from linearity in the curve are seen to increase with time, it is not clear whether those larger fluctuations are intrinsic to these parameter points leading to the disappearance of a straight Lyapunov region, or if increasing the number of initial conditions used in our Monte Carlo estimate of the average over phase space would provide an exponential region that extends the early time part with the same exponent. These large fluctuations therefore lead to large error bars on the classical Lyapunov exponent. Since the observed curve is not quite linear and that its slope increases with time, the fact that the quantum curve only follows the early part of the Poisson bracket means that the estimate using the commutator squared is lower than the computed classical exponent.  

At small values of the magnetic field, the curve of the classical Poisson bracket has a clear Lyapunov region. However, there are damped oscillations at early times and the period of these oscillations as well as the time they take to decay 
increases as $h_x \rightarrow 0$. Although the time range for which the quantum curve tracks the classical curve also increases a little, as seen in figure \ref{Pbvst2}, it does not increase enough to avoid the situation where we have only one oscillation or less left to which we can fit in the quantum procedure. The average over different time intervals is therefore not effective in smoothing out the effects of these oscillations and the Lyapunov exponents cannot be reliably extracted. For this reason, the quantum Lyapunov exponents are only shown for $h_x \geq 0.2$ on figure \ref{fig:lambda_vs_hx_clasvsquant} while the classical Lyapunov exponents are shown for smaller $h_x$ to demonstrate that they continue to decrease. 

We conclude that the classical limit of the quantum Lyapunov exponent can be matched to (a corresponding notion of) the classical Lyapunov exponent in regimes where the dynamics is sufficiently strongly chaotic.

\section{Discussion}
In this work, we studied different probes of quantum chaos in a few-body spin chain. By analysing a classical limit of this model, we showed how this matches onto notions of classical chaos and demonstrated how a notion of a Lyapunov exponent can be matched across this classical limit.

In the quantum model, it was demonstrated that the commutator squared undergoes a region of exponential growth and that the size of this region grows as we move towards the classical limit. A Lyapunov exponent can be extracted from this region of exponential growth and we found that how closely the level spacing statistics follow the Wigner-Dyson surmise, a measure of quantum chaos, is correlated with the size of this Lyapunov exponent, a measure of classical chaos. 

In the classical model, the quantity which is the classical limit of the commutator squared was identified and analysed. From it a Lyapunov exponent was extracted that matches onto the limit of the corresponding quantum quantity. Moreover, the quantum Lyapunov growth was observed to coincide with the averaged Poisson bracket squared for a time related to the size of the local Hilbert space.

Unfortunately, due to numerical limitations, we were only able to study very short chains. Improved numerical methods, perhaps the use of matrix product operator or other tensor network techniques, may provide hope of studying longer chains, but since the emergence of chaos may require the dynamics to explore large regions of the Hilbert space including highly entangled states the ansatz used by these methods may not be very efficient.\footnote{We would like to thank Maarten Van Damme for collaboration on a preliminary investigation of this topic.}

Nonetheless, studying longer chains would be very interesting as it would allow the dynamics of operator spreading to be investigated. It would also be interesting to understand what happens to the observation made in \cite{1805.05376} that for long chains the commutator squared tends to obey the diffusive form of their ansatz. It may be that there is a competition between this long distance physics and the large $j$ physics we identified in this work.   

In this work, we focused on identifying a regime of exponential growth in the commutator squared and extracting the rate of this growth, but there are other features in the behaviour of the commutator squared that could be studied. In particular, after the period of exponential growth, there is a regime before saturation is reached which does not appear to vanish in the classical limit. Understanding the shape of the operator wavefront would require us to understand this regime better. It would also be interesting to better understand the duration of the exponential regime. The separation between the scale of the start of the exponential growth and the eventual saturation is roughly $\frac12 \log j(j+1)$ as we expected, but since the exponential growth breaks down well before saturation, it still remains to understand what precisely controls this transition. In particular, we noticed that as we move away from the region in parameter space where the chaotic behaviour is strongest, the linear regime becomes more difficult to identify and the duration of the exponential growth is less clear as the transition from the exponential growth into the near saturation behaviour is less sharp. It seems that while the corrections to the classical limit are suppressed as we increase $j$, they are enhanced as we move towards integrability.

\section*{Acknowledgements}
We would like to thank Vijay Balasubramanian, Oleg Evnin and Laurens Vanderstraeten for discussions. We especially thank Tim De Jonckheere for collaboration in the early stages of this project and Maarten Van Damme for collaboration on preliminary investigations of the applicability of matrix product operator methods to the study of the model we have presented.

This work is supported in part by FWO-Vlaanderen through projects G044016N and G006918N and by Vrije Universiteit Brussel through the Strategic Research Program ``High-Energy Physics.'' M.~D.~C.~is supported by a PhD fellowship from the Research Foundation Flanders (FWO).
C.~R.~also acknowledges support from the Simons Foundation (\# 385592) through the It From Qubit Simons Collaboration, the US Department of Energy contract \# FG02-05ER-41367, from the Natural Sciences and Engineering Research Council of Canada (NSERC) funding reference number PDF-517316-2018 and from a Postdoctoral Fellowship 

\providecommand{\href}[2]{#2}\begingroup\raggedright
\endgroup

\end{document}